\documentclass{aa}  

\usepackage{graphicx}
\usepackage{booktabs}
\usepackage{siunitx}
\usepackage[varg]{txfonts}
\usepackage{caption}
\usepackage{subcaption}
\usepackage{multirow}
\usepackage{hyperref}
\usepackage{enumerate}
\usepackage[switch]{lineno}

\DeclareRobustCommand{\ion}[2]{%
\relax\ifmmode
\ifx\testbx\f@series
{\mathbf{#1\,\mathsc{#2}}}\else
{\mathrm{#1\,\mathsc{#2}}}\fi
\else\textup{#1\,{\mdseries\textsc{#2}}}%
\fi}

\newcommand{\lya}{Ly$\alpha$}
\newcommand{\oii}{[O~{\sc ii}]}
\newcommand{\feii}{Fe~{\sc ii}}

\newcommand{\mgii}{Mg~{\sc ii}}
\newcommand{\ciii}{C~{\sc iii}]}
\newcommand{\hi}{H~{\sc i}}
\newcommand{\neiii}{Ne~{\sc iii}}
\newcommand{\hei}{He~{\sc i} }

\newcommand{\cluster}{RCS~0224-0002 }
\newcommand{\ratio}{$\mathcal{R}_\text{\mgii{}}^\text{\feii{}}$}
\newcommand{\mgiiblue}{W$_r$(2796)}

\newcommand{\mgiiew}{W$_r$(2796)}
\newcommand{\feiired}{W$_r$(2600)}

\newcommand{\nodata}{\multicolumn{1}{c}{...}}
\newcommand{\kms}{km~s$^{-1}$}
\newcommand{\zcluster}{0.773}

\begin{document} 
    
   \title{A $z\sim1$ galactic-scale outflow transversally mapped to $\sim 50$~kpc
   \\ through gravitational-arc tomography\thanks{Based on observations collected at the European Southern Observatory under ESO programme 094.A-0141 (MUSE).}}
   \titlerunning{A $z\sim1$ galactic-scale outflow mapped through gravitational-arc tomography}

   \author{J. A. Hernández-Guajardo
          \inst{1}\thanks{Corresponding author: \url{joaquin.hernandezg@uc.cl}},
          L. F. Barrientos\inst{1},
          S. López\inst{2},
          E. J. Johnston\inst{3},
          C. Ledoux\inst{4},
          N. Tejos\inst{5},\\
          A. Afruni\inst{6,7},
          M. Solimano\inst{3},
          E. Jullo\inst{8},
          H. Cortés-Muñoz\inst{2},
          P. Noterdaeme\inst{9},\\
          J. González-López\inst{1},
          A. Ormazábal\inst{1},
          F. Muñoz-Olivares\inst{1}, and
          T. A. M. Berg\inst{10}
          }

       \institute{
            Instituto de Astrofísica, Facultad de Física, Pontificia Universidad Católica de Chile, Av. Vicuña Mackenna 4860, 7820436 Macul, Santiago, Chile.
         \and
             Departamento de Astronomía, Universidad de Chile, Casilla 36-D, Santiago, Chile.
         \and
            Instituto de Estudios Astrofísicos, Facultad de Ingeniería y Ciencias, Universidad Diego Portales, Av. Ejército Libertador 441,
            Santiago, Chile. 
         \and
            European Southern Observatory, Alonso de Córdova 3107, Vitacura, Casilla 19001, Santiago, Chile. 
         \and
            Instituto de Física, Pontificia Universidad Católica de Valparaíso, Casilla 4059, Valparaíso, Chile. 
         \and Dipartimento di Fisica e Astronomia, Università di
Firenze, Via G. Sansone 1, 50019 Sesto Fiorentino, Firenze, Italy. 
        \and INAF - Osservatorio Astrofisico di Arcetri, Largo E.
Fermi 5, Firenze, I-50125, Italy. 
         \and Aix-Marseille Univ., CNRS, CNES, LAM, Marseille, France. 
         \and Institut d’Astrophysique de Paris, CNRS-SU, UMR 7095, 98bis bd Arago, 75014 Paris, France.
         \and  Department of Physics and Astronomy, Camosun College, 3100 Foul Bay Rd, Victoria,  B.C., V8P 5J2, Canada.}
    \authorrunning{J. A. Hernández-Guajardo et al.}

   \date{Received 5 September 2025 / Accepted 19 February 2026}

  \abstract
   {
   We report spatially resolved measurements of cool gas traced by \mgii{} and \feii{} absorption in the circumgalactic medium (CGM) of a star-forming galaxy at $z\sim1$ (G1). The fortuitous alignment of a background gravitational arc at $z\sim2.4$ provides seven closely spaced ($\sim6$~kpc) transverse sightlines along the minor axis of G1, probing its CGM out to $\sim50$~kpc. This geometry allows us to detect a galactic-scale outflow simultaneously in down-the-barrel and transverse directions, where blue-shifted \mgii{} absorption is detected along both types of sightlines, revealing a large-scale, collimated wind. We measure blue-shifted line-of-sight velocities of {$v_{\mathrm{los}} \sim 62 \text{--} 239$~\kms{}} and line-of-sight velocity dispersions {$\sigma_{\mathrm{los}} \sim 53 \text{--} 133$~\kms{}}, suggesting a structure dominated by bulk motion. De-projection of $v_{\mathrm{los}}$ along the minor axis indicates that the outflow material barely approaches the escape velocity and is likely to be gravitationally bound to G1. We constrain an outflow opening angle $\theta_c\sim 18^\circ \text{--} 25^\circ$, and a mass outflow rate of $ \dot{M}_{\mathrm{out}} \!\gtrsim\!0.06$~$M_\odot\,\mathrm{yr}^{-1}$, corresponding to a mass loading factor $\eta\!\gtrsim\!0.004$, estimated within $\sim10\text{--}50$~kpc ($\sim0.05\text{--}0.3$~$R_\text{vir}$) of the galaxy centre. Our measurements, combined with previous arc tomography data along the major axis, indicate that normalizing impact parameters by galaxy B-band luminosity substantially reduces scatter in the established anti-correlation between \mgii{} equivalent width and impact parameter, while also diminishing possible excess of \mgii{} equivalent width towards the minor axis.
   }
    \keywords{galaxies: evolution --
             galaxies: halos --
             galaxies: high-redshift --
             intergalactic medium -- 
             quasars: absorption lines 
               }
               
\maketitle

\section{Introduction}
\label{sec:introduction}

The evolution of galaxies across cosmic time is driven by processes of inflows, outflows, and re-accretion of gas, collectively known as the baryon cycle. The circumgalactic medium (CGM) serves as the interface for these processes \citep[see reviews by][]{2017ARA&A..55..389T, 2020ARA&A..58..363P, 2023ARA&A..61..131F}. Outflows driven by supernovae, stellar winds, and active galactic nuclei (AGN) are key for regulating star formation, distributing metals and energy, and enriching the CGM with metals. Observationally, metal-enriched outflows are nearly ubiquitous in star-forming galaxies \citep[e.g.][]{2009ApJ...692..187W, 2010ApJ...719.1503R, 2012ApJ...760..127M,2014ApJ...794..156R}, yet it remains unclear whether, and how efficiently, these outflows carry baryons to the CGM or beyond \citep[e.g.][]{2020A&ARv..28....2V,2021MNRAS.501.5575A}. Answering these questions is crucial for understanding galaxy evolution, feedback, and the missing baryons problem.

One of the most widely used tracers of cool metal-enriched gas in the CGM is the \mgii{}~$\lambda\lambda 2796, 2803$ doublet, and is arguably one of the best tracers of the cool ($T\sim10^4$~K) metal-enriched CGM \citep{1991ApJS...77....1L,1995qal..conf..139S, 1996ApJ...471..164C, 2010ApJ...714.1521C, 2011ApJ...743...10B, 2013ApJ...776..114N, 2018ApJ...868..142R, 2018Natur.554..493L, 2020MNRAS.499.5022D, 2021ApJ...913...50L}. The \mgii{} doublet is efficiently observable from the ground over a wide redshift range, approximately $0.1 \lesssim z \lesssim 6$, even at moderate spectral resolution \citep[e.g.][]{2012ApJ...761..112M, 2013ApJ...776..114N, 2021MNRAS.502.4743H}. In emission, \mgii{} traces resonantly scattered photons, often revealing extended halos or outflows \citep[e.g.][]{2011ApJ...728...55R, 2021ApJ...909..151B, 2024A&A...691A...5P}. In absorption, strong ($W_0>0.3$~\r{AA}) systems are associated to $\sim100$~kpc halos around star-forming galaxies \citep[e.g.][]{2010ApJ...714.1521C, 2011ApJ...743...10B, 2013ApJ...776..114N}, showing slow evolution with redshift \citep{2025A&A...703A..21L}. Together, these properties establish the \mgii{} doublet as one of the most effective tracers for investigating the structure, kinematics, and evolution of the cool CGM throughout cosmic time.

Two complementary observational strategies have been historically used to probe gas flows in \mgii{}-bearing CGM, in absorption: down-the-barrel and quasar absorption lines. Down-the-barrel spectroscopy leverages the stellar continuum of the absorber galaxies themselves to probe gas along the line of sight toward the galaxy, and has revealed that outflows are ubiquitous and collimated \citep[e.g.][]{2014ApJ...794..156R}, having velocities of $100 \text{--}1\,000$~\kms{} \citep[e.g.][]{2007ApJ...663L..77T,2009ApJ...692..187W,2009ApJ...703.1394M,2010ApJ...719.1503R,2011ApJ...743...46C,2012ApJ...760..127M,2012MNRAS.426..801B,2014ApJ...794..156R,2015ApJ...809..147H,2016ApJ...833...39S}. Transverse absorption-line studies against background sources---usually quasars---probe gas at much larger galactocentric distances, typically tens to hundreds of kiloparsecs, at CGM scales \citep[e.g.][]{2013ApJ...776..114N, 2021MNRAS.502.4743H}. These reveal that \mgii{} absorbers are more common and stronger around blue, star-forming galaxies than red, quiescent ones \citep{2010ApJ...714.1521C}, and often align with either the major or minor axis of the host galaxy \citep[e.g.][]{2012ApJ...760L...7K,2019ApJ...878...84M,2021ApJ...913...50L}, a bi-modality that suggests a connection to inflows and outflows. However, both approaches have intrinsic limitations; down-the-barrel spectroscopy allows a straightforward discrimination between inflows and outflows, but provides no information about their spatial extent. Transverse absorption, on the other hand, constrains the spatial extent of the gas---through an impact parameter---but it is challenging to unambiguously associate the observed absorption kinematics with outflows, inflows, or other CGM processes. 

In the era of 10m-class telescopes, deep integral field spectroscopy has opened the way for the use of extended background sources to resolve the CGM in absorption. Gravitational-arc tomography \citep{2018Natur.554..493L} is a technique developed for studying the high-redshift CGM, leveraging bright gravitational arcs as extended background sources. This method effectively mimics many contiguous background sightlines, enabling spatially resolved spectroscopy of intervening gas associated with individual halos. Arc tomography results have shown---for the first time directly---that the CGM is clumpy and anisotropic \citep{2018Natur.554..493L}, and have revealed the presence of extended co-rotating disks \citep{2020MNRAS.491.4442L,2021MNRAS.507..663T} and signatures of outflows \citep{2021MNRAS.507..663T, 2021ApJ...914...92M,2022MNRAS.517.2214F,2025ApJ...986..190S}. Furthermore, the contiguous spatial information uniquely provided by arc tomography data has also provided constraints on the kpc-scale structure of the cool and warm CGM \citep{2023A&A...680A.112A, 2024A&A...691A.356L,2025ApJ...986..190S}. The rich spatial information provided by arc tomography offers a unique window to probe the structure and propagation of metal-enriched flows at CGM scales.

In this work, we present gravitational arc tomography along the projected minor axis of an intervening galaxy at $z\sim1$, traced simultaneously by down-the-barrel and transverse absorption. This configuration arises from the fortuitous alignment between a bright background gravitational arc and a foreground star-forming galaxy at $z\sim1$. It enables us to probe the possible outflowing gas both against the galaxy itself and, critically, along its minor axis—where outflows are expected to be most prominent and evident—over seven nearly contiguous sightlines spanning a $\sim50$~kpc projected line.

Our paper is organized as follows. In Sect.~\ref{sec:observations}, we describe observations. In Sect.~\ref{sec:lens-model}, we describe the construction of the lens model of the foreground galaxy cluster. In Sect.~\ref{sec:G1-properties} we present the emission properties of the absorbing galaxy, based on \textit{HST} imaging and MUSE spectroscopy. The absorption strength and kinematic measurements of the intervening gas are described in Sect.~\ref{sec:absorption-properties}. Our main results are presented in Sect.~\ref{sec:results} and discussed in Sect.~\ref{sec:discussion} . Throughout this paper, we assume a $\Lambda{\rm CDM}$ cosmology with the following parameters: $H_0=70~{\rm km}~{\rm s}^{-1}~{\rm Mpc}^{-1}$, $\Omega_m=0.3$, and $\Omega_\Lambda=0.7$. Magnitudes are reported in the AB system.

\begin{figure*}[h!]
    \centering
    \includegraphics[width=\linewidth]{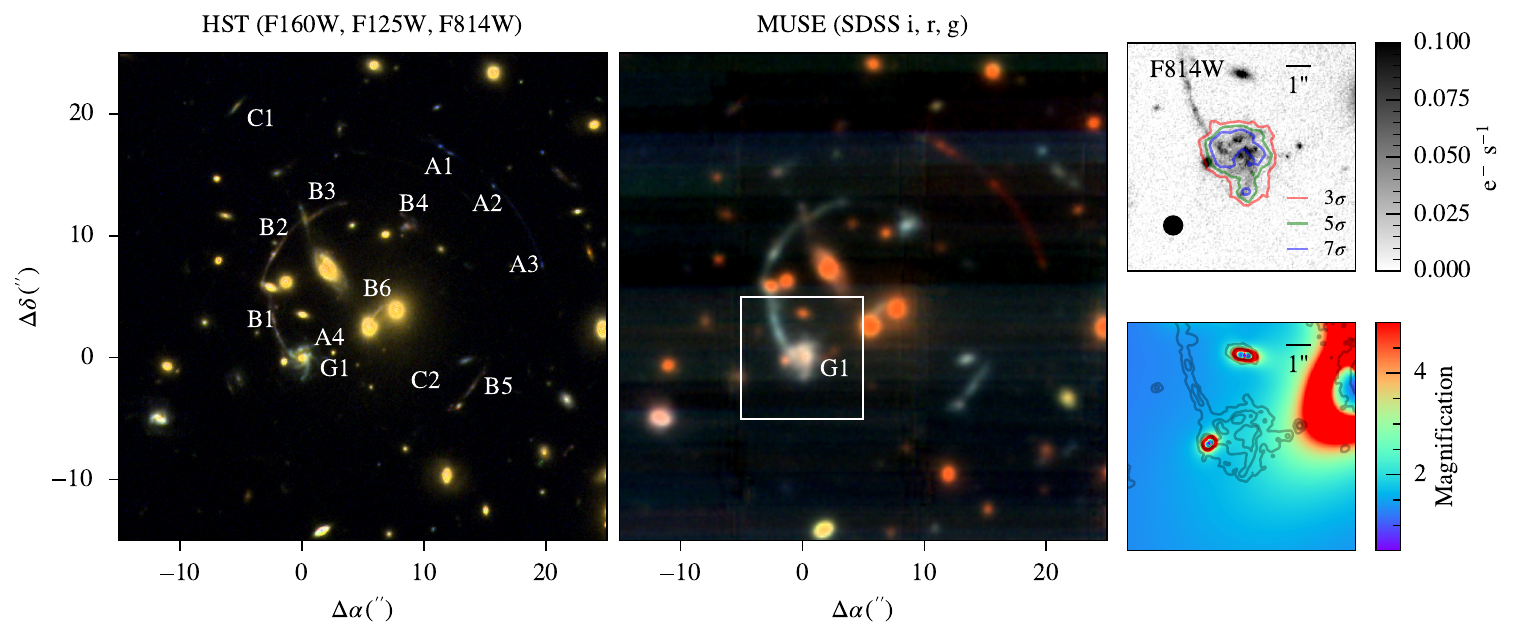}
    \caption{{\it Left}: \textit{HST} RGB (F160W, F125W, and F814W) image centred on the lensing cluster \cluster. Labels indicate the position of G1, and the lensed sources reported by \cite{2017MNRAS.467.3306S}, used for constraining the lens model. Redshifts of the lensed galaxies A, B, and C are $4.877$, $2.395$, and $5.498$, respectively. {\it Centre}: MUSE RGB image constructed by convolving the cube with the transmission curves of the broad-band SDSS \textit{i}, \textit{r}, \textit{g} filters. The white square, centred on G1, shows the size of the rightmost panels. {\it Top right}: \textit{HST}/F814W cutout centred on G1. Red, green, and blue contours show the $3\sigma$, $5\sigma$, and $7\sigma$ levels where its \oii{} emission is detected on the MUSE data. The typical PSF FWHM of 0.8\arcsec of the MUSE observations is represented by the diameter of the solid black circle. {\it Bottom right}: Magnification map centred on G1 showing the differential lensing effect at redshift of G1. Overlaid \textit{HST}/F814W contours delineate the distribution of light from surrounding sources. Sources enclosing areas of higher magnification correspond to cluster member galaxies.}
    \label{fig:G1-images}
\end{figure*}

\section{Observations}
\label{sec:observations}

\cluster is a galaxy cluster at $z_\text{cluster}=\text{\zcluster}$ discovered by the Red Cluster Sequence Survey \citep{2002AJ....123....1G}. The cluster acts as a gravitational lens for two background galaxies observed as extended and bright gravitational arcs at $z_A=4.877$ and $z_B=2.395$ \citep{2002AJ....123....1G,2007MNRAS.376..479S,2017MNRAS.467.3306S}, hereafter referred to as ``arc A'' and ``arc B'', respectively, and shown in Fig.~\ref{fig:G1-images}. A star-forming galaxy at $z\approx0.987$ (G1 from now on), is foreground and close in projection to arc B (see Fig.~\ref{fig:G1-images}). Absorption features of \mgii{} and \feii{} are detected at the redshift of G1 in the spectrum of arc B, which we interpret as cool CGM gas associated with G1, assuming the system is isolated.

\subsection{\textit{HST} imaging}

We used archival Hubble Space Telescope (\textit{HST}) imaging available in the Mikulski Archive for Space Telescopes (MAST), obtained under programs GO 14497 (PI: Smit) and GO 9135 (PI: Gladders). The system was observed with the Wide Field Planetary Camera 2 (WFPC2) in F606W ($6\times1.1$~ks), the Advanced Camera for Surveys (ACS) in F814W ($12\times1.1$~ks), and the Wide Field Camera 3 (WFC3) in F125W and F160W (each $2.6$~ks). All images were reduced using Drizzlepac \citep{2025drzp.book....3A}
, resampling onto a common 0.03~\arcsec~${\rm pix}^{-1}$ grid. A colour composite image using the F160W, F125W, and F814W filters is shown in Fig.~\ref{fig:G1-images}.

\subsection{MUSE spectroscopy}

In this paper, we used archival integral-field spectroscopy from VLT/MUSE \citep{2010SPIE.7735E..08B} (Programme 094.A-0141, PI: Swinbank), observed between UT November 13, 2014, and September 16, 2016. We used the ESO MUSE pipeline \citep{Weilbacher_2012} in the ESO Recipe Execution Tool (EsoRex v2.8.3) environment \citep{ESOREX} to reduce the data. As part of this routine, we created master bias, flat-field, and vacuum wavelength calibrations for each night using the associated raw calibrations provided by the Archive Facility. These master calibrations were applied to the raw science and standard-star observations as part of the pre-processing steps. 

The reduced standard star observations were used to create a flux calibration solution for each night, which was applied to the science frames. The sky background was directly measured and subtracted from the individual science frames using spaxels identified as representing the sky. The flux-calibrated and sky-subtracted exposures were then combined to produce a single data-cube. The residual sky contamination was removed using the Zurich Atmosphere Purge code \citep[ZAP, ][]{Soto_2016}. The final combined cube has a total exposure time of 3.75 hours with an average point spread function (PSF) full width at half maximum (FWHM) of $\approx0.8$\arcsec at 5500~\r{A}, and a 1$\sigma$ sensitivity of $1.6 \times 10^{-20}$ erg~s$^{-1}$~cm$^{-2}$~\r{A}$^{-1}$ at $\sim 5500$~\r{A}.

\section{Strong lensing model}
\label{sec:lens-model}

We constructed a strong lensing model of the foreground galaxy cluster to recover intrinsic spatial scales and physical properties in the absorber plane (i.e. at the redshift of G1). We guided the construction of our lens model by the one presented by \citet{2017MNRAS.467.3306S}. The model was constrained using the positions and redshifts of multiple lensed sources identified in \textit{HST} imaging and MUSE spectroscopy, including arcs A, B, and C, shown in Fig.~\ref{fig:G1-images}. The redshift of arc A ($z_A$=$z_\text{\oii{}}$ = $4.8760$), and arc B ($z_B$=$z_\text{\ciii{}}$ = $2.396$), were measured based on the nebular \oii{} and \ciii{} emission lines, respectively, by \citet{2017MNRAS.467.3306S}, on their MUSE observations, where they also performed a blind survey of \lya{} emitters, identifying the lensed source C, with $z_C$=$z_\text{\lya{}}$=$
5.498$. 

The details of the strong lens modelling are presented in Appendix \ref{sec:appendix-lens-modeling}. Our best-fit model achieves an image-plane RMS of $0.79$\arcsec, with a $\chi^2=13.7$ with $11$ free parameters. We use this model to reconstruct the geometry in the absorber-plane and to correct the observed fluxes for lensing magnification. Due to its proximity along the line of sight to the lensing galaxy cluster, G1 is subject to small (but non-negligible) distortion and magnification from the strong lensing effect. In Fig.~\ref{fig:G1-images} we show the magnification map around G1, from which we measure an average magnification factor of $\mu=1.89 \pm 0.07$ (areal average within the $3\sigma$ contour of the \oii{} emission, shown in Fig.~\ref{fig:G1-images}), a value that is robust to the specific lens model parametrization chosen for the analysis. For reconstructing the absorber-plane geometry, we applied the deflection matrices to the image-plane (i.e. as observed) positions, through the lens equation \citep[see, e.g.,][]{2018Natur.554..493L}. Impact parameters ($\rho$) are defined as the absorber-plane projected distances to the centre of G1. By sampling the deflection matrices from the posterior distributions of the model parameters, we estimate a $1\sigma$ confidence interval of the deflections, which propagates into an uncertainty of $~0.5$~kpc on $\rho$ in the absorber plane. Following \citet{2020MNRAS.491.4442L}, we arbitrarily adopt a total uncertainty of $1.5$~kpc on the impact parameters, corresponding to half of the binned spaxel size in the absorber plane (see Sect.~\ref{sec:absorption-properties}), in order to account for systematics (astrometry and lens model degeneracies).

\section{G1 properties}
\label{sec:G1-properties}

This section details the photometric, kinematic, and stellar population properties of G1, which are summarized in Table~\ref{tab:G1_properties}.

\subsection{Photometry and morphology}
\label{sec:G1_photometry}

We performed aperture photometry on the \textit{HST} imaging using \texttt{SExtractor} \citep{1996A&AS..117..393B}. The measured fluxes were corrected for lensing magnification, dividing the integrated fluxes by the average magnification, and for Galactic extinction, using the recalibration from  \citet{2011ApJ...737..103S} to the dust extinction maps from \citet{1998ApJ...500..525S}, assuming an F99 \citep{1999PASP..111...63F} dust extinction law. The rest-frame B-band luminosity was estimated from the \textit{HST} F814W band, corresponding to the B-band at $z\sim1$. We assumed a characteristic rest-frame B-band absolute magnitude $M_{B}^\star=-21.416$ at $z\sim1$ by interpolating the B-band luminosity function from \citet{2006ApJ...647..853W}. We find that G1 has a luminosity of $L_B =2.6~L_B^*$, above the characteristic luminosity at $z\sim1$.

\begin{figure}[h]
    \centering
    \includegraphics[width=\linewidth]{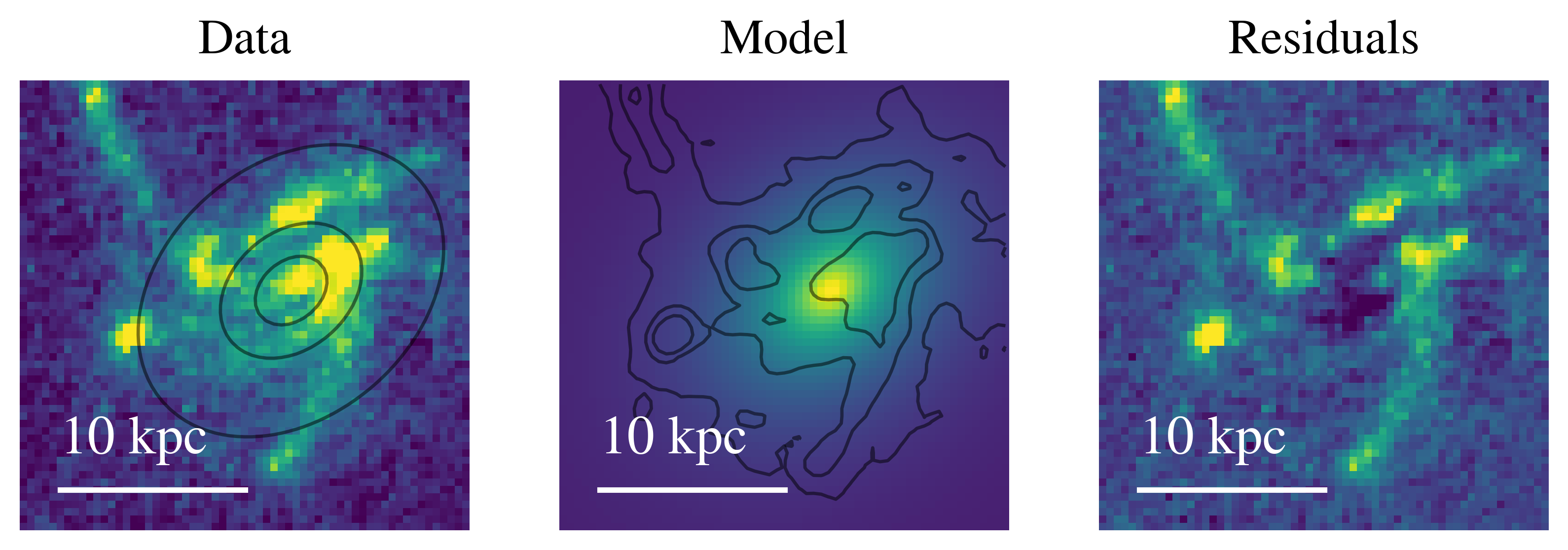}
    \caption{
    {\it Left:} De-lensed F814W cutout centred on G1. Black contours show representative flux levels of the best-fit Sérsic model. 
    {\it Centre:} Best-fit single-component Sérsic model obtained with \texttt{GALFIT}. Black contours indicate the same flux levels as in the left panel. Best-fit $R_e$, PA, $q$, and $n$ values are listed in Table~\ref{tab:G1_properties}. 
    {\it Right:} Flux residuals from the best-fit model.}
    \label{fig:galfit-G1}
\end{figure}

We used \texttt{GALFIT} \citep{2002AJ....124..266P} on the de-lensed F814W data to analyse the rest-frame B-band flux profile of G1, fitting a single Sérsic component with all parameters free. The best-fit Sérsic index is $n \approx 1.2\pm0.2$, consistent with an exponential disk profile. We estimated the inclination angle from the axis ratio $q$, assuming an infinitely thin disk ($\cos i = q$). The Sérsic index, axis ratio, position angle (PA), and effective radius ($R_e$) are listed in Table~\ref{tab:G1_properties}, while the best-fit model and residuals are shown in Fig.~\ref{fig:galfit-G1}. The residuals reveal spiral features and compact star-forming clumps, indicative of a clumpy, gas-rich disk. Such structures are typical of high-redshift disks undergoing active star formation \citep{2005ApJ...631...85E,2007ApJ...656....1L,2009ApJ...706.1364F,2015ApJ...800...39G, 2016ApJ...821...72S,2025MNRAS.536.3090K}.

\subsection{Stellar and halo properties}
\label{sec:G1_mass}

We used \texttt{BAGPIPES} \citep{2018MNRAS.480.4379C, 2019MNRAS.490..417C} to model the spectral energy distribution (SED) of G1, jointly fitting its MUSE and \textit{HST} photometry. The resulting stellar mass and SFR, and their respective errors were inferred from the posterior distributions. Additionally, the star formation rate (SFR) was independently estimated via the \oii{} luminosity, using the calibration from \citet{1998ARA&A..36..189K}. Both SFR estimates agree within $1\sigma$ and are listed in Table~\ref{tab:G1_properties}. With a specific SFR of $7.58 \times 10^{-10}$~yr$^{-1}$, G1 lies on top of the star-forming main sequence at $z \sim 1$ \citep{2023MNRAS.519.1526P}. 

Next, from the inferred stellar mass, we estimate a total halo mass $M_h$ using the stellar-mass-halo-mass (SMHM) relation from \citet{2010ApJ...710..903M}. Finally, we estimated the respective virial radius $R_\text{vir}$, maximum circular velocity $V_\text{circ}$, escape velocities $V_\text{esc}(r)$, and dark matter (DM) halo velocity dispersion $\sigma_\text{DM}$. All these properties are listed in Table~\ref{tab:G1_properties}, and the details of the methodology are described in more detail in Appendix~\ref{sec:appendix-sed-halo-properties}.

\subsection{Morpho-kinematic modelling}
\label{sec:G1_kinematics}

\begin{figure}[h]
    \centering
    \includegraphics[width=0.8\linewidth]{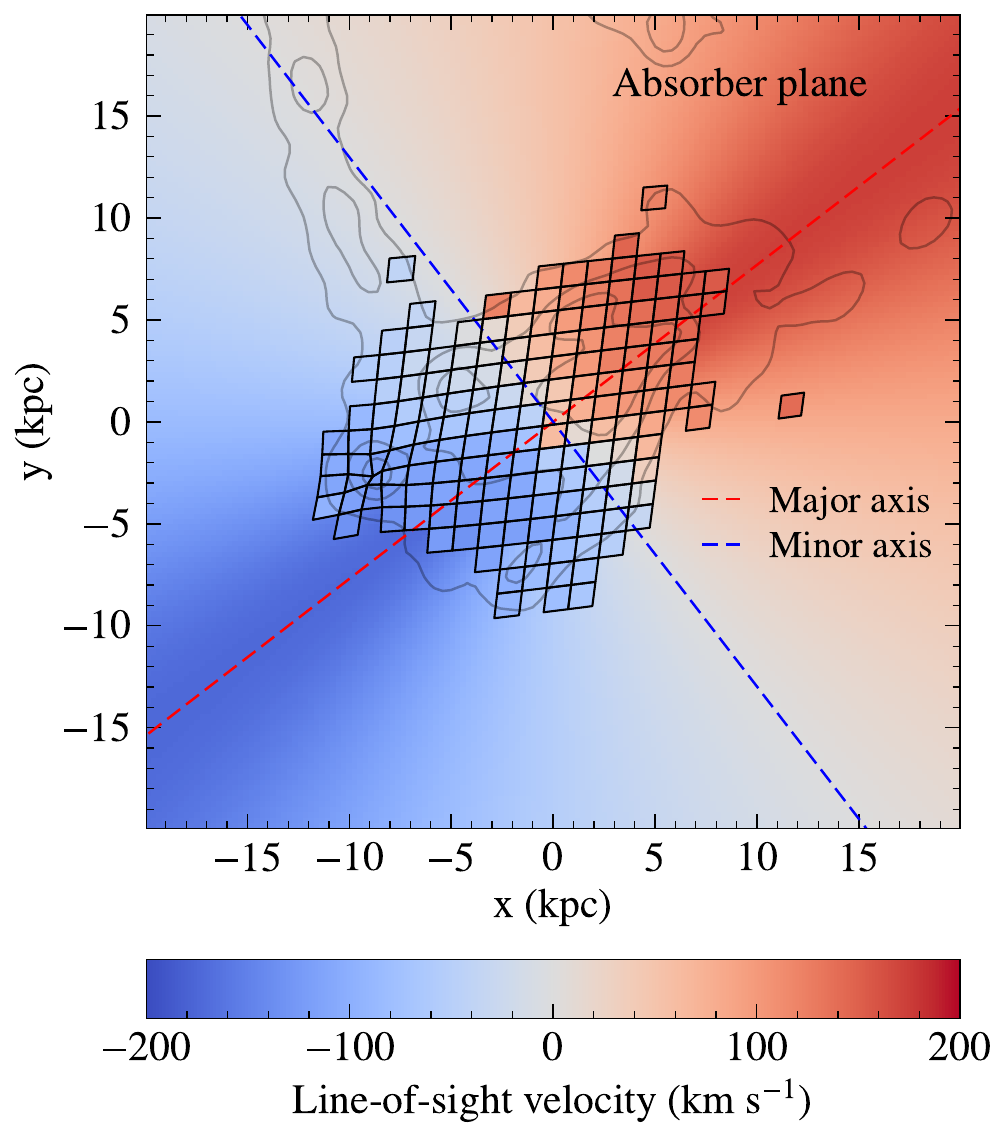}
    \caption{
    De-lensed (absorber-plane) centroid-velocity-map of the \oii{} emission from G1 and ERD model kinematics. Zero velocity is set at $z_\text{G1}$. Open rhomboids indicate native MUSE spaxels mapped into the absorber plane, with colours representing the measured line-of-sight \oii{} velocity centroid. The background colourmap shows the line-of-sight velocity field from the best-fit arctan disk model obtained with $\texttt{GalPak}^{\text{3D}}$. Dashed lines mark the kinematics major (red) and minor (blue) axes of G1. Grey contours trace the \textit{HST}/F814W flux distribution. Note that the spaxel-based velocities are shown for illustrative purposes only: $\texttt{GalPak}^{\text{3D}}$ does not extract velocities per spaxel, but fits a synthetic data-cube—convolved with the instrumental line spread function and PSF—which is directly compared to the observed cube.}
    \label{fig:oii-velocity}
\end{figure}

As shown in Fig.~\ref{fig:G1-images} (top right), the \oii{} emission of G1 is spatially resolved in the MUSE data, given the average $\sim 0.8$\arcsec PSF FWHM of the MUSE data, and the optical diameter of G1 being $\sim 2\arcsec$. We do not detect extended \oii{} emission beyond the stellar disk, which is expected given the exponential decrease in the surface brightness profile of ionized gas emission \citep[e.g.][]{2024NatAs...8.1602N}. Given the spectral resolution of MUSE (FWHM of $\sim50$~\kms{} at the rest-frame of \oii{} at $z\sim1$), we are able to resolve the \oii{} doublet and recover spatially resolved kinematics, as observed in the centroid-velocity-map of the \oii{} emission shown in Fig.~\ref{fig:oii-velocity} (black rhomboids). The \oii{} velocity field clearly resembles a typical rotating disk \citep[e.g.][]{2017A&A...608A...5G,2018MNRAS.477...18P}. 

We modelled the disk kinematics of G1 in the absorber plane using $\texttt{GalPak}^{3D}$ \citep{2015ascl...soft01014B}. We first de-lensed the MUSE data by applying the deflections from the lens model, reconstructing the absorber-plane emission on a regular grid with the native MUSE spaxel size of 0.2\arcsec, following the method described in \citet{2021MNRAS.507..663T}. The kinematics was modelled assuming an arctan rotation curve and the surface brightness with an exponential profile. The best-fit model parameters; systemic redshift ($z_\text{ \oii{}}$), maximum rotation velocity ($v_\text{max}$), turnover radius ($r_\text{to}$), effective radius ($R_e$), intrinsic gas velocity dispersion ($\sigma_\text{gas}$), inclination ($i$), and position angle (PA), are listed in Table~\ref{tab:G1_properties}. Hereafter, we adopt $z_\text{ \oii{}}$ as the systemic redshift of G1 ($z_\text{G1}=z_\text{ \oii{}}$).

The best-fit PA and $R_e$ of the \oii{} emission agree with those obtained from GALFIT within $2\sigma$. We use the $\texttt{GalPak}^{3D}$ PA to define the kinematic major and minor axes of G1, shown in Figs.~\ref{fig:oii-velocity} and \ref{fig:ew-map}. However, the $i$ derived from $\texttt{GalPak}^{3D}$ ($65.5^\circ$) differs by $\sim50\%$ from that inferred via the axis ratio ($42.3^\circ$). Since the kinematic model incorporates both morphological and kinematic constraints, it provides a more reliable estimate of the inclination; thus, we adopt the $\texttt{GalPak}^{3D}$--derived $i$ as the fiducial value. 

The $r_\mathrm{to}$ value is unusually small, likely because the \oii{} emission is weaker in the central regions of G1 and dominated by the clumps (Fig.~\ref{fig:G1-images}), limiting the model’s sensitivity to the inner velocity gradient. As a result, the modelling of the inner velocity structure may be more uncertain; however, the large-scale kinematic parameters—$v_\text{max}$, PA, and $i$—remain robust, as they are primarily constrained by the outer regions with stronger \oii{} emission. In the case of $\sigma_\mathrm{gas}$ the small value of $\sim10$~\kms{} is consistent with expectations for cool ionized gas in star-forming disks. 

Altogether, the morphological, kinematic, and stellar population properties of G1 point to a typical main-sequence disk galaxy at $z\sim1$, unperturbed by major interactions, undergoing sustained star formation. Indeed, we do not find any companion galaxy within $3000$~\kms{} of $z_{\rm G1}$ in the MUSE field of view ($\sim500~\text{kpc}\times500~\text{kpc}$ at $z_{\rm G1}$), and thus, we assume that G1 is an isolated galaxy. These properties provide essential context for interpreting the intervening metal absorption lines.

\begin{table}
    \tiny
    \caption{\label{tab:G1_properties} G1
 properties }
    \centering
    \renewcommand{\arraystretch}{1.2} 
    \begin{tabular}{l l}
        \toprule
        Magnification & $\mu$ = $1.89 \pm 0.07$ \\
        \toprule
        \multicolumn{2}{c}{\it From \oii{} emission and broad-band imaging (see Sect. \ref{sec:G1-properties})} \\
        \toprule
        Right Ascension &  ${\rm RA} = 02\text{h}24{\rm m}34.10{\rm s}$ \\
        Declination & ${\rm DEC} = {-}00{\rm d}02{\rm m}30.9{\rm s}$ \\
        Effective radius (Stars)$^{a,e}$ & ${\rm R}_e$ = $8.1 \pm 1.2~{\rm kpc}$ \\
        Position angle (Stars)$^{a,e}$ & PA = $130.7\pm2.3^\circ$ \\
        Axis ratio (Stars)$^{a,e}$ & q = $0.739\pm0.015$ \\
        Inclination (Stars)$^{a,e,f}$ & i = $42.3\pm1.3^\circ$ \\
        Sérsic index (Stars)$^{a,e}$ &  n = $1.2\pm0.2$ \\
        Apparent magnitude (F814W)$^{b,e}$ & $m_B$ = $20.97 \pm 0.16$ \\
        Absolute magnitude ($\sim$B band)$^{b,e}$ & $M_B$ = $-22.47 \pm 0.16$ \\
        Luminosity ($\sim$B band)$^{b,e}$ & $L_B$ = $2.6 \pm 0.4$ $L^\star_B$\\
        Stellar mass$^{b}$ & $\log (M_\star/M_\odot)$ = $10.29 \pm 0.01$\\
        SFR from SED$^{b}$ & $\mathrm{SFR}_{\mathrm{SED}}$ = $15.05 \pm 0.27$ ${\rm M}_\odot$ yr$^{-1}$ \\
        SFR from \oii{}$^{b,c}$ & $\mathrm{SFR}_\text{\oii{}}$  = $15 \pm 4$ ${\rm M}_\odot$ yr$^{-1}$\\ 
        Specific SFR$^{b}$ & $\mathrm{sSFR}_{\mathrm{SED}}$ = $\left( 7.58\pm0.29\right) \times 10^{-10}$ yr $^{-1}$\\
        \oii{} Luminosity$^{b}$ & $L_\text{\oii{}}$ = $\left(9.45\pm0.05\right)\times 10^{41}$ erg s$^{-1}$\\
        \toprule
        \multicolumn{2}{c}{\it From morpho-kinematical analysis of \oii{} emission (see Sect. \ref{sec:G1_kinematics})}\\
        \toprule
        Systemic redshift & $z_{\text G1}$ = $z_\text{ \oii{} }$ = $0.986790 \pm \left(4\times10^{-6}\right)$\\
        Maximum rotation velocity (Gas) & $v_{\mathrm{max}}$ = $191\pm2~{\rm km}\,{\rm s}^{-1}$  \\
        Turnover radius (gas)$^{a}$ & $r_{\text{to}}$ = $1.96 \pm 0.08~{\rm kpc}$  \\
        Position angle (gas)$^{a}$ & PA = $127.6 \pm 0.3^\circ$ \\
        Inclination (gas)$^{a}$ & $i$ = $65.5 \pm 0.3^\circ$ \\
        Velocity dispersion (gas)$^{a}$ & $\sigma_{\text{gas}}$ = $12.4 \pm 1.2\,{\rm km}\,{\rm s}^{-1} $ \\
        Effective radius (gas)$^{a}$ & ${\rm R}_e$ = $8.38 \pm 0.05~{\rm kpc}$ \\
        \toprule 
        \multicolumn{2}{c}{\it Halo properties}\\
        \toprule
        Halo mass & $\log( M_{\rm h}/M_\odot)$ = $12.1 \pm 0.1$\\
        Virial radius & $R_{\mathrm{vir}}$ = $159\pm13$ kpc \\
        Dark matter halo velocity dispersion & $\sigma_{\mathrm{DM}}\approx$104~\kms{} \\
        Halo maximum circular velocity$^{g}$ & $V_{\mathrm{circ}}$ $\approx200$~\kms{} \\
        Escape velocity$^{g}$ & $V_{\mathrm{esc}}(r=10~\text{kpc})$  $\approx560$~\kms{} \\
        Escape velocity$^{g}$ & $V_{\mathrm{esc}}(r=50~\text{kpc})$  $\approx460$~\kms{}\\
        \toprule
    \end{tabular}
        \tablefoot{\\
    \tablefoottext{a}{In the reconstructed absorber plane.}\\
    \tablefoottext{b}{De-magnified quantity using $\mu=1.89$.}\\
    \tablefoottext{c}{Obscured.}\\
    \tablefoottext{d}{Defined from the arctan rotation curve: $v(r)=v_\text{max} \arctan{(r/r_\text{to})}$.}\\
    \tablefoottext{e}{From \textit{HST}/F814W imaging.}\\
    \tablefoottext{f}{Assuming negligible disk intrinsic thickness.} \\
    \tablefoottext{g}{Assuming a NFW profile with concentration parameter 5.}
    }
\end{table}

\section{Intervening absorption properties}
\label{sec:absorption-properties}

Intervening \mgii{} $\lambda\lambda2796,2803$ and \feii{} $\lambda\lambda2586,2600$ absorptions at $z_\text{abs}\sim z_\text{G1}$ are detected over the B1 and B2 images of arc B, and over G1. To measure the absorption-lines properties, we apply $4\times4$ spatial binning to the native 0.2\arcsec\,MUSE spaxels, following \citet{2018Natur.554..493L}. This yields binned spaxels of 0.8\arcsec, spaced by at least one PSF FWHM, balancing the need to boost S/N and minimize spaxel cross-talk, while preserving sufficient spatial resolution to map the intervening absorptions. We define a mask enclosing G1, and the B1–B6 arc images, and retain only those binned spaxels with continuum $\text{S/N}\geq2$ near the \mgii{} spectral region for the subsequent analysis.

We fit the combined \mgii{} and \feii{} absorption spectrum using four Gaussian components, one per transition. The velocity centroid and dispersion were tied for all four lines, under the assumption that they trace gas with similar properties (e.g., \citealt{2001AJ....122..679C}). For each ion, the relative amplitudes of the two transitions were allowed to vary between the optically thin limit (set by the ratio of their oscillator strengths) and the optically thick limit (unity ratio). With $R\sim2000$ at $5500~\text{\r{A}}$, MUSE lacks the resolution to de-blend individual narrow kinematic components; hence, each transition is modelled with a single Gaussian component. The line-of-sight velocity ($v_{\rm los}$) and velocity dispersion ($\sigma_{\rm los}$) are then derived from the Gaussian centroid and width, respectively, with the latter corrected for instrumental broadening by subtracting the line spread function in quadrature. Rest-frame equivalent widths ($W_r = \sqrt{2\pi}A\sigma/(1+z_\mathrm{abs})$) were calculated in the continuum-normalized spectrum from the Gaussian parameters $\sigma$ and $A$ (standard deviation and amplitude, respectively) at the rest-frame of G1. The continuum level was independently measured for each transition on $100~\text{\r{A}}$ windows on both the blue and red sides of the absorption features. The $1\sigma$ uncertainties in $W_r$ ($\sigma_{W_r}$)  were derived through the propagation of the uncertainties in $A$ and $\sigma$. 

We define a significant detection as having a rest-frame equivalent width significance $W_r/\sigma_{W_r} \geq 2$ in at least two transitions. For spaxels meeting the S/N but lacking significant detections, we report 2$\sigma$ upper limits: $ 2 \sigma_{W_r}$, where $\sigma_{W_r}(1 + z_{\rm abs}) = \mathrm{FWHM}/\langle\mathrm{S/N}\rangle$, following \citet{2018Natur.554..493L}. We detect significant absorption on seven transverse and 13 down-the-barrel sightlines. While the rather loose S/N pre-selection ($\mathrm{S/N}\geq2$) ensured scanning of all arc spaxels, the final classification of detections is automated and based on the above significance criterion. In Figs.~\ref{fig:spatial_distrib_fits} and \ref{fig:spatial_distrib_fits_feii} we show the image-plane binned spaxels and their fitted Gaussian profiles (when $\text{S/N} \geq 2$), for \mgii{} and \feii{}, respectively. Only spaxels in arcs B1, B2, and over G1 show significant absorptions. The derived $v_{\rm los}$ values are measured with respect to $z_{\rm G1}$, and span a range of  $\approx\left[-60, -240\right]$~\kms{} across all sightlines.

\begin{figure*}[h!]
    \centering
    \includegraphics[width=0.9\linewidth]{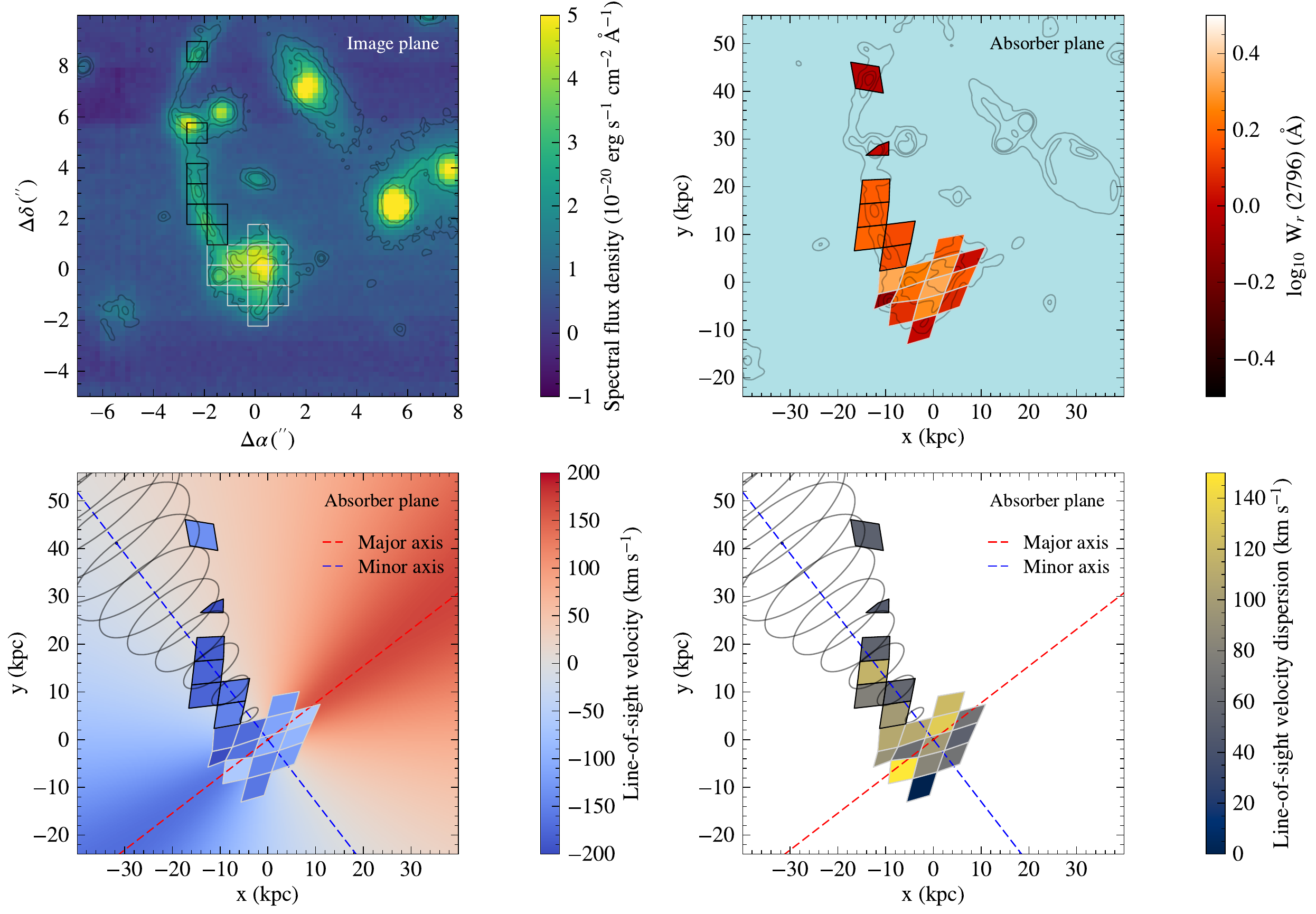}
    \caption{ {\it Top left}: Image-plane (i.e., as observed) MUSE pseudo-broad-band image (SDSS {\it r}) around G1 and the background gravitational arc B. Open squares show the $4\times4$ binned spaxels with detected \mgii{} absorption. Spaxels with black (white) edges correspond to transverse (down-the-barrel) sightlines. {\it Top right}: Absorber-plane (de-lensed) distribution of the measured \mgii{} equivalent widths $W_r(2796)$. Contours of the de-lensed \textit{HST}/F814W image of the field are shown for spatial reference. Cluster galaxies and other non-lensed sources should not appear in the absorber plane, but we included them in the contours to provide a spatial reference. {\it Bottom left}: Absorber-plane distribution of the  \mgii{} line-of-sight velocities with respect to $z_\text{G1}$. The background colourmap shows the projected line-of-sight velocity of a rotating extended disk model for G1. The kinematic major and minor axes of G1 are indicated with the red and blue dashed lines, respectively. Black solid ellipses show a schematic model of a conical outflow along the minor axis of G1, with opening angle $\theta_c=17.8^\circ$ and inclination $i=65.5^\circ$. {\it Bottom right}: Absorber-plane distribution of the \mgii{} line-of-sight velocity dispersion after subtracting the MUSE instrumental line spread function. Black curves and red and blue dashed lines indicate the same outflow model and kinematics axes of G1 as in the bottom left panel. }
    \label{fig:ew-map}
\end{figure*}

Absorber-plane impact parameters $\rho$ and azimuthal angles $\phi$ are defined relative to G1’s de-lensed centroid and receding major axis, respectively, following \citet{2021MNRAS.507..663T}. Table~\ref{table:ew_mg} details the absorption properties measured on each spaxel and Fig.~\ref{fig:ew-map} shows the spatial distribution of the spaxels with detected absorptions, in the image plane, and \mgiiblue{}, $v_\text{los}$, and $\sigma_\text{los}$, in the absorber plane, for the \mgii{} absorptions. These spatially resolved absorption maps serve as the basis for our results presented in the following section.

\section{Results}
\label{sec:results}

The absorptions detected over arc B, and G1, exhibit similar velocities. Given that G1 is an isolated galaxy, we hereafter assume G1 is responsible for the intervening \mgii{} and \feii{} absorption detected in the background arc B (see Sect. \ref{sec:absorption-properties}). Thus, spaxels probing the continuum of arc B provide transverse sightlines, probing the CGM of G1, whereas those intersecting G1's own stellar continuum provide down-the-barrel measurements of the intervening gas directly in front of G1. The transverse sightlines span impact parameters in the range $\sim10\text{--}50$~kpc, corresponding to $\sim0.05\text{--}0.3$~$R_\text{vir}$, and are closely aligned with the minor axis of G1 ($\phi\sim90^\circ$).

\subsection{Absorption strength}
\label{sec:results_absorption}

\begin{figure*}
    \centering
    \includegraphics[width=\linewidth]{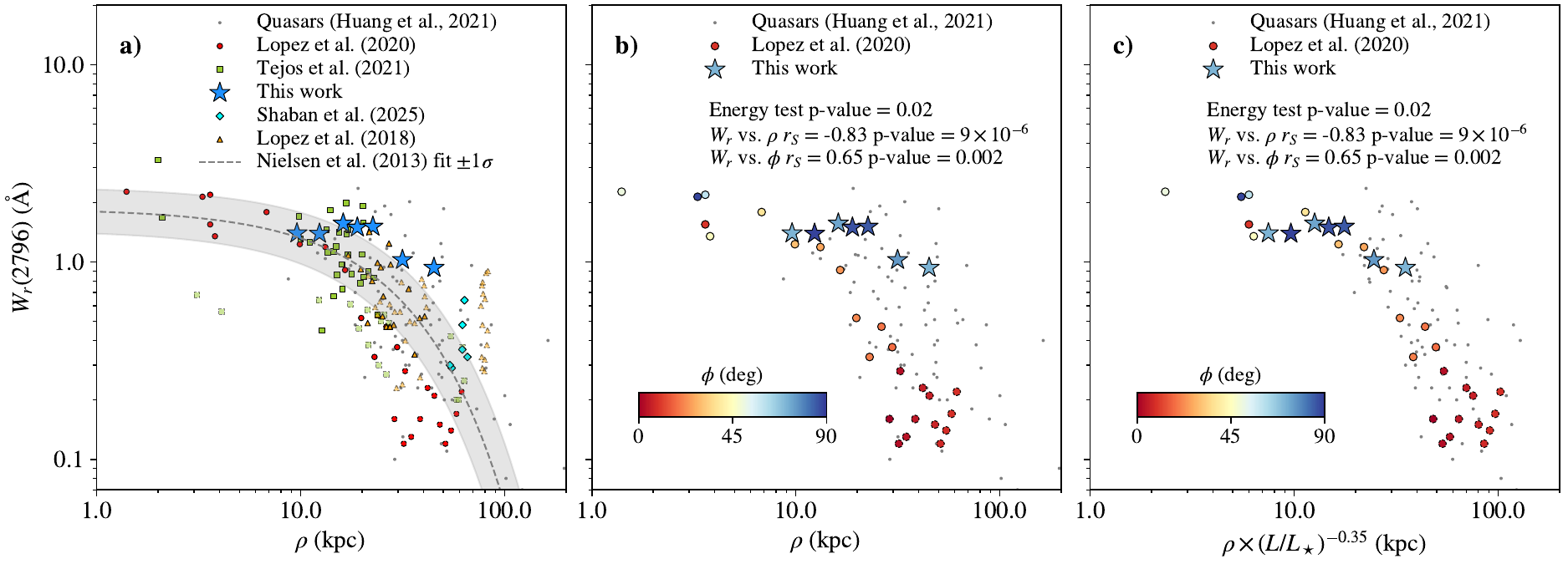}
     \caption{{\it Panel a}: Rest-frame \mgii{} equivalent width, $W_r(2796)$, as a function of impact parameter, $\rho$, to the absorber galaxy. Our arc tomography measurements are shown as blue stars. Black dots represent quasar-galaxy pairs at $z \lesssim 0.5$ from \citet{2021MNRAS.502.4743H}. The gray dashed line and shaded region correspond to the log-linear fit ($\pm1\sigma$) from \citet{2013ApJ...776..114N} based on the MAGIICAT sample at $0.07 \leq z \leq 1.1$. Coloured and semi-transparent symbols show individual measurements and upper limits, respectively, from gravitational arc systems from \citet{2018Natur.554..493L}, \citet{2020MNRAS.491.4442L}, \citet{2021MNRAS.507..663T}, and \citet{2025ApJ...986..190S}. {\it Panel b}: Comparison between the system presented in this work (probed primarily along the minor axis) and the system from \citet{2020MNRAS.491.4442L} (probed along the major axis). Symbols are coloured by azimuthal angle, $\phi$, of each sightline relative to the galaxy’s major axis. Overlaid are the results of statistical analyses, including the Energy test statistic and partial Pearson correlation coefficients between $W_r(2796)$ and $\rho$, and $W_r(2796)$ and $\phi$. {\it Panel c}: Same as Panel b, but with impact parameters scaled by galaxy B-band absolute magnitude using Eq.~\ref{eq:rho_scaled}, following \citet{2010ApJ...714.1521C} }
     \label{fig:ew_impact_parameter}
\end{figure*}

Figure~\ref{fig:ew_impact_parameter} (panel a) shows \mgiiew{} as a function of $\rho$. We compare our transverse \mgiiew{} measurements to both quasar-galaxy pairs \citep{2013ApJ...776..114N,2021MNRAS.502.4743H} and other arc tomography systems \citep{2018Natur.554..493L,2020MNRAS.491.4442L,2021MNRAS.507..663T,2025ApJ...986..190S}. Our measurements fall within the scatter of the quasar statistics \citep{2021MNRAS.502.4743H} and the log-linear fit from \citet{2013ApJ...776..114N}, and are also encompassed by the scatter of the ensemble of arc tomography systems \citep{2018Natur.554..493L, 2020MNRAS.491.4442L, 2021MNRAS.507..663T, 2025ApJ...986..190S}. We exclude the down-the-barrel measurement from the \mgiiew{}–$\rho$ analysis because impact parameters are not defined for this geometry and these sightlines cross half of the CGM from the inside out. 

A Pearson correlation analysis for \mgiiew{} vs. $\rho$  yields a Pearson correlation coefficient $r_P = -0.75$, and p-value $p = 0.026$, confirming a statistically significant anti-correlation, a trend that aligns with previous quasar and arc-based measurements \citep[e.g.][]{2013ApJ...776..115N, 2021MNRAS.502.4743H, 2018Natur.554..493L}. In terms of normalization, our \mgiiblue{}, measured along the minor axis, lie near the upper envelope of the \mgiiblue{}–$\rho$ space of the combined arc tomography systems over the $\sim10\text{--}50$~kpc range. 

\subsection{Absorption kinematics}
\label{sec:results_kinematics}

Both down-the-barrel and transverse absorptions along G1’s minor axis exhibit comparable velocities with down-the-barrel line-of-sight velocities spanning $v_{\text{los}, \text{dtb}}=\left[-63, -239\right]$~\kms{} and transverse velocities spanning $v_{\text{los}, \perp}=\left[-134, -197\right]$~\kms{} (Fig.~\ref{fig:ew-map} and Table~\ref{table:ew_mg}). The blue-shifted down-the-barrel absorptions alone already indicate the presence of outflowing gas \citep[e.g.][]{2014ApJ...794..156R}, and the coherent blueshifts along G1's minor axis sightlines indicate this outflow extends to tens of kpc (i.e. CGM scales; alternative scenarios are discussed in Sect.~\ref{sec:other-scenarios}). The velocity range probed by $v_{\text{los}, \text{dtb}}$ is comparable to that spanned $v_{\text{los}, \perp}$, although $v_{\text{los}, \text{dtb}}$ extends to slightly lower velocities. This likely reflects contributions from absorbing components at or near the systemic velocity of G1, which are unresolved in our data.

Interpreting the CGM kinematics requires placing the absorption velocities in the context of the host galaxy dynamics. Previous studies have shown that strong \mgii{} absorbers may either co-rotate with the stellar disk of the host galaxy \citep[e.g.][]{2017ApJ...835..267H,2019MNRAS.485.1961Z} or trace collimated outflows at kiloparsec scales \citep[e.g.][]{2014ApJ...792L..12K,2020MNRAS.492.4576Z}. In most galaxy–quasar pair studies, however, these scenarios remain difficult to distinguish because each galaxy is usually probed by a single transverse sightline. Gravitational-arc tomography overcomes this limitation by providing multiple contiguous sightlines through the CGM. Indeed, previous arc tomography studies have revealed cases of CGM gas co-rotating with the stellar disk \citep{2020MNRAS.491.4442L,2021MNRAS.507..663T,2022MNRAS.517.2214F}.

To test the co-rotation scenario, we constructed an extended rotating disk (ERD) model following \cite{2021MNRAS.507..663T}, based on the kinematic model of G1. Starting from the parameters of the arctan model (PA, $r_t$, $v_{\rm max}$, $i$), we extrapolated the projected velocity field beyond the stellar disk. The resulting ERD line-of-sight velocity field is shown along with the \oii{} centroid-velocity-map in Fig.~\ref{fig:oii-velocity}, and with the absorption kinematics in Fig.~\ref{fig:ew-map} (panel c). The absorption velocities clearly differ from those predicted by the ERD model, indicating that the gas kinematics is not explained by disk co-rotation. Although the ERD model does not account for beam-smearing effects introduced by the PSF, this limitation does not affect the qualitative conclusion, as the velocity offsets between the data and the model are large and systematic.

Finally, the absorption kinematics exhibit a remarkable degree of coherence. $v_\text{los}$ and $\sigma_\text{los}$ show little scatter across the transverse sightlines, with the scatter in $v_\text{los}$ of  $\sigma_{v,\text{ los}}\approx20$~\kms{}, and the scatter in $\sigma_{\text{los}}$ of $\sigma_{\sigma, \text{los}}\approx24$~\kms{}. Notably, the $\sigma_{v,\text{los}}$ value, is a factor of $\sim4$ smaller than the mean  velocity dispersion $\bar{\sigma}_\text{los}\approx72$~\kms{}. Together, these properties point to a coherent, large-scale, blue-shifted structure extending across tens of kpc, consistent with the presence of a galactic-scale outflow pointing toward the observer.

\subsection{Outflow interpretation}

\subsubsection{Outflow geometry}
\label{sec:geometry}

The predominantly blue-shifted \mgii{} absorptions relative to G1’s systemic velocity argue strongly against an isotropic outflow. In such a scenario, we would expect symmetric absorption profiles in the transverse direction—combining contributions from both the approaching and receding sides of the outflow bubble—and velocity centroids close to the systemic redshift. However, none of the transverse sightlines show significant absorption at redshifted velocities (Fig.~\ref{fig:spatial_distrib_fits}). This, together with the blue-shifted down-the-barrel absorptions,  indicates that the absorbing material lies entirely on the near side of the galaxy and that the outflow is collimated rather than spherical.

To understand the spatial configuration of the outflowing gas, we model the outflow as a bi-conical structure originating from the galaxy centre \citep[e.g.,][]{2012MNRAS.424.1952G}. We use the azimuthal angles of the transverse sightlines (Table~\ref{table:ew_mg}) to constrain the outflow opening angle. To intercept all transverse absorptions, the cone must have a minimum half-opening angle of $\theta_\text{c,min} = 17.8^\circ$. Assuming uniform gas density within the cone \citep[following][]{2014MNRAS.438.1435C} and accounting for galaxy inclination, the geometry must satisfy $i + \theta_c \lesssim 90^\circ$ to avoid redshifted transverse absorptions, which restricts the maximum opening angle to $\theta_\text{c,max} \sim 24.5^\circ$. A schematic of the inferred geometry is shown in Fig.~\ref{fig:ew-map}.

This narrow opening angle ($\sim18^\circ\text{--}25^\circ$) lies below typical statistical averages of $\sim50^\circ$ from outflow detection rates in inclined galaxies \citep{2012ApJ...760..127M, 2014ApJ...794..156R}, but aligns with individual quasar-based constraints of $\sim6^\circ\text{--}56^\circ$ \citep{2012MNRAS.424.1952G, 2019MNRAS.490.4368S} and arc tomography estimates ($\theta \sim 36^\circ$; \citealt{2022MNRAS.517.2214F}). Extended background sources sampling a wider range of azimuthal angles \citep[e.g.,][]{2021MNRAS.507..663T, 2022MNRAS.517.2214F} will be essential to constrain outflow geometries more tightly.

\subsubsection{Outflow velocity profile}
\label{sec:outflow-kinematics}

The unique alignment of the background arc with G1’s minor axis allows us to trace outflow velocities all the way from the disk of G1 to CGM scales. Assuming a collimated wind \citep{2014ApJ...794..156R,2005ApJ...621..227M,2009AIPC.1201..142W,2012ApJ...760..127M} outflowing perpendicular to the disk of G1, transverse impact parameters were de-projected into a radial distance, $r=\rho/\sin{i}\approx 1.1\times\rho$, to the centre of G1. Line-of-sight velocities were likewise de-projected as $v_\text{depr}=v_\text{los}/\cos{i}\approx2.4\times v_\text{los}$.

\begin{figure}[h!]
    \centering
    \includegraphics[width=0.9\linewidth]{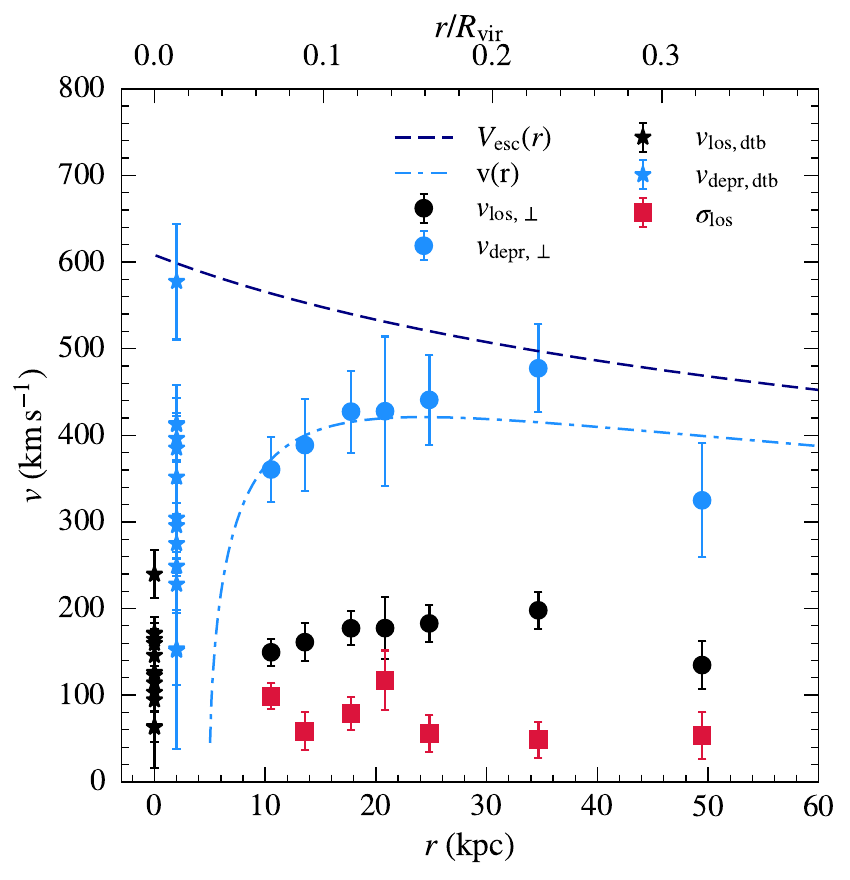}
    \caption{
    Velocity profile of the \mgii{}-traced outflow in G1 as a function of de-projected radius. Black circles show the absolute line-of-sight velocity centroids, $|v_{\text{los}, \perp}|$, measured along transverse sightlines probing the projected minor axis of G1, plotted as a function of de-projected galactocentric radius $r=\rho/\sin i$. Light-blue circles show the corresponding de-projected velocities, $v_{\text{depr},\perp}=|v_{\text{los},\perp}|/\cos i$, assuming a collimated outflow perpendicular to the disk with inclination $i=65.5^\circ$. Red squares indicate the line-of-sight velocity dispersions along the transverse sightlines, $\sigma_{\text{los}, \perp}$. Black stars at $r=0$ show the absolute line-of-sight down-the-barrel velocities, $|v_{\text{los, dtb}}|$, placed at $r=0$ for visualization purposes only owing to the unknown physical scales probed by these sightlines; light-blue stars offset slightly in $r$ indicate the corresponding de-projected down-the-barrel velocities, $v_{\text{depr, dtb}}=|v_{\text{los, dtb}}|/\cos i$. The navy dashed curve shows the escape-velocity profile $V_{\rm esc}(r)$ (Sect.~\ref{sec:G1_mass}). The light-blue dash-dotted curve shows the best-fit momentum-driven wind model $v(r)$ fitted to $v_{\text{depr},\perp}$ vs. $r$, with best-fit parameters $r_0=5.25$~kpc, $v_0=116$~\kms{}, and $\log(N_{\rm H}/\mathrm{cm}^{-2})=19.96$. The top axis indicates the corresponding radial distance in units of the virial radius, $r/R_{\rm vir}$.
    }
    \label{fig:velocity-profile}
\end{figure}

With these assumptions, we obtain de-projected velocities of $v_\text{depr,dtb}=150\text{--}576$~\kms{} for the down-the-barrel sightlines, and $v_{\text{depr},\perp}=318\text{--}477$~\kms{} for the transverse ones. In Fig.~\ref{fig:velocity-profile}, we show   $v_{\text{los}, \perp}$, $v_{\text{los, dtb}}$, $\sigma_\text{los}$, $v_{\text{depr},\perp}$, and $v_{\text{depr},\text{dtb}}$ as a function of $r$. For the down-the-barrel sightlines, we arbitrarily set $r=0$ because the physical scales probed by these sightlines are unknown, while explicitly not assuming that the absorption originates at this distance. 

Using the dark-matter halo model described in Sect.~\ref{sec:G1_mass} and Appendix~\ref{sec:appendix-sed-halo-properties}, we computed the expected escape-velocity profile $V_\text{esc}(r)$, also shown in Fig.~\ref{fig:velocity-profile}. The $v_{\text{depr},\perp}$ values approach but do not exceed $V_\text{esc}(r)$, suggesting that the outflowing material is consistent with being gravitationally bound to G1. Similarly, the $v_{\text{depr},\text{dtb}}$ lie below $V_\text{esc}(r)$ for $r<0.3~R_{\rm vir}$, with the exception of a single sightline that may exceed the escape velocity depending on the (unknown) radius at which the absorption is produced.

The $v_{\text{depr},\perp}$ (and $v_{\text{los},\perp}$) values increase with radius up to $\sim0.2~R_{\rm vir}$ before declining near $\approx0.3~R_{\rm vir}$, suggesting an initially accelerating and then decelerating wind structure. The apparent decline in velocity observed in Fig.~\ref{fig:velocity-profile}, although supported by only one point, is consistent with previous observations. For example, \citet{2014ApJ...792L..12K} report blue-shifted down-the-barrel velocities of $45\text{--}255$~\kms{} in a $z\sim0.2$ galaxy with comparable stellar mass and SFR, but significantly lower transverse velocities ($40\text{--}80$~\kms{}) at $\rho\sim58$~kpc, suggesting outflow deceleration at large radii.

To interpret the velocity profile, we adopt a simple momentum-driven wind model \citep{2005ApJ...618..569M}, which has been shown to provide a reasonable description of the kinematics of cool outflows traced by low-ionization absorption lines \citep[e.g.][]{2015ApJ...809..147H}. Following \citet{2015ApJ...809..147H}, the inward force is given by the gravitational potential $\phi(r)$ and the cloud mass $M_c$ as $F_\text{in}=-M_c\nabla\phi(r)$, while the outward force is $F_\text{out}=A_c \dot{p}_\star / 4 \pi r^2$. Here, $\dot{p}_\star=4.8\times10^{33}\times\left(\text{SFR}/ M_\odot \text{yr}^{-1}\right)$~dyne, is the momentum injection rate from star formation, $M_c=A_c N_\text{H}\,\mu \,m_p$ is the mass of an outflowing cloud, with $A_c$ its cross section, $N_\text{H}$ its hydrogen column density, $\mu$ the mean particle mass in units of the proton mass $m_p$ \citep[we adopt $\mu\simeq1.6$, ][]{2020A&ARv..28....2V}. The net force is thus given by $F=F_\text{out}-F_\text{in}=v(r)\left[dv(r)/dt\right]$. Solving for the radial velocity profile

\begin{equation}
    v(r)= \sqrt{ v_0^2+2\left[\phi(r_0)-\phi(r) \right] + 2\frac{\dot{p}_\star}{4\pi\mu m_p N_\text{H}}\left[\frac{1}{r_0}- \frac{1}{r}\right]   }
\end{equation}

\noindent where the only free parameters are a normalization $v_0$ at a given $r_0$, and $N_\text{H}$, for a given gravitational potential model. Again, we assume the same dark-matter halo model used for estimating $V_\text{esc}(r)$. The parameters $r_0$ and $v_0$ should not be interpreted as the physical launch radius or initial velocity of the outflow, but rather as normalization parameters of the adopted equation of motion.

A least-squares fit to $v_{\text{depr}, \perp} $ vs. $r$ results in $r_0=5.25$~kpc, $v_0=116$~\kms{}, and $\log( N_{\text{H}}/\text{cm}^{-2}) = 19.96$. The best-fit momentum-driven model trajectory is shown in Fig.~\ref{fig:velocity-profile}, and is consistent at a $<1.5\sigma$ level with all the transverse data points, capturing the initial acceleration, the subsequent decline in velocity, and the sub-escape velocities. We discuss the implications of the outflow radial velocity profile in Sect.~\ref{sec:recycling}.

\subsubsection{Mass outflow rate}

With the geometry and kinematics of the outflow characterized, we estimated the associated mass outflow rate. Assuming a thin-shell bi-conical geometry, the instantaneous mass outflow rate at radius $r$ is $\dot{{\rm M}}_{\rm out} = \Omega \mu m_p N_H v r$,  where $\Omega = \Omega_\text{outflow} C_f$ is the effective solid angle subtended by the outflow, modified by the cloud covering fraction $C_f$. As seen from the galaxy, $\Omega_\text{outflow}=4\pi\left(1-\cos{\theta}\right)$ for a bi-conical structure. Here, $\mu \simeq 1.6$ is the mean mass per hydrogen atom \citep[accounting for the helium fraction;][]{2020A&ARv..28....2V}, $m_p$ is the proton mass, $N_H$ is the total hydrogen column density, and $v$ is the outflow velocity at radius $r$ \citep{2005ApJS..160..115R,2020A&ARv..28....2V}.

We adopted a conservative lower limit on ${N_\text{\hi{}}}$ derived in Sect.~\ref{sec:column-density} using Eq.~\ref{eq:logN} and Eq.~\ref{eq:logNHI}, and assume $N_H=N_\text{\hi{}}$. With our constraint on the minimum opening angle, we obtained $\Omega_\text{outflow, min} \approx 0.6$~sr. Also, we assumed $C_f\simeq0.7$ \citep{2025A&A...693A.200B} measured for \mgii{} absorptions against gravitational arcs. Considering the $r$ and $v_{\text{depr}, \perp}$ values from Sect.~\ref{sec:geometry}, we estimate a minimum mass outflow rate $\dot{{\rm M}}_{\rm out, min}\left(r\right)$, which ranges from $\approx 0.06~M_\odot~{\rm yr}^{-1}$ at $\rho\approx$10~kpc, to $\approx 0.21~M_\odot~{\rm yr}^{-1}$ at $\rho\approx45$~kpc. 

From our lower limits on $\dot{{\rm M}}_{\rm out, min}$ we estimate  minimum mass loading factors of $\eta_{\rm min} = \dot{\rm M}_\text{out, min} / \text{SFR} \approx \left(0.4\text{--}1.5\right)\times10^{-2}$. Similarly, for the maximum opening angle, the minimum mass outflow rate increases to $\dot{{\rm M}}_{\rm out, min} \approx \left(0.12\text{--}0.4\right)~M_\odot~{\rm yr}^{-1}$ and $\eta_{\rm min} \approx \left(0.8\text{--}2.7\right)\times10^{-2}$.  These lower limits are broadly consistent with quasar-traced galactic outflows from MEGAFLOW \citep{2019MNRAS.490.4368S} and similar studies \citep[e.g.,][]{2012MNRAS.426..801B, 2015ApJ...809..147H, 2019ApJ...878...84M}. Such works typically report $\dot{M}_{\rm out} \sim 0.1\text{--}10~M_\odot\,\text{yr}^{-1}$ and $ \eta \sim 0.1\text{--}10 $, albeit with significant scatter and methodological differences.

Our inferred mass loading factors, $\eta_\text{min} \approx \left(0.4\text{--}2.7\right)\times10^{-2}$, are significantly lower than the predictions from high-resolution zoom-in simulations such as FIRE-2 \citep{2021MNRAS.508.2979P}, which suggest $\eta \sim 0.3$ (and typically not exceeding 1) for galaxies with stellar masses $\log( M_\star/M_\odot) \sim 10.3$ at $z \sim 1$. Large-volume simulations such as TNG-50 \citep{2019MNRAS.490.3234N} and EAGLE \citep{2020MNRAS.494.3971M} predict even higher mass loading factors, $\eta \sim 10$, along with outflow velocities of $\sim 400$~\kms{} and bi-conical geometries with opening angles of $40^\circ\text{--}50^\circ$ at similar galactocentric distances. Our measurements therefore provide conservative lower limits that are compatible with these models, but do not allow us to discriminate among them given the uncertainties in geometry, ionization corrections, and column density.

Finally, recent high-resolution simulations of galactic winds \citep{2020ApJ...903L..34K, 2020ApJ...900...61K, 2020ApJ...895...43S, 2020ApJ...890L..30L} explicitly modelling the multiphase nature of outflows, indicate that most of the outflowing mass resides in cold and warm ($T \sim 10^4\text{--}10^5$~K), while most of the energy is carried by hot ($T \sim 10^6\text{--}10^7$~K) gas. Our \mgii{}-based measurements trace only the cool component, potentially missing a substantial (and in some models dominant) contribution from the warm and hot phases. A complete census of the mass and energy budget would require constraints on these higher-temperature components, which remain inaccessible with the current data.

\section{Discussion}
\label{sec:discussion}

\subsection{The smoking gun of a galactic-scale outflow}
\label{sec:discussion-outflow}

The collective kinematic signatures observed in both down-the-barrel and transverse sightlines provide compelling evidence for a large-scale, collimated outflow originating from G1. The presence of spatially coherent blue-shifted absorption along the minor axis and down-the-barrel sightlines is difficult to reconcile with alternative mechanisms (see Sect.~\ref{sec:other-scenarios}). The ratio $v_{\rm los}/\sigma_{\rm los}\sim2.5$ confirms that bulk motions dominate, and the low scatter in both velocity centroids and dispersions across tens of kpc suggests a high degree of coherence in the gas dynamics.

Our de-projected velocities inferred along the arc fall within the expected range for outflow velocities \citep{2009AIPC.1201..142W,2009ApJ...692..187W,2012ApJ...760..127M,2014ApJ...794..156R}, and are consistent with momentum-driven wind models, which predict outflow velocities of $2\text{--}3\times V_{\rm circ}$, and close to $V_{\rm esc}$ \citep[e.g.,][]{2013MNRAS.430.3213B,2017MNRAS.464.4977K}. Our tomographic data also provide, for the first time, a direct observational characterization of the radial velocity profile of a cool outflow at kpc scales.  The geometry of the background arc allows us to map $v_{\text{depr}, \perp}(r)$ up to $\sim50$~kpc, revealing a clear acceleration followed by possible deceleration. A simple momentum-driven cloud model (Sect.~\ref{sec:outflow-kinematics}) seems to reproduce this behaviour: early acceleration arises from the momentum injected by the ongoing star formation, while the deceleration reflects the increasing relative influence of the galaxy’s gravitational potential. The model requires column densities of $\log(N_{\text{\hi{}}}/\text{cm}^{-2})\approx20$ to reproduce the measured velocities, implying that only relatively dense cool clouds can reach several tens of kiloparsecs through this mechanism. However, this column density is as seen from the galaxy radially outwards, and does not necessarily need to match the line-of-sight $N_\text{{\hi{}}}$.

The values of $\sigma_{\rm los}\approx 50\text{--}120$~\kms{} are comparable to those seen in \mgii{} absorbers toward background quasars \citep[e.g.,][]{2018ApJ...866...36L}, and similar to values reported in other arc-based studies \citep{2020MNRAS.491.4442L,2021MNRAS.507..663T,2021ApJ...914...92M}. We remark that the observed velocity dispersions are not expected to trace the thermal broadening of the gas. Instead, they likely reflect the kinematics of ensembles of cool, coherence complexes at the kpc scales \citep[e.g.][]{2023A&A...680A.112A, 2024A&A...691A.356L}, that are unresolved at the spectral resolution of our observations. High-resolution observations of background quasars generally reveal the presence of multiple narrow components \citep[e.g.][]{2001AJ....122..679C}, generally interpreted as discrete clumps, which are expected to be smaller than the spatial sampling of our observations of $\sim6~{\rm kpc}\times6~{\rm kpc}$ \citep[see e.g.][]{2014MNRAS.438.1435C, 2023A&A...680A.112A}. In this case, the integrated line profiles likely represent the collective motion of such clouds within each spaxel. 

These $\sigma_{\rm los}$ values can be interpreted in the context of a bi-conical outflow geometry with opening angle $\theta_c$, which naturally produces a variation in $v_{\rm los}$ in the range $\cos{(i-\theta_c)}\lesssim v_{\rm los}/V\lesssim\cos{(i+\theta_c)}$ along a sightline intercepting the axis of the outflow cone given a radial velocity $V$ and galaxy inclination $i$. For an opening angle $\theta_c=17.8^\circ$ (see Sect.~\ref{sec:geometry}), and considering G1's inclination, an outflow velocity of  $V=400$~\kms{} results in a spread $\Delta v_{\rm los}\approx200$~\kms{}, which may explain the observed line-widths considering $\sigma_{\rm los}$ only characterizes $1\sigma$ of the full spread of $v_{\rm los}$, and is expected to remain constant if $V$ and $\theta_c$ $\sim$ constant. Measuring an azimuthal dependence of $\sigma_{\rm los}$ within an outflow cone will be essential for confirming this projection effect.

The relatively low scatter in $v_\text{los}$ and $\sigma_\text{los}$ across tens of kpc implies a remarkable degree of kinematic coherence of the kinematics of these clouds, both in bulk velocity and integrated motions along the line of sight, further supporting the outflow interpretation if projection effects dominate $\sigma_\text{los}$. Taken together, the large-scale kinematic coherence, the minor-axis orientation of the sightlines, and the consistency with momentum-driven wind models strongly support a scenario in which the observed gas traces the bulk kinematics of a collimated, galactic-scale outflow, with projection effects likely dominating the observed velocity dispersions. 

\subsubsection{Signatures of gas recycling}
\label{sec:recycling}

Our results show that the outflow plays a key role in sustaining cool, metal-enriched gas within the CGM of G1, injecting metals out to at least $0.3~R_\text{vir}$. The de-projected velocities remain below the halo escape velocity, indicating that most of the ejected material is gravitationally bound and might eventually recycle back onto the galaxy. Indeed, single-sightline-probed galactic winds \citep{2012MNRAS.426..801B, 2019MNRAS.490.4368S} indicate that cool outflows traced by \mgii{} absorption hardly reach the escape velocities of halos for galaxies between $10<\log{(M_\star/M_\odot)}<11$. In the broader CGM context, \mgii{} absorption is typically concentrated within the inner halo, with the bulk of absorbers residing at $\lesssim 0.3~R_\text{vir}$, and with rapidly declining covering fractions beyond $\sim 0.3\text{--}0.5~R_\text{vir}$ \citep{2013ApJ...776..114N, 2021MNRAS.502.4743H}. Our detection of cool gas out to $0.3~R_\text{vir}$ therefore places G1 at the characteristic outer edge where strong \mgii{}-bearing clouds remain common in star-forming halos at $z\sim1$.

The sub-escape velocities and the apparent deceleration at $\sim0.3~R_\text{vir}$ suggest that the bulk of the cool outflowing gas is confined within the inner halo, where it may mix with the ambient CGM or be recycled back onto the galaxy. The suppression of \mgii{} absorption beyond this radius, as seen in galaxy–quasar pair studies \citep{2013ApJ...763L..42C}, further supports this interpretation. However, since the deceleration in our system is inferred from a single outer sightline, additional tomographic constraints at larger impact parameters are required to confirm this behaviour in individual systems.

The presence of cool gas at $0.3~R_\text{vir}$ implies a travel time of $\sim100$~Myr, corresponding to a recycling timescale of at least $\sim200$~Myr, which is consistent with cosmological simulations predicting that most of the recycling occurs within $0.3$~$R_\text{vir}$ with median recycling timescales of $100\text{--}350$~Myr at these redshifts. Assuming a constant $\dot{M}_\text{out}=0.2~M_\odot~\text{yr}^{-1}$, which is reasonable because the UV and emission-line-based SFRs are consistent  \citep[an indicator of steady star formation in the last few hundred Myr][]{2012ARA&A..50..531K}, we obtain a minimum processed gas mass of $0.4\times10^{8}~M_\odot$ stored within the CGM of G1. We emphasize that this result is based on a lower limit on $\dot{M}_\text{out}$, and that \mgii{} traces only the cool phase; therefore, this reservoir mass should be regarded as a lower limit to the total recycled mass in the halo. 

\subsubsection{Alternative scenarios}
\label{sec:other-scenarios}

To assess whether the absorbing gas could originate from mechanisms other than outflows, we consider several alternatives. First, we disfavour ram-pressure stripping and tidal tails. G1 is isolated, with no neighbouring galaxies within $3000$~\kms{} in redshift and $500~\text{kpc} \times 500~\text{kpc}$ in projected area. Moreover, tidal tails are typically narrow \citep[A few kpc, see e.g.][]{2020ApJ...889...70B}, making it unlikely for our extended arc—spanning tens of kpc in projection—to align by chance with such a structure.

A pressure-supported CGM is also disfavoured. Although the measured $\sigma_\text{los}$ values are comparable to the expected dark matter halo velocity dispersion of $\sigma_{\rm DM}\approx104$~\kms{}, our data show that velocity offsets clearly dominate over the observed dispersions, with a mean ratio $v_\text{los}/\sigma_\text{los}\sim2.5$. In a pressure-supported medium, one would instead expect dispersion-dominated kinematics with no systematic velocity shifts, whereas our measurements favour ordered bulk motions.

We further disfavour accretion in the form of an extended rotating disk. The transverse absorptions arise along the projected minor axis and exhibit kinematics inconsistent with co-planar rotation (see Sect.~\ref{sec:G1_kinematics}). A contribution from co-rotating ISM embedded within the down-the-barrel absorption cannot be completely excluded; however, isolating such a component would require higher spectral resolution and S/N.

Another interpretation is that the transverse sightlines probe radial inflows on the far side of the galaxy. While we cannot entirely rule out this possibility, down-the-barrel absorptions unambiguously indicate a galactic outflow towards the observer. Considering that $v_\text{dtb}$ and $v_{\perp}$ have similar values and are both oriented along the minor axis, the simplest and most self-consistent explanation is that the transverse absorptions probe the same outflow.

\subsection{Azimuthal vs. luminosity dependence on \mgiiew{}}
\label{sec:discussion-scatter}

Several studies have found strong evidence a bi-modality in the azimuthal distribution of \mgii{} absorbers using both individual background sources \citep[e.g.,][]{2012ApJ...760L...7K,2019MNRAS.490.4368S, 2019ApJ...878...84M,2021ApJ...913...50L}, and stacked spectroscopy \citep{2011ApJ...743...10B}, particularly in inclined blue star-forming galaxies, and stronger \mgii{} equivalent widths are preferentially found along the minor axis of galaxies \citep{2019ApJ...878...84M,2019MNRAS.490.4368S, 2021ApJ...913...50L}. The usual interpretation is that the bi-modality arises because of outflows and co-planar accretion in the form of extended disks, which are expected to emerge along the minor and major axes, respectively. Arc tomography results support this picture: \citet{2021MNRAS.507..663T} and \citet{2022MNRAS.517.2214F} identified a mild azimuthal correlation in \mgii{}, consistent with a composite CGM made of bi-conical outflows and extended disks, while \citet{2020MNRAS.491.4442L} found evidence of an extended co-rotating along the major axis of their absorber galaxy (G1-0311 hereafter). In our study, our galaxy (G1-0224, for the following discussion) is probed predominantly along the minor axis, where we found strong evidence for the presence of an outflow. Although our transverse sightlines do not sample the full range of azimuthal angles, the fact that G1-0224 and G1-0311 are probed close to the minor and major axes, respectively, provides a useful comparison to assess how \mgiiew{} varies with geometry.

Despite their different orientations, both systems show low intrinsic scatter in their individual $W_r$–$\rho$ relations, indicating coherent cool gas structures over scales of tens of kpc. When considered together, however, they span much of the scatter seen in quasar-based \mgii{} samples (Fig.~\ref{fig:ew_impact_parameter}, panel b). Importantly, the two galaxies differ substantially in $M_\star$, $M_B$, and SFR: G1-0224 is $\sim4$ times more massive and $\sim20$ times more luminous in $B$-band than G1-0311, which is transitioning toward quiescence.

The \mgii{} absorption is well-established to strongly correlate with host galaxy properties such as stellar mass and luminosity \citep[e.g.,][]{2008ApJ...687..745C, 2010ApJ...714.1521C, 2011ApJ...743...10B}. \citet{2010ApJ...714.1521C} showed that much of the $W_r$–$\rho$ scatter can be reduced by scaling the impact parameters with B-band luminosity:

\begin{equation}
\log \rho_{L_\star} = \log \rho \times (L_B/L_{B_\star})^{-0.35}
\label{eq:rho_scaled}
\end{equation}

\noindent implying that the extent of the cool, \mgii{}-bearing CGM grows with galaxy luminosity. This is one manifestation of the self-similar CGM proposed by \citet{2013ApJ...779...87C}, in which galaxies of different luminosities and masses follow a common $W_r$–($\rho/R_{\rm vir})$ relation.

To disentangle the effects of azimuth and luminosity on \mgiiew{}, we calculated partial Spearman rank correlations while accounting for the natural degeneracy between $\phi$ and $\rho$ introduced by the arc geometries. Considering both G1-0224 and G1-0311, we find a strong anti-correlation between \mgiiew{} and $\rho$ (Spearman rank correlation coefficient $r_{\rm S}$ and associated $p$-value; $r_S = -0.83$, $p = 9\times10^{-6}$), as expected from quasar-based studies \citep{2013ApJ...776..114N, 2021MNRAS.502.4743H}, but only a moderate but statistically significant correlation with $\phi$ ($r_S = 0.65$, $p = 0.002$). Scaling $\rho$ using Eq.~\ref{eq:rho_scaled} weakens the \mgiiew{}-$\phi$ correlation ($r_S=0.31$, $p=0.18$), whereas the $W_r$–$\rho_{L_\star}$ remains highly significant ($r_S=-0.87$, $p=10^{-6}$). A multivariate Energy Test shows that the original $W_r$–$\rho$ distributions of the two galaxies differ ($p=0.03$), but become statistically consistent after luminosity scaling ($p=0.22$). This confirms that B-band luminosity reduces the scatter in the $W_r$–$\rho$ relation, implying that part of the large dispersion in quasar absorber statistics has an important contribution from the heterogeneity of the galaxy population, and also that the $W_r$–$\rho$ anti-correlation has a similar shape along both major and minor axis directions once host properties are accounted for.

This result must be interpreted with care. The apparent similarity of the $W_r$–$\rho$ profiles for G1-0224 and G1-0311 does not contradict the well-established azimuthal trends derived from large \mgii{} absorber samples, which demonstrate a statistical excess of absorbers along galaxy minor axes. Instead, our analysis shows that in two well-characterised halos—one probed near the minor axis (G1-0224) and the other near the major axis (G1-0311)—the $W_r$–$\rho$ relations are essentially indistinguishable once scaled by host luminosity (and therefore indirectly by halo mass/size). Despite sampling geometries traditionally associated with different physical processes (outflows along the minor axis vs. extended disks/inflows along the major axis), the cool gas absorption strength declines with radius in nearly the same way in these two systems. This does not refute the existence of azimuthal bi-modality; rather, it shows that projected geometry alone does not guarantee a significant difference in $W_r$ at fixed $\rho$. Host galaxy properties—such as luminosity, stellar mass, and star-formation rate—may exert a stronger influence on the absorber strength than the projected orientation of the sightlines. Determining whether this is a general feature of individual systems or specific to the particular evolutionary states of G1-0224 and G1-0311, will require a larger sample of arc tomography systems.

\section{Summary and conclusions}
\label{sec:conclusions}
We have conducted a spatially resolved study of the cool CGM surrounding G1, a luminous ($2.6~L^*_B$), star-forming ($\mathrm{SFR}\approx15~{\rm M}_\odot\,{\rm yr}^{-1}$), and relatively massive ($\log{(M_\star/M_\odot)} \approx 10.3$) main-sequence galaxy at $z \sim 1$. The fortuitous alignment of a background gravitational arc enabled us to probe metal absorption from the CGM of G1 in both down-the-barrel and transverse sightlines, spanning projected distances from the galactic centre out to $\sim50$~kpc along the minor axis. This unique configuration revealed a galactic-scale outflow emerging along the minor axis of G1, allowing us to characterize its geometry and kinematics with unprecedented spatial resolution for absorption-line studies. Our main findings are summarized as follows:

\begin{enumerate}[(I)]
    \item Spatially coherent, blue-shifted metal absorption lines detected along the minor axis of a star-forming galaxy, extending out to projected distances of $\sim50$~kpc, provide strong evidence for the detection of a large-scale outflow traced in absorption using gravitational-arc tomography. 
    
    \item The data disfavours a spherical or isotropic outflow geometry, and instead, support a collimated wind emerging along the minor axis with opening angle $\sim18^\circ\text{--}25^\circ$, minimum mass outflow rates $\dot{{\rm M}}_\text{ min, out} \sim 0.06\text{--}0.4~M_\odot \,\text{ yr}^{-1}$, and mass loading factor of $\eta_\text{min}\sim0.004\text{--}0.027$, measured within projected distances of $10\text{--}50$~kpc. These values are broadly consistent with expectations from cosmological simulations.
    
    \item The transverse and down-the-barrel absorptions show strong blueshifts, with values up to $\sim240$~\kms{}. Combined with relatively low $\sigma_\text{los}$ values, this implies that the gas is dominated by bulk motion ($v_\text{los}/\sigma_\text{los} \sim 2.5$). The small scatter in $v_\text{los}$ and $\sigma_\text{los}$ points to high kinematic coherence over tens of kpc.
    
    \item A momentum-driven wind model appears to describe the radial profile of the outflowing cool gas. An initial acceleration of $v_{\text{depr},\perp}$ can be explained by momentum injection from the star-formation, and a subsequent deceleration dominated by the gravitational potential. Both the measured $v_{\text{depr},\perp}$ values and the momentum-driven cloud model indicate that the gas barely reaches the escape velocity of the halo, and is likely to be retained within the halo. We estimate a recycling timescale of $\sim200$~Myr, implying a minimum processed gas mass of $0.4\times10^8~M_\odot$, in the CGM of G1, within $0.3~R_\text{vir}$.
    
    \item Our measurements of \mgiiew{} are statistically consistent with the canonical $W_r$-$\rho$ anti-correlation established from quasar-galaxy pairs. Notably, the scatter in our measurements is small and comparable to that in previous arc tomography studies, supporting the scenario that a significant fraction of the scatter in the quasar statistics is driven by heterogeneity in the host-galaxy properties.
    
    \item Compared with a previously studied arc system probing the major axis \citep{2020MNRAS.491.4442L}, our minor-axis sightlines yield systematically higher \mgiiew{} values. However, once the impact parameters are scaled by the B-band luminosity of the host galaxy, this difference diminishes, and the $W_r$–$\rho$ distributions become statistically consistent, indicating that major/minor axis alignment does not necessarily guarantee smaller/larger \mgiiew{} after accounting for host-galaxy properties.

\end{enumerate}

Our findings provide direct evidence for the physical connection between galactic outflows and the cool, metal-bearing CGM. By resolving its structure and kinematics across tens of kpc, we offer new constraints on the geometry, extent, and kinematics of feedback-driven winds.

\begin{acknowledgements}
We thank the anonymous referee for constructive comments and suggestions that significantly improved the clarity and quality of the manuscript. S.L. and N.T. acknowledge support by FONDECYT grant 1231187. J.H. acknowledges support from Beca ANID Doctorado Nacional Folio 21230522, J.H. and L.F.B. acknowledge support from ANID BASAL project FB210003, FONDECYT projects 1230231, and CASSACA project CCJRF1906. Based on observations collected at the European Southern Observatory under ESO programme 094.A-0141 (PI: SWINBANK). This research is partly based on observations made with the NASA/ESA Hubble Space Telescope obtained from the Space Telescope Science Institute, which is operated by the Association of Universities for Research in Astronomy, Inc., under NASA contract NAS 5–26555. These observations are associated with program GO 14497 (PI Smit), and GO 9135 (PI Gladders).  This research has made use of the Astrophysics Data System, funded by NASA under Cooperative Agreement 80NSSC21M00561.

This work made use of Astropy:
\footnote{\url{http://www.astropy.org} \citep{2013A&A...558A..33A, 2018AJ....156..123A, 2022ApJ...935..167A}}
a community-developed core Python package and an ecosystem of tools and resources for astronomy. This work also thanks contributors to  $\mathrm{\texttt{GalPak}}^{\mathrm{3D}}$ \footnote{\url{https://galpak3d.univ-lyon1.fr} \citep{2015ascl...soft01014B}}
, \texttt{MPDAF}
\footnote{\url{https://mpdaf.readthedocs.io} \citep{2016ascl.soft11003B}}
, \texttt{BAGPIPES} \footnote{\url{https://bagpipes.readthedocs.io/} \citep{2018MNRAS.480.4379C, 2019MNRAS.490..417C}}
, \texttt{NumPy}
\footnote{\url{https://numpy.org/} \citep{harris2020array}}
, \texttt{SciPy}
\footnote{\url{https://scipy.org/} \citep{2020SciPy-NMeth}}
, \texttt{Matplotlib}
\footnote{ \url{https://matplotlib.org/} \citep{Hunter:2007}}
 \texttt{LINETOOLS} \footnote{\url{https://linetools.readthedocs.io} \citep{2017zndo...1036773P}}
, the \texttt{PYTHON} programming language
\footnote{\url{https://www.python.org/}}
, and the free and open-source community.

\end{acknowledgements}

\bibliographystyle{aa}
\bibliography{export-bibtex}

@ARTICLE{2021ApJ...914...92M,
       author = {{Mortensen}, Kris and {Keerthi Vasan}, G.~C. and {Jones}, Tucker and {Faucher-Gigu{\`e}re}, Claude-Andr{\'e} and {Sanders}, Ryan L. and {Ellis}, Richard S. and {Leethochawalit}, Nicha and {Stark}, Daniel P.},
        title = "{Kinematics of the Circumgalactic Medium of a z = 0.77 Galaxy from Mg II Tomography}",
      journal = {\apj},
         year = 2021,
        month = jun,
       volume = {914},
       number = {2},
          eid = {92},
        pages = {92},
          doi = {10.3847/1538-4357/abfa11},
archivePrefix = {arXiv},
       eprint = {2006.00006},
 primaryClass = {astro-ph.GA},
       adsurl = {https://ui.adsabs.harvard.edu/abs/2021ApJ...914...92M}
}

@ARTICLE{2005ApJ...631...85E,
       author = {{Elmegreen}, Debra Meloy and {Elmegreen}, Bruce G. and {Rubin}, Douglas S. and {Schaffer}, Meredith A.},
        title = "{Galaxy Morphologies in the Hubble Ultra Deep Field: Dominance of Linear Structures at the Detection Limit}",
      journal = {\apj},
         year = 2005,
        month = sep,
       volume = {631},
       number = {1},
        pages = {85-100},
          doi = {10.1086/432502},
archivePrefix = {arXiv},
       eprint = {astro-ph/0508216},
 primaryClass = {astro-ph},
       adsurl = {https://ui.adsabs.harvard.edu/abs/2005ApJ...631...85E}
}

@ARTICLE{2007ApJ...656....1L,
       author = {{Law}, David R. and {Steidel}, Charles C. and {Erb}, Dawn K. and {Pettini}, Max and {Reddy}, Naveen A. and {Shapley}, Alice E. and {Adelberger}, Kurt L. and {Simenc}, David J.},
        title = "{The Physical Nature of Rest-UV Galaxy Morphology during the Peak Epoch of Galaxy Formation}",
      journal = {\apj},
         year = 2007,
        month = feb,
       volume = {656},
       number = {1},
        pages = {1-26},
          doi = {10.1086/510357},
archivePrefix = {arXiv},
       eprint = {astro-ph/0610693},
 primaryClass = {astro-ph},
       adsurl = {https://ui.adsabs.harvard.edu/abs/2007ApJ...656....1L}
}

@ARTICLE{2009ApJ...706.1364F,
       author = {{F{\"o}rster Schreiber}, N.~M. and {Genzel}, R. and {Bouch{\'e}}, N. and {Cresci}, G. and {Davies}, R. and {Buschkamp}, P. and {Shapiro}, K. and {Tacconi}, L.~J. and {Hicks}, E.~K.~S. and {Genel}, S. and et al.},
        title = "{The SINS Survey: SINFONI Integral Field Spectroscopy of z \raisebox{-0.5ex}\textasciitilde 2 Star-forming Galaxies}",
      journal = {\apj},
         year = 2009,
        month = dec,
       volume = {706},
       number = {2},
        pages = {1364-1428},
          doi = {10.1088/0004-637X/706/2/1364},
archivePrefix = {arXiv},
       eprint = {0903.1872},
 primaryClass = {astro-ph.CO},
       adsurl = {https://ui.adsabs.harvard.edu/abs/2009ApJ...706.1364F}
}

@ARTICLE{2015ApJ...800...39G,
       author = {{Guo}, Yicheng and {Ferguson}, Henry C. and {Bell}, Eric F. and {Koo}, David C. and {Conselice}, Christopher J. and {Giavalisco}, Mauro and {Kassin}, Susan and {Lu}, Yu and {Lucas}, Ray and {Mandelker}, Nir and et al.},
        title = "{Clumpy Galaxies in CANDELS. I. The Definition of UV Clumps and the Fraction of Clumpy Galaxies at 0.5 < z < 3}",
      journal = {\apj},
         year = 2015,
        month = feb,
       volume = {800},
       number = {1},
          eid = {39},
        pages = {39},
          doi = {10.1088/0004-637X/800/1/39},
archivePrefix = {arXiv},
       eprint = {1410.7398},
 primaryClass = {astro-ph.GA},
       adsurl = {https://ui.adsabs.harvard.edu/abs/2015ApJ...800...39G}
}

@ARTICLE{2018ApJ...859...55C,
       author = {{Child}, Hillary L. and {Habib}, Salman and {Heitmann}, Katrin and {Frontiere}, Nicholas and {Finkel}, Hal and {Pope}, Adrian and {Morozov}, Vitali},
        title = "{Halo Profiles and the Concentration-Mass Relation for a {\ensuremath{\Lambda}}CDM Universe}",
      journal = {\apj},
         year = 2018,
        month = may,
       volume = {859},
       number = {1},
          eid = {55},
        pages = {55},
          doi = {10.3847/1538-4357/aabf95},
archivePrefix = {arXiv},
       eprint = {1804.10199},
 primaryClass = {astro-ph.CO},
       adsurl = {https://ui.adsabs.harvard.edu/abs/2018ApJ...859...55C}
}

@ARTICLE{1996ApJ...462..563N,
       author = {{Navarro}, Julio F. and {Frenk}, Carlos S. and {White}, Simon D.~M.},
        title = "{The Structure of Cold Dark Matter Halos}",
      journal = {\apj},
         year = 1996,
        month = may,
       volume = {462},
        pages = {563},
          doi = {10.1086/177173},
archivePrefix = {arXiv},
       eprint = {astro-ph/9508025},
 primaryClass = {astro-ph},
       adsurl = {https://ui.adsabs.harvard.edu/abs/1996ApJ...462..563N}
}

@BOOK{2011piim.book.....D,
       author = {{Draine}, Bruce T.},
        title = "{Physics of the Interstellar and Intergalactic Medium}",
         year = 2011,
       adsurl = {https://ui.adsabs.harvard.edu/abs/2011piim.book.....D}
}

@ARTICLE{2020ApJ...890L..30L,
       author = {{Li}, Miao and {Bryan}, Greg L.},
        title = "{Simple Yet Powerful: Hot Galactic Outflows Driven by Supernovae}",
      journal = {\apjl},
         year = 2020,
        month = feb,
       volume = {890},
       number = {2},
          eid = {L30},
        pages = {L30},
          doi = {10.3847/2041-8213/ab7304},
archivePrefix = {arXiv},
       eprint = {1910.09554},
 primaryClass = {astro-ph.GA},
       adsurl = {https://ui.adsabs.harvard.edu/abs/2020ApJ...890L..30L}
}

@ARTICLE{2020ApJ...895...43S,
       author = {{Schneider}, Evan E. and {Ostriker}, Eve C. and {Robertson}, Brant E. and {Thompson}, Todd A.},
        title = "{The Physical Nature of Starburst-driven Galactic Outflows}",
      journal = {\apj},
         year = 2020,
        month = may,
       volume = {895},
       number = {1},
          eid = {43},
        pages = {43},
          doi = {10.3847/1538-4357/ab8ae8},
archivePrefix = {arXiv},
       eprint = {2002.10468},
 primaryClass = {astro-ph.GA},
       adsurl = {https://ui.adsabs.harvard.edu/abs/2020ApJ...895...43S}
}

@ARTICLE{2020ApJ...900...61K,
       author = {{Kim}, Chang-Goo and {Ostriker}, Eve C. and {Somerville}, Rachel S. and {Bryan}, Greg L. and {Fielding}, Drummond B. and {Forbes}, John C. and {Hayward}, Christopher C. and {Hernquist}, Lars and {Pandya}, Viraj},
        title = "{First Results from SMAUG: Characterization of Multiphase Galactic Outflows from a Suite of Local Star-forming Galactic Disk Simulations}",
      journal = {\apj},
         year = 2020,
        month = sep,
       volume = {900},
       number = {1},
          eid = {61},
        pages = {61},
          doi = {10.3847/1538-4357/aba962},
archivePrefix = {arXiv},
       eprint = {2006.16315},
 primaryClass = {astro-ph.GA},
       adsurl = {https://ui.adsabs.harvard.edu/abs/2020ApJ...900...61K}
}

@ARTICLE{2020ApJ...903L..34K,
       author = {{Kim}, Chang-Goo and {Ostriker}, Eve C. and {Fielding}, Drummond B. and {Smith}, Matthew C. and {Bryan}, Greg L. and {Somerville}, Rachel S. and {Forbes}, John C. and {Genel}, Shy and {Hernquist}, Lars},
        title = "{A Framework for Multiphase Galactic Wind Launching Using TIGRESS}",
      journal = {\apjl},
         year = 2020,
        month = nov,
       volume = {903},
       number = {2},
          eid = {L34},
        pages = {L34},
          doi = {10.3847/2041-8213/abc252},
archivePrefix = {arXiv},
       eprint = {2010.09090},
 primaryClass = {astro-ph.GA},
       adsurl = {https://ui.adsabs.harvard.edu/abs/2020ApJ...903L..34K}
}

@ARTICLE{2016ApJ...821...72S,
       author = {{Shibuya}, Takatoshi and {Ouchi}, Masami and {Kubo}, Mariko and {Harikane}, Yuichi},
        title = "{Morphologies of \raisebox{-0.5ex}\textasciitilde190,000 Galaxies at z = 0-10 Revealed with HST Legacy Data. II. Evolution of Clumpy Galaxies}",
      journal = {\apj},
         year = 2016,
        month = apr,
       volume = {821},
       number = {2},
          eid = {72},
        pages = {72},
          doi = {10.3847/0004-637X/821/2/72},
archivePrefix = {arXiv},
       eprint = {1511.07054},
 primaryClass = {astro-ph.GA},
       adsurl = {https://ui.adsabs.harvard.edu/abs/2016ApJ...821...72S}
}

@ARTICLE{2025MNRAS.536.3090K,
       author = {{Kalita}, Boris S. and {Suzuki}, Tomoko L. and {Kashino}, Daichi and {Silverman}, John D. and {Daddi}, Emanuele and {Ho}, Luis C. and {Ding}, Xuheng and {Mercier}, Wilfried and {Faisst}, Andreas L. and {Sheth}, Kartik and {Valentino}, Francesco and {Puglisi}, Annagrazia and {Saito}, Toshiki and {Kakkad}, Darshan and {Ilbert}, Olivier and {Khostovan}, Ali Ahmad and {Liu}, Zhaoxuan and {Tanaka}, Takumi and {Magdis}, Georgios and {Zavala}, Jorge A. and {Tan}, Qinghua and {Kartaltepe}, Jeyhan S. and {Yang}, Lilan and {Koekemoer}, Anton M. and {McKinney}, Jed and {Robertson}, Brant E. and {Jin}, Shuowen and {Hayward}, Christopher C. and {Hirschmann}, Michaela and {Franco}, Maximilien and {Shuntov}, Marko and {Gozaliasl}, Ghassem and {Kaminsky}, Aidan and {Rich}, R. Michael},
        title = "{Clumps as multiscale structures in cosmic noon galaxies}",
      journal = {\mnras},
         year = 2025,
        month = jan,
       volume = {536},
       number = {3},
        pages = {3090-3111},
          doi = {10.1093/mnras/stae2781},
archivePrefix = {arXiv},
       eprint = {2501.03328},
 primaryClass = {astro-ph.GA},
       adsurl = {https://ui.adsabs.harvard.edu/abs/2025MNRAS.536.3090K}
}

@ARTICLE{2024NatAs...8.1602N,
       author = {{Nielsen}, Nikole M. and {Fisher}, Deanne B. and {Kacprzak}, Glenn G. and {Chisholm}, John and {Martin}, D. Christopher and {Reichardt Chu}, Bronwyn and {Sandstrom}, Karin M. and {Rickards Vaught}, Ryan J.},
        title = "{An emission map of the disk-circumgalactic medium transition in starburst IRAS 08339+6517}",
      journal = {Nature Astronomy},
         year = 2024,
        month = dec,
       volume = {8},
        pages = {1602-1609},
          doi = {10.1038/s41550-024-02365-x},
archivePrefix = {arXiv},
       eprint = {2311.00856},
 primaryClass = {astro-ph.GA},
       adsurl = {https://ui.adsabs.harvard.edu/abs/2024NatAs...8.1602N}
}

@ARTICLE{2024A&A...691A...5P,
       author = {{Pessa}, Ismael and {Wisotzki}, Lutz and {Urrutia}, Tanya and {Pharo}, John and {Augustin}, Ramona and {Bouch{\'e}}, Nicolas F. and {Feltre}, Anna and {Guo}, Yucheng and {Kozlova}, Daria and {Krajnovic}, Davor and {Kusakabe}, Haruka and {Leclercq}, Floriane and {Salas}, H{\'e}ctor and {Schaye}, Joop and {Verhamme}, Anne},
        title = "{A galactic outflow traced by its extended Mg II emission out to a {\ensuremath{\sim}}30 kpc radius in the Hubble Ultra Deep Field with MUSE}",
      journal = {\aap},
         year = 2024,
        month = nov,
       volume = {691},
          eid = {A5},
        pages = {A5},
          doi = {10.1051/0004-6361/202450547},
archivePrefix = {arXiv},
       eprint = {2408.16067},
 primaryClass = {astro-ph.GA},
       adsurl = {https://ui.adsabs.harvard.edu/abs/2024A&A...691A...5P}
}

@ARTICLE{2013MNRAS.430.3213B,
       author = {{Barai}, Paramita and {Viel}, Matteo and {Borgani}, Stefano and {Tescari}, Edoardo and {Tornatore}, Luca and {Dolag}, Klaus and {Killedar}, Madhura and {Monaco}, Pierluigi and {D'Odorico}, Valentina and {Cristiani}, Stefano},
        title = "{Galactic winds in cosmological simulations of the circumgalactic medium}",
      journal = {\mnras},
         year = 2013,
        month = apr,
       volume = {430},
       number = {4},
        pages = {3213-3234},
          doi = {10.1093/mnras/stt125},
archivePrefix = {arXiv},
       eprint = {1210.3582},
 primaryClass = {astro-ph.CO},
       adsurl = {https://ui.adsabs.harvard.edu/abs/2013MNRAS.430.3213B}
}

@ARTICLE{2017MNRAS.464.4977K,
       author = {{Katsianis}, A. and {Tescari}, E. and {Blanc}, G. and {Sargent}, M.},
        title = "{The evolution of the star formation rate function and cosmic star formation rate density of galaxies at z {\ensuremath{\sim}} 1-4}",
      journal = {\mnras},
         year = 2017,
        month = feb,
       volume = {464},
       number = {4},
        pages = {4977-4994},
          doi = {10.1093/mnras/stw2680},
archivePrefix = {arXiv},
       eprint = {1610.03441},
 primaryClass = {astro-ph.GA},
       adsurl = {https://ui.adsabs.harvard.edu/abs/2017MNRAS.464.4977K}
}

@ARTICLE{2022ApJ...935..167A,
       author = {{Astropy Collaboration} and {Price-Whelan}, Adrian M. and {Lim}, Pey Lian and {Earl}, Nicholas and {Starkman}, Nathaniel and {Bradley}, Larry and {Shupe}, David L. and {Patil}, Aarya A. and {Corrales}, Lia and {Brasseur}, C.~E. and {N{\"o}the}, Maximilian and {Donath}, Axel and {Tollerud}, Erik and {Morris}, Brett M. and {Ginsburg}, Adam and {Vaher}, Eero and {Weaver}, Benjamin A. and {Tocknell}, James and {Jamieson}, William and {van Kerkwijk}, Marten H. and {Robitaille}, Thomas P. and {Merry}, Bruce and {Bachetti}, Matteo and {G{\"u}nther}, H. Moritz and {Aldcroft}, Thomas L. and {Alvarado-Montes}, Jaime A. and {Archibald}, Anne M. and {B{\'o}di}, Attila and {Bapat}, Shreyas and {Barentsen}, Geert and {Baz{\'a}n}, Juanjo and {Biswas}, Manish and {Boquien}, M{\'e}d{\'e}ric and {Burke}, D.~J. and {Cara}, Daria and {Cara}, Mihai and {Conroy}, Kyle E. and {Conseil}, Simon and {Craig}, Matthew W. and {Cross}, Robert M. and {Cruz}, Kelle L. and {D'Eugenio}, Francesco and {Dencheva}, Nadia and {Devillepoix}, Hadrien A.~R. and {Dietrich}, J{\"o}rg P. and {Eigenbrot}, Arthur Davis and {Erben}, Thomas and {Ferreira}, Leonardo and {Foreman-Mackey}, Daniel and {Fox}, Ryan and {Freij}, Nabil and {Garg}, Suyog and {Geda}, Robel and {Glattly}, Lauren and {Gondhalekar}, Yash and {Gordon}, Karl D. and {Grant}, David and {Greenfield}, Perry and {Groener}, Austen M. and {Guest}, Steve and {Gurovich}, Sebastian and {Handberg}, Rasmus and {Hart}, Akeem and {Hatfield-Dodds}, Zac and {Homeier}, Derek and {Hosseinzadeh}, Griffin and {Jenness}, Tim and {Jones}, Craig K. and {Joseph}, Prajwel and {Kalmbach}, J. Bryce and {Karamehmetoglu}, Emir and {Ka{\l}uszy{\'n}ski}, Miko{\l}aj and {Kelley}, Michael S.~P. and {Kern}, Nicholas and {Kerzendorf}, Wolfgang E. and {Koch}, Eric W. and {Kulumani}, Shankar and {Lee}, Antony and {Ly}, Chun and {Ma}, Zhiyuan and {MacBride}, Conor and {Maljaars}, Jakob M. and {Muna}, Demitri and {Murphy}, N.~A. and {Norman}, Henrik and {O'Steen}, Richard and {Oman}, Kyle A. and {Pacifici}, Camilla and {Pascual}, Sergio and {Pascual-Granado}, J. and {Patil}, Rohit R. and {Perren}, Gabriel I. and {Pickering}, Timothy E. and {Rastogi}, Tanuj and {Roulston}, Benjamin R. and {Ryan}, Daniel F. and {Rykoff}, Eli S. and {Sabater}, Jose and {Sakurikar}, Parikshit and {Salgado}, Jes{\'u}s and {Sanghi}, Aniket and {Saunders}, Nicholas and {Savchenko}, Volodymyr and {Schwardt}, Ludwig and {Seifert-Eckert}, Michael and {Shih}, Albert Y. and {Jain}, Anany Shrey and {Shukla}, Gyanendra and {Sick}, Jonathan and {Simpson}, Chris and {Singanamalla}, Sudheesh and {Singer}, Leo P. and {Singhal}, Jaladh and {Sinha}, Manodeep and {Sip{\H{o}}cz}, Brigitta M. and {Spitler}, Lee R. and {Stansby}, David and {Streicher}, Ole and {{\v{S}}umak}, Jani and {Swinbank}, John D. and {Taranu}, Dan S. and {Tewary}, Nikita and {Tremblay}, Grant R. and {de Val-Borro}, Miguel and {Van Kooten}, Samuel J. and {Vasovi{\'c}}, Zlatan and {Verma}, Shresth and {de Miranda Cardoso}, Jos{\'e} Vin{\'\i}cius and {Williams}, Peter K.~G. and {Wilson}, Tom J. and {Winkel}, Benjamin and {Wood-Vasey}, W.~M. and {Xue}, Rui and {Yoachim}, Peter and {Zhang}, Chen and {Zonca}, Andrea and {Astropy Project Contributors}},
        title = "{The Astropy Project: Sustaining and Growing a Community-oriented Open-source Project and the Latest Major Release (v5.0) of the Core Package}",
      journal = {\apj},
         year = 2022,
        month = aug,
       volume = {935},
       number = {2},
          eid = {167},
        pages = {167},
          doi = {10.3847/1538-4357/ac7c74},
archivePrefix = {arXiv},
       eprint = {2206.14220},
 primaryClass = {astro-ph.IM},
       adsurl = {https://ui.adsabs.harvard.edu/abs/2022ApJ...935..167A}
}

@ARTICLE{2018AJ....156..123A,
       author = {{Astropy Collaboration} and {Price-Whelan}, A.~M. and {Sip{\H{o}}cz}, B.~M. and {G{\"u}nther}, H.~M. and {Lim}, P.~L. and {Crawford}, S.~M. and {Conseil}, S. and {Shupe}, D.~L. and {Craig}, M.~W. and {Dencheva}, N. and {Ginsburg}, A. and {VanderPlas}, J.~T. and {Bradley}, L.~D. and {P{\'e}rez-Su{\'a}rez}, D. and {de Val-Borro}, M. and {Aldcroft}, T.~L. and {Cruz}, K.~L. and {Robitaille}, T.~P. and {Tollerud}, E.~J. and {Ardelean}, C. and {Babej}, T. and {Bach}, Y.~P. and {Bachetti}, M. and {Bakanov}, A.~V. and {Bamford}, S.~P. and {Barentsen}, G. and {Barmby}, P. and {Baumbach}, A. and {Berry}, K.~L. and {Biscani}, F. and {Boquien}, M. and {Bostroem}, K.~A. and {Bouma}, L.~G. and {Brammer}, G.~B. and {Bray}, E.~M. and {Breytenbach}, H. and {Buddelmeijer}, H. and {Burke}, D.~J. and {Calderone}, G. and {Cano Rodr{\'\i}guez}, J.~L. and {Cara}, M. and {Cardoso}, J.~V.~M. and {Cheedella}, S. and {Copin}, Y. and {Corrales}, L. and {Crichton}, D. and {D'Avella}, D. and {Deil}, C. and {Depagne}, {\'E}. and {Dietrich}, J.~P. and {Donath}, A. and {Droettboom}, M. and {Earl}, N. and {Erben}, T. and {Fabbro}, S. and {Ferreira}, L.~A. and {Finethy}, T. and {Fox}, R.~T. and {Garrison}, L.~H. and {Gibbons}, S.~L.~J. and {Goldstein}, D.~A. and {Gommers}, R. and {Greco}, J.~P. and {Greenfield}, P. and {Groener}, A.~M. and {Grollier}, F. and {Hagen}, A. and {Hirst}, P. and {Homeier}, D. and {Horton}, A.~J. and {Hosseinzadeh}, G. and {Hu}, L. and {Hunkeler}, J.~S. and {Ivezi{\'c}}, {\v{Z}}. and {Jain}, A. and {Jenness}, T. and {Kanarek}, G. and {Kendrew}, S. and {Kern}, N.~S. and {Kerzendorf}, W.~E. and {Khvalko}, A. and {King}, J. and {Kirkby}, D. and {Kulkarni}, A.~M. and {Kumar}, A. and {Lee}, A. and {Lenz}, D. and {Littlefair}, S.~P. and {Ma}, Z. and {Macleod}, D.~M. and {Mastropietro}, M. and {McCully}, C. and {Montagnac}, S. and {Morris}, B.~M. and {Mueller}, M. and {Mumford}, S.~J. and {Muna}, D. and {Murphy}, N.~A. and {Nelson}, S. and {Nguyen}, G.~H. and {Ninan}, J.~P. and {N{\"o}the}, M. and {Ogaz}, S. and {Oh}, S. and {Parejko}, J.~K. and {Parley}, N. and {Pascual}, S. and {Patil}, R. and {Patil}, A.~A. and {Plunkett}, A.~L. and {Prochaska}, J.~X. and {Rastogi}, T. and {Reddy Janga}, V. and {Sabater}, J. and {Sakurikar}, P. and {Seifert}, M. and {Sherbert}, L.~E. and {Sherwood-Taylor}, H. and {Shih}, A.~Y. and {Sick}, J. and {Silbiger}, M.~T. and {Singanamalla}, S. and {Singer}, L.~P. and {Sladen}, P.~H. and {Sooley}, K.~A. and {Sornarajah}, S. and {Streicher}, O. and {Teuben}, P. and {Thomas}, S.~W. and {Tremblay}, G.~R. and {Turner}, J.~E.~H. and {Terr{\'o}n}, V. and {van Kerkwijk}, M.~H. and {de la Vega}, A. and {Watkins}, L.~L. and {Weaver}, B.~A. and {Whitmore}, J.~B. and {Woillez}, J. and {Zabalza}, V. and {Astropy Contributors}},
        title = "{The Astropy Project: Building an Open-science Project and Status of the v2.0 Core Package}",
      journal = {\aj},
         year = 2018,
        month = sep,
       volume = {156},
       number = {3},
          eid = {123},
        pages = {123},
          doi = {10.3847/1538-3881/aabc4f},
archivePrefix = {arXiv},
       eprint = {1801.02634},
 primaryClass = {astro-ph.IM},
       adsurl = {https://ui.adsabs.harvard.edu/abs/2018AJ....156..123A}
}

@ARTICLE{2020ApJ...889...70B,
       author = {{Bonaca}, Ana and {Pearson}, Sarah and {Price-Whelan}, Adrian M. and {Dey}, Arjun and {Geha}, Marla and {Kallivayalil}, Nitya and {Moustakas}, John and {Mu{\~n}oz}, Ricardo and {Myers}, Adam D. and {Schlegel}, David J. and {Valdes}, Francisco},
        title = "{Variations in the Width, Density, and Direction of the Palomar 5 Tidal Tails}",
      journal = {\apj},
         year = 2020,
        month = jan,
       volume = {889},
       number = {1},
          eid = {70},
        pages = {70},
          doi = {10.3847/1538-4357/ab5afe},
archivePrefix = {arXiv},
       eprint = {1910.00592},
 primaryClass = {astro-ph.GA},
       adsurl = {https://ui.adsabs.harvard.edu/abs/2020ApJ...889...70B}
}

@ARTICLE{2020MNRAS.492.4576Z,
       author = {{Zabl}, Johannes and {Bouch{\'e}}, Nicolas F. and {Schroetter}, Ilane and {Wendt}, Martin and {Contini}, Thierry and {Schaye}, Joop and {Marino}, Raffaella A. and {Muzahid}, Sowgat and {Pezzulli}, Gabriele and {Verhamme}, Anne and {Wisotzki}, Lutz},
        title = "{MusE GAs FLOw and Wind (MEGAFLOW) IV. A two sightline tomography of a galactic wind}",
      journal = {\mnras},
         year = 2020,
        month = mar,
       volume = {492},
       number = {3},
        pages = {4576-4588},
          doi = {10.1093/mnras/stz3607},
archivePrefix = {arXiv},
       eprint = {1911.09342},
 primaryClass = {astro-ph.GA},
       adsurl = {https://ui.adsabs.harvard.edu/abs/2020MNRAS.492.4576Z}
}

@ARTICLE{2021ApJ...909..151B,
       author = {{Burchett}, Joseph N. and {Rubin}, Kate H.~R. and {Prochaska}, J. Xavier and {Coil}, Alison L. and {Vaught}, Ryan Rickards and {Hennawi}, Joseph F.},
        title = "{Circumgalactic Mg II Emission from an Isotropic Starburst Galaxy Outflow Mapped by KCWI}",
      journal = {\apj},
         year = 2021,
        month = mar,
       volume = {909},
       number = {2},
          eid = {151},
        pages = {151},
          doi = {10.3847/1538-4357/abd4e0},
archivePrefix = {arXiv},
       eprint = {2005.03017},
 primaryClass = {astro-ph.GA},
       adsurl = {https://ui.adsabs.harvard.edu/abs/2021ApJ...909..151B}
}

@ARTICLE{2014MNRAS.438.1435C,
       author = {{Chen}, Hsiao-Wen and {Gauthier}, Jean-Ren{\'e} and {Sharon}, Keren and {Johnson}, Sean D. and {Nair}, Preethi and {Liang}, Cameron J.},
        title = "{Spatially resolved velocity maps of halo gas around two intermediate-redshift galaxies}",
      journal = {\mnras},
         year = 2014,
        month = feb,
       volume = {438},
       number = {2},
        pages = {1435-1450},
          doi = {10.1093/mnras/stt2288},
archivePrefix = {arXiv},
       eprint = {1312.0016},
 primaryClass = {astro-ph.GA},
       adsurl = {https://ui.adsabs.harvard.edu/abs/2014MNRAS.438.1435C}
}

@ARTICLE{2013A&A...558A..33A,
       author = {{Astropy Collaboration} and {Robitaille}, Thomas P. and {Tollerud}, Erik J. and {Greenfield}, Perry and {Droettboom}, Michael and {Bray}, Erik and {Aldcroft}, Tom and {Davis}, Matt and {Ginsburg}, Adam and {Price-Whelan}, Adrian M. and {Kerzendorf}, Wolfgang E. and {Conley}, Alexander and {Crighton}, Neil and {Barbary}, Kyle and {Muna}, Demitri and {Ferguson}, Henry and {Grollier}, Fr{\'e}d{\'e}ric and {Parikh}, Madhura M. and {Nair}, Prasanth H. and {Unther}, Hans M. and {Deil}, Christoph and {Woillez}, Julien and {Conseil}, Simon and {Kramer}, Roban and {Turner}, James E.~H. and {Singer}, Leo and {Fox}, Ryan and {Weaver}, Benjamin A. and {Zabalza}, Victor and {Edwards}, Zachary I. and {Azalee Bostroem}, K. and {Burke}, D.~J. and {Casey}, Andrew R. and {Crawford}, Steven M. and {Dencheva}, Nadia and {Ely}, Justin and {Jenness}, Tim and {Labrie}, Kathleen and {Lim}, Pey Lian and {Pierfederici}, Francesco and {Pontzen}, Andrew and {Ptak}, Andy and {Refsdal}, Brian and {Servillat}, Mathieu and {Streicher}, Ole},
        title = "{Astropy: A community Python package for astronomy}",
      journal = {\aap},
         year = 2013,
        month = oct,
       volume = {558},
          eid = {A33},
        pages = {A33},
          doi = {10.1051/0004-6361/201322068},
archivePrefix = {arXiv},
       eprint = {1307.6212},
 primaryClass = {astro-ph.IM},
       adsurl = {https://ui.adsabs.harvard.edu/abs/2013A&A...558A..33A}
}

@ARTICLE{2013ApJ...763L..42C,
       author = {{Churchill}, Christopher W. and {Nielsen}, Nikole M. and {Kacprzak}, Glenn G. and {Trujillo-Gomez}, Sebastian},
        title = "{The Self-similarity of the Circumgalactic Medium with Galaxy Virial Mass: Implications for Cold-mode Accretion}",
      journal = {\apjl},
         year = 2013,
        month = feb,
       volume = {763},
       number = {2},
          eid = {L42},
        pages = {L42},
          doi = {10.1088/2041-8205/763/2/L42},
archivePrefix = {arXiv},
       eprint = {1211.1008},
 primaryClass = {astro-ph.CO},
       adsurl = {https://ui.adsabs.harvard.edu/abs/2013ApJ...763L..42C}
}

@ARTICLE{2020SciPy-NMeth,
  author  = {Virtanen, Pauli and Gommers, Ralf and Oliphant, Travis E. and
            Haberland, Matt and Reddy, Tyler and Cournapeau, David and
            Burovski, Evgeni and Peterson, Pearu and Weckesser, Warren and
            Bright, Jonathan and {van der Walt}, St{\'e}fan J. and
            Brett, Matthew and Wilson, Joshua and Millman, K. Jarrod and
            Mayorov, Nikolay and Nelson, Andrew R. J. and Jones, Eric and
            Kern, Robert and Larson, Eric and Carey, C J and
            Polat, {\.I}lhan and Feng, Yu and Moore, Eric W. and
            {VanderPlas}, Jake and Laxalde, Denis and Perktold, Josef and
            Cimrman, Robert and Henriksen, Ian and Quintero, E. A. and
            Harris, Charles R. and Archibald, Anne M. and
            Ribeiro, Ant{\^o}nio H. and Pedregosa, Fabian and
            {van Mulbregt}, Paul and {SciPy 1.0 Contributors}},
  title   = {{{SciPy} 1.0: Fundamental Algorithms for Scientific
            Computing in Python}},
  journal = {Nature Methods},
  year    = {2020},
  volume  = {17},
  pages   = {261--272},
  adsurl  = {https://rdcu.be/b08Wh},
  doi     = {10.1038/s41592-019-0686-2},
}

@ARTICLE{2021MNRAS.501.5575A,
       author = {{Afruni}, Andrea and {Fraternali}, Filippo and {Pezzulli}, Gabriele},
        title = "{Most of the cool CGM of star-forming galaxies is not produced by supernova feedback}",
      journal = {\mnras},
         year = 2021,
        month = mar,
       volume = {501},
       number = {4},
        pages = {5575-5596},
          doi = {10.1093/mnras/staa3759},
archivePrefix = {arXiv},
       eprint = {2012.00770},
 primaryClass = {astro-ph.GA},
       adsurl = {https://ui.adsabs.harvard.edu/abs/2021MNRAS.501.5575A}
}

@ARTICLE{2018MNRAS.477..616E,
       author = {{Elahi}, Pascal J. and {Power}, Chris and {Lagos}, Claudia del P. and {Poulton}, Rhys and {Robotham}, Aaron S.~G.},
        title = "{Using velocity dispersion to estimate halo mass: Is the Local Group in tension with {\ensuremath{\Lambda}}CDM?}",
      journal = {\mnras},
         year = 2018,
        month = jun,
       volume = {477},
       number = {1},
        pages = {616-623},
          doi = {10.1093/mnras/sty590},
archivePrefix = {arXiv},
       eprint = {1712.01989},
 primaryClass = {astro-ph.GA},
       adsurl = {https://ui.adsabs.harvard.edu/abs/2018MNRAS.477..616E}
}

@ARTICLE{2019ApJ...878...84M,
       author = {{Martin}, Crystal L. and {Ho}, Stephanie H. and {Kacprzak}, Glenn G. and {Churchill}, Christopher W.},
        title = "{Kinematics of Circumgalactic Gas: Feeding Galaxies and Feedback}",
      journal = {\apj},
         year = 2019,
        month = jun,
       volume = {878},
       number = {2},
          eid = {84},
        pages = {84},
          doi = {10.3847/1538-4357/ab18ac},
archivePrefix = {arXiv},
       eprint = {1901.09123},
 primaryClass = {astro-ph.GA},
       adsurl = {https://ui.adsabs.harvard.edu/abs/2019ApJ...878...84M}
}

@ARTICLE{2015ApJ...809..147H,
       author = {{Heckman}, Timothy M. and {Alexandroff}, Rachel M. and {Borthakur}, Sanchayeeta and {Overzier}, Roderik and {Leitherer}, Claus},
        title = "{The Systematic Properties of the Warm Phase of Starburst-Driven Galactic Winds}",
      journal = {\apj},
         year = 2015,
        month = aug,
       volume = {809},
       number = {2},
          eid = {147},
        pages = {147},
          doi = {10.1088/0004-637X/809/2/147},
archivePrefix = {arXiv},
       eprint = {1507.05622},
 primaryClass = {astro-ph.GA},
       adsurl = {https://ui.adsabs.harvard.edu/abs/2015ApJ...809..147H}
}

@ARTICLE{2017ApJ...850..156L,
       author = {{Lan}, Ting-Wen and {Fukugita}, Masataka},
        title = "{Mg II Absorbers: Metallicity Evolution and Cloud Morphology}",
      journal = {\apj},
         year = 2017,
        month = dec,
       volume = {850},
       number = {2},
          eid = {156},
        pages = {156},
          doi = {10.3847/1538-4357/aa93eb},
archivePrefix = {arXiv},
       eprint = {1707.09830},
 primaryClass = {astro-ph.GA},
       adsurl = {https://ui.adsabs.harvard.edu/abs/2017ApJ...850..156L}
}

@ARTICLE{2009MNRAS.393..808M,
       author = {{M{\'e}nard}, Brice and {Chelouche}, Doron},
        title = "{On the HI content, dust-to-gas ratio and nature of MgII absorbers}",
      journal = {\mnras},
         year = 2009,
        month = mar,
       volume = {393},
       number = {3},
        pages = {808-815},
          doi = {10.1111/j.1365-2966.2008.14225.x},
archivePrefix = {arXiv},
       eprint = {0803.0745},
 primaryClass = {astro-ph},
       adsurl = {https://ui.adsabs.harvard.edu/abs/2009MNRAS.393..808M}
}

@software{2017zndo...1036773P,
       author = {{Prochaska}, J. Xavier and {Tejos}, Nicolas and {Crighton}, Neil and {Jnburchett} and {Tiffanyhsyu} and {Tuo-Ji} and {Marijana777} and {Ktirimba} and {Jhennawi} and {Cooke}, Ryan and {O'Meara}, John and {Werk}, Jessica},
        title = "{linetools/linetools: Third Minor Release}",
         year = 2017,
        month = oct,
          eid = {10.5281/zenodo.1036773},
          doi = {10.5281/zenodo.1036773},
      version = {v0.3},
    publisher = {Zenodo},
       adsurl = {https://ui.adsabs.harvard.edu/abs/2017zndo...1036773P}
}

@ARTICLE{2024A&A...691A.356L,
       author = {{Lopez}, S. and {Afruni}, A. and {Zamora}, D. and {Tejos}, N. and {Ledoux}, C. and {Hernandez}, J. and {Berg}, T.~A.~M. and {Cortes}, H. and {Urbina}, F. and {Johnston}, E.~J. and {Barrientos}, L.~F. and {Bayliss}, M.~B. and {Cuellar}, R. and {Krogager}, J.~K. and {Noterdaeme}, P. and {Solimano}, M.},
        title = "{Transverse clues on the kiloparsec-scale structure of the circumgalactic medium as traced by C IV absorption}",
      journal = {\aap},
         year = 2024,
        month = nov,
       volume = {691},
          eid = {A356},
        pages = {A356},
          doi = {10.1051/0004-6361/202451200},
archivePrefix = {arXiv},
       eprint = {2410.03029},
 primaryClass = {astro-ph.GA},
       adsurl = {https://ui.adsabs.harvard.edu/abs/2024A&A...691A.356L}
}

@Article{Hunter:2007,
  Author    = {Hunter, J. D.},
  Title     = {Matplotlib: A 2D graphics environment},
  Journal   = {Computing in Science \& Engineering},
  Volume    = {9},
  Number    = {3},
  Pages     = {90--95},
  publisher = {IEEE COMPUTER SOC},
  doi       = {10.1109/MCSE.2007.55},
  year      = 2007
}

@ARTICLE{2025A&A...693A.200B,
       author = {{Berg}, Trystyn A.~M. and {Afruni}, Andrea and {Ledoux}, C{\'e}dric and {Lopez}, Sebastian and {Noterdaeme}, Pasquier and {Tejos}, Nicolas and {Hernandez}, Joaquin and {Barrientos}, Felipe and {Johnston}, Evelyn J.},
        title = "{Mapping the spatial extent of H I-rich absorbers using Mg II absorption along gravitational arcs}",
      journal = {\aap},
         year = 2025,
        month = jan,
       volume = {693},
          eid = {A200},
        pages = {A200},
          doi = {10.1051/0004-6361/202452199},
archivePrefix = {arXiv},
       eprint = {2412.07652},
 primaryClass = {astro-ph.GA},
       adsurl = {https://ui.adsabs.harvard.edu/abs/2025A&A...693A.200B}
}

@ARTICLE{2019MNRAS.490..417C,
       author = {{Carnall}, A.~C. and {McLure}, R.~J. and {Dunlop}, J.~S. and {Cullen}, F. and {McLeod}, D.~J. and {Wild}, V. and {Johnson}, B.~D. and {Appleby}, S. and {Dav{\'e}}, R. and {Amorin}, R. and {Bolzonella}, M. and {Castellano}, M. and {Cimatti}, A. and {Cucciati}, O. and {Gargiulo}, A. and {Garilli}, B. and {Marchi}, F. and {Pentericci}, L. and {Pozzetti}, L. and {Schreiber}, C. and {Talia}, M. and {Zamorani}, G.},
        title = "{The VANDELS survey: the star-formation histories of massive quiescent galaxies at 1.0 < z < 1.3}",
      journal = {\mnras},
         year = 2019,
        month = nov,
       volume = {490},
       number = {1},
        pages = {417-439},
          doi = {10.1093/mnras/stz2544},
archivePrefix = {arXiv},
       eprint = {1903.11082},
 primaryClass = {astro-ph.GA},
       adsurl = {https://ui.adsabs.harvard.edu/abs/2019MNRAS.490..417C}
}

@Article{         harris2020array,
 title         = {Array programming with {NumPy}},
 author        = {Charles R. Harris and K. Jarrod Millman and St{\'{e}}fan J.
                 van der Walt and Ralf Gommers and Pauli Virtanen and David
                 Cournapeau and Eric Wieser and Julian Taylor and Sebastian
                 Berg and Nathaniel J. Smith and Robert Kern and Matti Picus
                 and Stephan Hoyer and Marten H. van Kerkwijk and Matthew
                 Brett and Allan Haldane and Jaime Fern{\'{a}}ndez del
                 R{\'{i}}o and Mark Wiebe and Pearu Peterson and Pierre
                 G{\'{e}}rard-Marchant and Kevin Sheppard and Tyler Reddy and
                 Warren Weckesser and Hameer Abbasi and Christoph Gohlke and
                 Travis E. Oliphant},
 year          = {2020},
 month         = sep,
 journal       = {Nature},
 volume        = {585},
 number        = {7825},
 pages         = {357--362},
 doi           = {10.1038/s41586-020-2649-2},
 publisher     = {Springer Science and Business Media {LLC}},
 url           = {https://doi.org/10.1038/s41586-020-2649-2}
}

@software{2016ascl.soft11003B,
       author = {{Bacon}, Roland and {Piqueras}, Laure and {Conseil}, Simon and {Richard}, Johan and {Shepherd}, Martin},
        title = "{MPDAF: MUSE Python Data Analysis Framework}",
 howpublished = {Astrophysics Source Code Library, record ascl:1611.003},
         year = 2016,
        month = nov,
          eid = {ascl:1611.003},
       adsurl = {https://ui.adsabs.harvard.edu/abs/2016ascl.soft11003B}
}

@ARTICLE{2019MNRAS.490.4368S,
       author = {{Schroetter}, Ilane and {Bouch{\'e}}, Nicolas F. and {Zabl}, Johannes and {Contini}, Thierry and {Wendt}, Martin and {Schaye}, Joop and {Mitchell}, Peter and {Muzahid}, Sowgat and {Marino}, Raffaella A. and {Bacon}, Roland and {Lilly}, Simon J. and {Richard}, Johan and {Wisotzki}, Lutz},
        title = "{MusE GAs FLOw and Wind (MEGAFLOW) - III. Galactic wind properties using background quasars}",
      journal = {\mnras},
         year = 2019,
        month = dec,
       volume = {490},
       number = {3},
        pages = {4368-4381},
          doi = {10.1093/mnras/stz2822},
archivePrefix = {arXiv},
       eprint = {1907.09967},
 primaryClass = {astro-ph.GA},
       adsurl = {https://ui.adsabs.harvard.edu/abs/2019MNRAS.490.4368S}
}

@INPROCEEDINGS{1995qal..conf..139S,
       author = {{Steidel}, C.~C.},
        title = "{The Nature and Evolution of Absorption-Selected Galaxies}",
    booktitle = {QSO Absorption Lines},
         year = 1995,
       editor = {{Meylan}, Georges},
        month = jan,
        pages = {139},
          doi = {10.48550/arXiv.astro-ph/9509098},
archivePrefix = {arXiv},
       eprint = {astro-ph/9509098},
 primaryClass = {astro-ph},
       adsurl = {https://ui.adsabs.harvard.edu/abs/1995qal..conf..139S}
}

@INCOLLECTION{2025drzp.book....3A,
       author = {{Anand}, G.~S. and {Mack}, J. and {Avila}, R.~J. and {Bajaj}, V. and {Kuhn}, B. and {Revalski}, M. and {Som}, D.},
        title = "{The DrizzlePac Handbook, v. 3}",
    booktitle = {The DrizzlePac Handbook},
         year = 2025,
        pages = {3},
       adsurl = {https://ui.adsabs.harvard.edu/abs/2025drzp.book....3A}
}

@ARTICLE{2005ApJ...618..569M,
       author = {{Murray}, Norman and {Quataert}, Eliot and {Thompson}, Todd A.},
        title = "{On the Maximum Luminosity of Galaxies and Their Central Black Holes: Feedback from Momentum-driven Winds}",
      journal = {\apj},
         year = 2005,
        month = jan,
       volume = {618},
       number = {2},
        pages = {569-585},
          doi = {10.1086/426067},
archivePrefix = {arXiv},
       eprint = {astro-ph/0406070},
 primaryClass = {astro-ph},
       adsurl = {https://ui.adsabs.harvard.edu/abs/2005ApJ...618..569M}
}

@ARTICLE{2025A&A...703A..21L,
       author = {{Lopez}, Sebastian and {Krogager}, Jens-Kristian},
        title = "{How far have metals reached? Reconciling statistical constraints and enrichment models at reionization}",
      journal = {\aap},
         year = 2025,
        month = nov,
       volume = {703},
          eid = {A21},
        pages = {A21},
          doi = {10.1051/0004-6361/202555546},
archivePrefix = {arXiv},
       eprint = {2509.02434},
 primaryClass = {astro-ph.GA},
       adsurl = {https://ui.adsabs.harvard.edu/abs/2025A&A...703A..21L}
}

@ARTICLE{2012ApJ...761..112M,
       author = {{Matejek}, Michael S. and {Simcoe}, Robert A.},
        title = "{A Survey of Mg II Absorption at 2 < z < 6 with Magellan/FIRE. I. Sample and Evolution of the Mg II Frequency}",
      journal = {\apj},
         year = 2012,
        month = dec,
       volume = {761},
       number = {2},
          eid = {112},
        pages = {112},
          doi = {10.1088/0004-637X/761/2/112},
archivePrefix = {arXiv},
       eprint = {1201.3919},
 primaryClass = {astro-ph.CO},
       adsurl = {https://ui.adsabs.harvard.edu/abs/2012ApJ...761..112M}
}

@ARTICLE{2011ApJ...728...55R,
       author = {{Rubin}, Kate H.~R. and {Prochaska}, J. Xavier and {M{\'e}nard}, Brice and {Murray}, Norman and {Kasen}, Daniel and {Koo}, David C. and {Phillips}, Andrew C.},
        title = "{Low-ionization Line Emission from a Starburst Galaxy: A New Probe of a Galactic-scale Outflow}",
      journal = {\apj},
         year = 2011,
        month = feb,
       volume = {728},
       number = {1},
          eid = {55},
        pages = {55},
          doi = {10.1088/0004-637X/728/1/55},
archivePrefix = {arXiv},
       eprint = {1008.3397},
 primaryClass = {astro-ph.CO},
       adsurl = {https://ui.adsabs.harvard.edu/abs/2011ApJ...728...55R}
}

@ARTICLE{2018ApJ...868..142R,
       author = {{Rubin}, Kate H.~R. and {Diamond-Stanic}, Aleksandar M. and {Coil}, Alison L. and {Crighton}, Neil H.~M. and {Stewart}, Kyle R.},
        title = "{Galaxies Probing Galaxies in PRIMUS. II. The Coherence Scale of the Cool Circumgalactic Medium}",
      journal = {\apj},
         year = 2018,
        month = dec,
       volume = {868},
       number = {2},
          eid = {142},
        pages = {142},
          doi = {10.3847/1538-4357/aad566},
archivePrefix = {arXiv},
       eprint = {1806.08801},
 primaryClass = {astro-ph.GA},
       adsurl = {https://ui.adsabs.harvard.edu/abs/2018ApJ...868..142R}
}

@ARTICLE{1996ApJ...471..164C,
       author = {{Churchill}, Christopher W. and {Steidel}, Charles C. and {Vogt}, Steven S.},
        title = "{On the Spatial and Kinematic Distributions of Mg II Absorbing Gas in < z< approximately 0.7 Galaxies}",
      journal = {\apj},
         year = 1996,
        month = nov,
       volume = {471},
        pages = {164},
          doi = {10.1086/177960},
archivePrefix = {arXiv},
       eprint = {astro-ph/9605180},
 primaryClass = {astro-ph},
       adsurl = {https://ui.adsabs.harvard.edu/abs/1996ApJ...471..164C}
}

@ARTICLE{1991ApJS...77....1L,
       author = {{Lanzetta}, Kenneth M. and {Wolfe}, Arthur M. and {Turnshek}, David A. and {Lu}, Limin and {McMahon}, Richard G. and {Hazard}, Cyril},
        title = "{A New Spectroscopic Survey for Damped LY alpha Absorption Lines from High-Redshift Galaxies}",
      journal = {\apjs},
         year = 1991,
        month = sep,
       volume = {77},
        pages = {1},
          doi = {10.1086/191596},
       adsurl = {https://ui.adsabs.harvard.edu/abs/1991ApJS...77....1L}
}

@ARTICLE{2009ApJ...700.1299J,
       author = {{Jenkins}, Edward B.},
        title = "{A Unified Representation of Gas-Phase Element Depletions in the Interstellar Medium}",
      journal = {\apj},
         year = 2009,
        month = aug,
       volume = {700},
       number = {2},
        pages = {1299-1348},
          doi = {10.1088/0004-637X/700/2/1299},
archivePrefix = {arXiv},
       eprint = {0905.3173},
 primaryClass = {astro-ph.GA},
       adsurl = {https://ui.adsabs.harvard.edu/abs/2009ApJ...700.1299J}
}

@ARTICLE{2012ARA&A..50..531K,
       author = {{Kennicutt}, Robert C. and {Evans}, Neal J.},
        title = "{Star Formation in the Milky Way and Nearby Galaxies}",
      journal = {\araa},
         year = 2012,
        month = sep,
       volume = {50},
        pages = {531-608},
          doi = {10.1146/annurev-astro-081811-125610},
archivePrefix = {arXiv},
       eprint = {1204.3552},
 primaryClass = {astro-ph.GA},
       adsurl = {https://ui.adsabs.harvard.edu/abs/2012ARA&A..50..531K}
}

@ARTICLE{2023A&A...680A.112A,
       author = {{Afruni}, A. and {Lopez}, S. and {Anshul}, P. and {Tejos}, N. and {Noterdaeme}, P. and {Berg}, T.~A.~M. and {Ledoux}, C. and {Solimano}, M. and {Gonzalez-Lopez}, J. and {Gronke}, M. and {Barrientos}, F. and {Johnston}, E.~J.},
        title = "{Directly constraining the spatial coherence of the z {\ensuremath{\sim}} 1 circumgalactic medium}",
      journal = {\aap},
         year = 2023,
        month = dec,
       volume = {680},
          eid = {A112},
        pages = {A112},
          doi = {10.1051/0004-6361/202347867},
archivePrefix = {arXiv},
       eprint = {2310.13732},
 primaryClass = {astro-ph.GA},
       adsurl = {https://ui.adsabs.harvard.edu/abs/2023A&A...680A.112A}
}

@ARTICLE{2023MNRAS.519.1526P,
       author = {{Popesso}, P. and {Concas}, A. and {Cresci}, G. and {Belli}, S. and {Rodighiero}, G. and {Inami}, H. and {Dickinson}, M. and {Ilbert}, O. and {Pannella}, M. and {Elbaz}, D.},
        title = "{The main sequence of star-forming galaxies across cosmic times}",
      journal = {\mnras},
         year = 2023,
        month = feb,
       volume = {519},
       number = {1},
        pages = {1526-1544},
          doi = {10.1093/mnras/stac3214},
archivePrefix = {arXiv},
       eprint = {2203.10487},
 primaryClass = {astro-ph.GA},
       adsurl = {https://ui.adsabs.harvard.edu/abs/2023MNRAS.519.1526P}
}

@ARTICLE{1986A&A...169....1B,
       author = {{Bergeron}, J. and {Stasi{\'n}ska}, G.},
        title = "{Absorption line systems in QSO spectra : properties derived from observations and from photoionization models.}",
      journal = {\aap},
         year = 1986,
        month = nov,
       volume = {169},
        pages = {1-3},
       adsurl = {https://ui.adsabs.harvard.edu/abs/1986A&A...169....1B}
}

@ARTICLE{1996ARA&A..34..279S,
       author = {{Savage}, Blair D. and {Sembach}, Kenneth R.},
        title = "{Interstellar Abundances from Absorption-Line Observations with the Hubble Space Telescope}",
      journal = {\araa},
         year = 1996,
        month = jan,
       volume = {34},
        pages = {279-330},
          doi = {10.1146/annurev.astro.34.1.279},
       adsurl = {https://ui.adsabs.harvard.edu/abs/1996ARA&A..34..279S}
}

@ARTICLE{2002AJ....124..266P,
       author = {{Peng}, Chien Y. and {Ho}, Luis C. and {Impey}, Chris D. and {Rix}, Hans-Walter},
        title = "{Detailed Structural Decomposition of Galaxy Images}",
      journal = {\aj},
         year = 2002,
        month = jul,
       volume = {124},
       number = {1},
        pages = {266-293},
          doi = {10.1086/340952},
archivePrefix = {arXiv},
       eprint = {astro-ph/0204182},
 primaryClass = {astro-ph},
       adsurl = {https://ui.adsabs.harvard.edu/abs/2002AJ....124..266P}
}

@ARTICLE{2017ApJ...835..267H,
       author = {{Ho}, Stephanie H. and {Martin}, Crystal L. and {Kacprzak}, Glenn G. and {Churchill}, Christopher W.},
        title = "{Quasars Probing Galaxies. I. Signatures of Gas Accretion at Redshift Approximately 0.2}",
      journal = {\apj},
         year = 2017,
        month = feb,
       volume = {835},
       number = {2},
          eid = {267},
        pages = {267},
          doi = {10.3847/1538-4357/835/2/267},
archivePrefix = {arXiv},
       eprint = {1611.04579},
 primaryClass = {astro-ph.GA},
       adsurl = {https://ui.adsabs.harvard.edu/abs/2017ApJ...835..267H}
}

@ARTICLE{2018MNRAS.477...18P,
       author = {{Patr{\'\i}cio}, V. and {Richard}, J. and {Carton}, D. and {Contini}, T. and {Epinat}, B. and {Brinchmann}, J. and {Schmidt}, K.~B. and {Krajnovi{\'c}}, D. and {Bouch{\'e}}, N. and {Weilbacher}, P.~M. and {Pell{\'o}}, R. and {Caruana}, J. and {Maseda}, M. and {Finley}, H. and {Bauer}, F.~E. and {Martinez}, J. and {Mahler}, G. and {Lagattuta}, D. and {Cl{\'e}ment}, B. and {Soucail}, G. and {Wisotzki}, L.},
        title = "{Kinematics, turbulence, and star formation of z {\ensuremath{\sim}} 1 strongly lensed galaxies seen with MUSE}",
      journal = {\mnras},
         year = 2018,
        month = jun,
       volume = {477},
       number = {1},
        pages = {18-44},
          doi = {10.1093/mnras/sty555},
archivePrefix = {arXiv},
       eprint = {1802.08451},
 primaryClass = {astro-ph.GA},
       adsurl = {https://ui.adsabs.harvard.edu/abs/2018MNRAS.477...18P}
}

@ARTICLE{2017A&A...608A...5G,
       author = {{Gu{\'e}rou}, Adrien and {Krajnovi{\'c}}, Davor and {Epinat}, Benoit and {Contini}, Thierry and {Emsellem}, Eric and {Bouch{\'e}}, Nicolas and {Bacon}, Roland and {Michel-Dansac}, Leo and {Richard}, Johan and {Weilbacher}, Peter M. and {Schaye}, Joop and {Marino}, Raffaella Anna and {den Brok}, Mark and {Erroz-Ferrer}, Santiago},
        title = "{The MUSE Hubble Ultra Deep Field Survey. V. Spatially resolved stellar kinematics of galaxies at redshift 0.2 {\ensuremath{\lesssim}} z {\ensuremath{\lesssim}} 0.8}",
      journal = {\aap},
         year = 2017,
        month = nov,
       volume = {608},
          eid = {A5},
        pages = {A5},
          doi = {10.1051/0004-6361/201730905},
archivePrefix = {arXiv},
       eprint = {1710.07694},
 primaryClass = {astro-ph.GA},
       adsurl = {https://ui.adsabs.harvard.edu/abs/2017A&A...608A...5G}
}

@ARTICLE{2019MNRAS.490.3234N,
       author = {{Nelson}, Dylan and {Pillepich}, Annalisa and {Springel}, Volker and {Pakmor}, R{\"u}diger and {Weinberger}, Rainer and {Genel}, Shy and {Torrey}, Paul and {Vogelsberger}, Mark and {Marinacci}, Federico and {Hernquist}, Lars},
        title = "{First results from the TNG50 simulation: galactic outflows driven by supernovae and black hole feedback}",
      journal = {\mnras},
         year = 2019,
        month = dec,
       volume = {490},
       number = {3},
        pages = {3234-3261},
          doi = {10.1093/mnras/stz2306},
archivePrefix = {arXiv},
       eprint = {1902.05554},
 primaryClass = {astro-ph.GA},
       adsurl = {https://ui.adsabs.harvard.edu/abs/2019MNRAS.490.3234N}
}

@ARTICLE{2020MNRAS.494.3971M,
       author = {{Mitchell}, Peter D. and {Schaye}, Joop and {Bower}, Richard G. and {Crain}, Robert A.},
        title = "{Galactic outflow rates in the EAGLE simulations}",
      journal = {\mnras},
         year = 2020,
        month = may,
       volume = {494},
       number = {3},
        pages = {3971-3997},
          doi = {10.1093/mnras/staa938},
archivePrefix = {arXiv},
       eprint = {1910.09566},
 primaryClass = {astro-ph.GA},
       adsurl = {https://ui.adsabs.harvard.edu/abs/2020MNRAS.494.3971M}
}

@ARTICLE{2025ApJ...986..190S,
       author = {{Shaban}, Ahmed and {Bordoloi}, Rongmon and {O'Meara}, John M. and {Sharon}, Keren and {Tejos}, Nicolas and {Lopez}, Sebastian and {Ledoux}, C{\'e}dric and {Barrientos}, L. Felipe and {Rigby}, Jane R.},
        title = "{Spatially Resolved Circumgalactic Medium around a Star-forming Galaxy Driving a Galactic Outflow at z {\ensuremath{\approx}} 0.8}",
      journal = {\apj},
         year = 2025,
        month = jun,
       volume = {986},
       number = {2},
          eid = {190},
        pages = {190},
          doi = {10.3847/1538-4357/add0b9},
archivePrefix = {arXiv},
       eprint = {2501.17940},
 primaryClass = {astro-ph.GA},
       adsurl = {https://ui.adsabs.harvard.edu/abs/2025ApJ...986..190S}
}

@ARTICLE{2021MNRAS.502.4743H,
       author = {{Huang}, Yun-Hsin and {Chen}, Hsiao-Wen and {Shectman}, Stephen A. and {Johnson}, Sean D. and {Zahedy}, Fakhri S. and {Helsby}, Jennifer E. and {Gauthier}, Jean-Ren{\'e} and {Thompson}, Ian B.},
        title = "{A complete census of circumgalactic Mg II at redshift z {\ensuremath{\lesssim}} 0.5}",
      journal = {\mnras},
         year = 2021,
        month = apr,
       volume = {502},
       number = {4},
        pages = {4743-4761},
          doi = {10.1093/mnras/stab360},
archivePrefix = {arXiv},
       eprint = {2009.12372},
 primaryClass = {astro-ph.GA},
       adsurl = {https://ui.adsabs.harvard.edu/abs/2021MNRAS.502.4743H}
}

@ARTICLE{2021MNRAS.508.2979P,
       author = {{Pandya}, Viraj and {Fielding}, Drummond B. and {Angl{\'e}s-Alc{\'a}zar}, Daniel and {Somerville}, Rachel S. and {Bryan}, Greg L. and {Hayward}, Christopher C. and {Stern}, Jonathan and {Kim}, Chang-Goo and {Quataert}, Eliot and {Forbes}, John C. and {Faucher-Gigu{\`e}re}, Claude-Andr{\'e} and {Feldmann}, Robert and {Hafen}, Zachary and {Hopkins}, Philip F. and {Kere{\v{s}}}, Du{\v{s}}an and {Murray}, Norman and {Wetzel}, Andrew},
        title = "{Characterizing mass, momentum, energy, and metal outflow rates of multiphase galactic winds in the FIRE-2 cosmological simulations}",
      journal = {\mnras},
         year = 2021,
        month = dec,
       volume = {508},
       number = {2},
        pages = {2979-3008},
          doi = {10.1093/mnras/stab2714},
archivePrefix = {arXiv},
       eprint = {2103.06891},
 primaryClass = {astro-ph.GA},
       adsurl = {https://ui.adsabs.harvard.edu/abs/2021MNRAS.508.2979P}
}

@ARTICLE{2022MNRAS.517.2214F,
       author = {{Fernandez-Figueroa}, A. and {Lopez}, S. and {Tejos}, N. and {Berg}, T.~A.~M. and {Ledoux}, C. and {Noterdaeme}, P. and {Afruni}, A. and {Barrientos}, L.~F. and {Gonzalez-Lopez}, J. and {Hamel}, M. and {Johnston}, E.~J. and {Katsianis}, A. and {Sharon}, K. and {Solimano}, M.},
        title = "{Orientation effects on cool gas absorption from gravitational-arc tomography of a z = 0.77 disc galaxy}",
      journal = {\mnras},
         year = 2022,
        month = dec,
       volume = {517},
       number = {2},
        pages = {2214-2220},
          doi = {10.1093/mnras/stac2851},
archivePrefix = {arXiv},
       eprint = {2209.14134},
 primaryClass = {astro-ph.GA},
       adsurl = {https://ui.adsabs.harvard.edu/abs/2022MNRAS.517.2214F}
}

@ARTICLE{2021MNRAS.507..663T,
       author = {{Tejos}, N. and {L{\'o}pez}, S. and {Ledoux}, C. and {Fern{\'a}ndez-Figueroa}, A. and {Rivas}, N. and {Sharon}, K. and {Johnston}, E.~J. and {Florian}, M.~K. and {D'Ago}, G. and {Katsianis}, A. and {Barrientos}, F. and {Berg}, T. and {Corro-Guerra}, F. and {Hamel}, M. and {Moya-Sierralta}, C. and {Poudel}, S. and {Rigby}, J.~R. and {Solimano}, M.},
        title = "{Telltale signs of metal recycling in the circumgalactic medium of a z   0.77 galaxy}",
      journal = {\mnras},
         year = 2021,
        month = oct,
       volume = {507},
       number = {1},
        pages = {663-679},
          doi = {10.1093/mnras/stab2147},
archivePrefix = {arXiv},
       eprint = {2105.01673},
 primaryClass = {astro-ph.GA},
       adsurl = {https://ui.adsabs.harvard.edu/abs/2021MNRAS.507..663T}
}

@ARTICLE{2005ApJ...621..227M,
       author = {{Martin}, Crystal L.},
        title = "{Mapping Large-Scale Gaseous Outflows in Ultraluminous Galaxies with Keck II ESI Spectra: Variations in Outflow Velocity with Galactic Mass}",
      journal = {\apj},
         year = 2005,
        month = mar,
       volume = {621},
       number = {1},
        pages = {227-245},
          doi = {10.1086/427277},
archivePrefix = {arXiv},
       eprint = {astro-ph/0410247},
 primaryClass = {astro-ph},
       adsurl = {https://ui.adsabs.harvard.edu/abs/2005ApJ...621..227M}
}

@ARTICLE{2020MNRAS.499.5022D,
       author = {{Dutta}, Rajeshwari and {Fumagalli}, Michele and {Fossati}, Matteo and {Lofthouse}, Emma K. and {Prochaska}, J. Xavier and {Arrigoni Battaia}, Fabrizio and {Bielby}, Richard M. and {Cantalupo}, Sebastiano and {Cooke}, Ryan J. and {Murphy}, Michael T. and {O'Meara}, John M.},
        title = "{MUSE Analysis of Gas around Galaxies (MAGG) - II: metal-enriched halo gas around z {\ensuremath{\sim}} 1 galaxies}",
      journal = {\mnras},
         year = 2020,
        month = dec,
       volume = {499},
       number = {4},
        pages = {5022-5046},
          doi = {10.1093/mnras/staa3147},
archivePrefix = {arXiv},
       eprint = {2009.14219},
 primaryClass = {astro-ph.GA},
       adsurl = {https://ui.adsabs.harvard.edu/abs/2020MNRAS.499.5022D}
}

@ARTICLE{2020MNRAS.491.4442L,
       author = {{Lopez}, S. and {Tejos}, N. and {Barrientos}, L.~F. and {Ledoux}, C. and {Sharon}, K. and {Katsianis}, A. and {Florian}, M.~K. and {Rivera-Thorsen}, E. and {Bayliss}, M.~B. and {Dahle}, H. and {Fernandez-Figueroa}, A. and {Gladders}, M.~D. and {Gronke}, M. and {Hamel}, M. and {Pessa}, I. and {Rigby}, J.~R.},
        title = "{Slicing the cool circumgalactic medium along the major axis of a star-forming galaxy at z = 0.7}",
      journal = {\mnras},
         year = 2020,
        month = jan,
       volume = {491},
       number = {3},
        pages = {4442-4461},
          doi = {10.1093/mnras/stz3183},
archivePrefix = {arXiv},
       eprint = {1911.04809},
 primaryClass = {astro-ph.GA},
       adsurl = {https://ui.adsabs.harvard.edu/abs/2020MNRAS.491.4442L}
}

@ARTICLE{2011ApJ...737..103S,
       author = {{Schlafly}, Edward F. and {Finkbeiner}, Douglas P.},
        title = "{Measuring Reddening with Sloan Digital Sky Survey Stellar Spectra and Recalibrating SFD}",
      journal = {\apj},
         year = 2011,
        month = aug,
       volume = {737},
       number = {2},
          eid = {103},
        pages = {103},
          doi = {10.1088/0004-637X/737/2/103},
archivePrefix = {arXiv},
       eprint = {1012.4804},
 primaryClass = {astro-ph.GA},
       adsurl = {https://ui.adsabs.harvard.edu/abs/2011ApJ...737..103S}
}

@ARTICLE{2018ApJ...866...36L,
       author = {{Lan}, Ting-Wen and {Mo}, Houjun},
        title = "{The Circumgalactic Medium of eBOSS Emission Line Galaxies: Signatures of Galactic Outflows in Gas Distribution and Kinematics}",
      journal = {\apj},
         year = 2018,
        month = oct,
       volume = {866},
       number = {1},
          eid = {36},
        pages = {36},
          doi = {10.3847/1538-4357/aadc08},
archivePrefix = {arXiv},
       eprint = {1806.05786},
 primaryClass = {astro-ph.GA},
       adsurl = {https://ui.adsabs.harvard.edu/abs/2018ApJ...866...36L}
}

@ARTICLE{2019MNRAS.485.1961Z,
       author = {{Zabl}, Johannes and {Bouch{\'e}}, Nicolas F. and {Schroetter}, Ilane and {Wendt}, Martin and {Finley}, Hayley and {Schaye}, Joop and {Conseil}, Simon and {Contini}, Thierry and {Marino}, Raffaella A. and {Mitchell}, Peter and {Muzahid}, Sowgat and {Pezzulli}, Gabriele and {Wisotzki}, Lutz},
        title = "{MusE GAs FLOw and Wind (MEGAFLOW) II. A study of gas accretion around z {\ensuremath{\approx}} 1 star-forming galaxies with background quasars}",
      journal = {\mnras},
         year = 2019,
        month = may,
       volume = {485},
       number = {2},
        pages = {1961-1980},
          doi = {10.1093/mnras/stz392},
archivePrefix = {arXiv},
       eprint = {1901.11416},
 primaryClass = {astro-ph.GA},
       adsurl = {https://ui.adsabs.harvard.edu/abs/2019MNRAS.485.1961Z}
}

@ARTICLE{2018MNRAS.480.4379C,
       author = {{Carnall}, A.~C. and {McLure}, R.~J. and {Dunlop}, J.~S. and {Dav{\'e}}, R.},
        title = "{Inferring the star formation histories of massive quiescent galaxies with BAGPIPES: evidence for multiple quenching mechanisms}",
      journal = {\mnras},
         year = 2018,
        month = nov,
       volume = {480},
       number = {4},
        pages = {4379-4401},
          doi = {10.1093/mnras/sty2169},
archivePrefix = {arXiv},
       eprint = {1712.04452},
 primaryClass = {astro-ph.GA},
       adsurl = {https://ui.adsabs.harvard.edu/abs/2018MNRAS.480.4379C}
}

@ARTICLE{2018Natur.554..493L,
       author = {{Lopez}, Sebastian and {Tejos}, Nicolas and {Ledoux}, C{\'e}dric and {Barrientos}, L. Felipe and {Sharon}, Keren and {Rigby}, Jane R. and {Gladders}, Michael D. and {Bayliss}, Matthew B. and {Pessa}, Ismael},
        title = "{A clumpy and anisotropic galaxy halo at redshift 1 from gravitational-arc tomography}",
      journal = {\nat},
         year = 2018,
        month = feb,
       volume = {554},
       number = {7693},
        pages = {493-496},
          doi = {10.1038/nature25436},
archivePrefix = {arXiv},
       eprint = {1801.10175},
 primaryClass = {astro-ph.GA},
       adsurl = {https://ui.adsabs.harvard.edu/abs/2018Natur.554..493L}
}

@ARTICLE{2017ARA&A..55..389T,
       author = {{Tumlinson}, Jason and {Peeples}, Molly S. and {Werk}, Jessica K.},
        title = "{The Circumgalactic Medium}",
      journal = {\araa},
         year = 2017,
        month = aug,
       volume = {55},
       number = {1},
        pages = {389-432},
          doi = {10.1146/annurev-astro-091916-055240},
archivePrefix = {arXiv},
       eprint = {1709.09180},
 primaryClass = {astro-ph.GA},
       adsurl = {https://ui.adsabs.harvard.edu/abs/2017ARA&A..55..389T}
}

@ARTICLE{2017MNRAS.467.3306S,
       author = {{Smit}, Renske and {Swinbank}, A.~M. and {Massey}, Richard and {Richard}, Johan and {Smail}, Ian and {Kneib}, J. -P.},
        title = "{A gravitationally boosted MUSE survey for emission-line galaxies at z {\ensuremath{\gtrsim}} 5 behind the massive cluster RCS 0224}",
      journal = {\mnras},
         year = 2017,
        month = may,
       volume = {467},
       number = {3},
        pages = {3306-3323},
          doi = {10.1093/mnras/stx245},
archivePrefix = {arXiv},
       eprint = {1701.08160},
 primaryClass = {astro-ph.GA},
       adsurl = {https://ui.adsabs.harvard.edu/abs/2017MNRAS.467.3306S}
}

@ARTICLE{2016ApJ...833...39S,
       author = {{Schroetter}, I. and {Bouch{\'e}}, N. and {Wendt}, M. and {Contini}, T. and {Finley}, H. and {Pell{\'o}}, R. and {Bacon}, R. and {Cantalupo}, S. and {Marino}, R.~A. and {Richard}, J. and {Lilly}, S.~J. and {Schaye}, J. and {Soto}, K. and {Steinmetz}, M. and {Straka}, L.~A. and {Wisotzki}, L.},
        title = "{Muse Gas Flow and Wind (MEGAFLOW). I. First MUSE Results on Background Quasars}",
      journal = {\apj},
         year = 2016,
        month = dec,
       volume = {833},
       number = {1},
          eid = {39},
        pages = {39},
          doi = {10.3847/1538-4357/833/1/39},
archivePrefix = {arXiv},
       eprint = {1605.03412},
 primaryClass = {astro-ph.GA},
       adsurl = {https://ui.adsabs.harvard.edu/abs/2016ApJ...833...39S}
}

@ARTICLE{2021ApJ...913...50L,
       author = {{Lundgren}, Britt F. and {Creech}, Samantha and {Brammer}, Gabriel and {Kirse}, Nathan and {Peek}, Matthew and {Wake}, David and {York}, Donald G. and {Chisholm}, John and {Erb}, Dawn K. and {Kulkarni}, Varsha P. and {Straka}, Lorrie and {Tremonti}, Christy and {van Dokkum}, Pieter},
        title = "{The Geometry of Cold, Metal-enriched Gas around Galaxies at z {\ensuremath{\sim}} 1.2}",
      journal = {\apj},
         year = 2021,
        month = may,
       volume = {913},
       number = {1},
          eid = {50},
        pages = {50},
          doi = {10.3847/1538-4357/abef6a},
archivePrefix = {arXiv},
       eprint = {2102.10117},
 primaryClass = {astro-ph.GA},
       adsurl = {https://ui.adsabs.harvard.edu/abs/2021ApJ...913...50L}
}

@ARTICLE{2010ApJ...714.1521C,
       author = {{Chen}, Hsiao-Wen and {Helsby}, Jennifer E. and {Gauthier}, Jean-Ren{\'e} and {Shectman}, Stephen A. and {Thompson}, Ian B. and {Tinker}, Jeremy L.},
        title = "{An Empirical Characterization of Extended Cool Gas Around Galaxies Using Mg II Absorption Features}",
      journal = {\apj},
         year = 2010,
        month = may,
       volume = {714},
       number = {2},
        pages = {1521-1541},
          doi = {10.1088/0004-637X/714/2/1521},
archivePrefix = {arXiv},
       eprint = {1004.0705},
 primaryClass = {astro-ph.CO},
       adsurl = {https://ui.adsabs.harvard.edu/abs/2010ApJ...714.1521C}
}

@ARTICLE{2008ApJ...687..745C,
       author = {{Chen}, Hsiao-Wen and {Tinker}, Jeremy L.},
        title = "{The Baryon Content of Dark Matter Halos: Empirical Constraints from Mg II Absorbers}",
      journal = {\apj},
         year = 2008,
        month = nov,
       volume = {687},
       number = {2},
        pages = {745-756},
          doi = {10.1086/591927},
archivePrefix = {arXiv},
       eprint = {0801.2169},
 primaryClass = {astro-ph},
       adsurl = {https://ui.adsabs.harvard.edu/abs/2008ApJ...687..745C}
}

@ARTICLE{2020ARA&A..58..363P,
       author = {{P{\'e}roux}, C{\'e}line and {Howk}, J. Christopher},
        title = "{The Cosmic Baryon and Metal Cycles}",
      journal = {\araa},
         year = 2020,
        month = aug,
       volume = {58},
        pages = {363-406},
          doi = {10.1146/annurev-astro-021820-120014},
archivePrefix = {arXiv},
       eprint = {2011.01935},
 primaryClass = {astro-ph.GA},
       adsurl = {https://ui.adsabs.harvard.edu/abs/2020ARA&A..58..363P}
}

@ARTICLE{2015ascl...soft01014B,
       author = {{Bouch{\'e}}, N. and {Carfantan}, H. and {Schroetter}, I. and {Michel-Dansac}, L. and {Contini}, T.},
        title = "{GalPaK 3D: Galaxy parameters and kinematics extraction from 3D data}",
 howpublished = {Astrophysics Source Code Library, record ascl:1501.014},
         year = 2015,
        month = jan,
          eid = {ascl:1501.014},
       adsurl = {https://ui.adsabs.harvard.edu/abs/2015ascl.soft01014B}
}

@ARTICLE{2014ApJ...794..156R,
       author = {{Rubin}, Kate H.~R. and {Prochaska}, J. Xavier and {Koo}, David C. and {Phillips}, Andrew C. and {Martin}, Crystal L. and {Winstrom}, Lucas O.},
        title = "{Evidence for Ubiquitous Collimated Galactic-scale Outflows along the Star-forming Sequence at z \raisebox{-0.5ex}\textasciitilde 0.5}",
      journal = {\apj},
         year = 2014,
        month = oct,
       volume = {794},
       number = {2},
          eid = {156},
        pages = {156},
          doi = {10.1088/0004-637X/794/2/156},
archivePrefix = {arXiv},
       eprint = {1307.1476},
 primaryClass = {astro-ph.CO},
       adsurl = {https://ui.adsabs.harvard.edu/abs/2014ApJ...794..156R}
}

@ARTICLE{2014ApJ...792L..12K,
       author = {{Kacprzak}, Glenn G. and {Martin}, Crystal L. and {Bouch{\'e}}, Nicolas and {Churchill}, Christopher W. and {Cooke}, Jeff and {LeReun}, Audrey and {Schroetter}, Ilane and {Ho}, Stephanie H. and {Klimek}, Elizabeth},
        title = "{New Perspective on Galaxy Outflows from the First Detection of Both Intrinsic and Traverse Metal-line Absorption}",
      journal = {\apjl},
         year = 2014,
        month = sep,
       volume = {792},
       number = {1},
          eid = {L12},
        pages = {L12},
          doi = {10.1088/2041-8205/792/1/L12},
archivePrefix = {arXiv},
       eprint = {1407.4126},
 primaryClass = {astro-ph.GA},
       adsurl = {https://ui.adsabs.harvard.edu/abs/2014ApJ...792L..12K}
}

@ARTICLE{2013ApJ...779...87C,
       author = {{Churchill}, Christopher W. and {Trujillo-Gomez}, Sebastian and {Nielsen}, Nikole M. and {Kacprzak}, Glenn G.},
        title = "{MAGIICAT III. Interpreting Self-similarity of the Circumgalactic Medium with Virial Mass Using Mg II Absorption}",
      journal = {\apj},
         year = 2013,
        month = dec,
       volume = {779},
       number = {1},
          eid = {87},
        pages = {87},
          doi = {10.1088/0004-637X/779/1/87},
archivePrefix = {arXiv},
       eprint = {1308.2618},
 primaryClass = {astro-ph.GA},
       adsurl = {https://ui.adsabs.harvard.edu/abs/2013ApJ...779...87C}
}

@ARTICLE{2013ApJ...776..115N,
       author = {{Nielsen}, Nikole M. and {Churchill}, Christopher W. and {Kacprzak}, Glenn G.},
        title = "{MAGIICAT II. General Characteristics of the Mg II Absorbing Circumgalactic Medium}",
      journal = {\apj},
         year = 2013,
        month = oct,
       volume = {776},
       number = {2},
          eid = {115},
        pages = {115},
          doi = {10.1088/0004-637X/776/2/115},
archivePrefix = {arXiv},
       eprint = {1211.1380},
 primaryClass = {astro-ph.CO},
       adsurl = {https://ui.adsabs.harvard.edu/abs/2013ApJ...776..115N}
}

@ARTICLE{2013ApJ...776..114N,
       author = {{Nielsen}, Nikole M. and {Churchill}, Christopher W. and {Kacprzak}, Glenn G. and {Murphy}, Michael T.},
        title = "{MAGIICAT I. The Mg II Absorber-Galaxy Catalog}",
      journal = {\apj},
         year = 2013,
        month = oct,
       volume = {776},
       number = {2},
          eid = {114},
        pages = {114},
          doi = {10.1088/0004-637X/776/2/114},
archivePrefix = {arXiv},
       eprint = {1304.6716},
 primaryClass = {astro-ph.CO},
       adsurl = {https://ui.adsabs.harvard.edu/abs/2013ApJ...776..114N}
}

@ARTICLE{2012ApJ...760..127M,
       author = {{Martin}, Crystal L. and {Shapley}, Alice E. and {Coil}, Alison L. and {Kornei}, Katherine A. and {Bundy}, Kevin and {Weiner}, Benjamin J. and {Noeske}, Kai G. and {Schiminovich}, David},
        title = "{Demographics and Physical Properties of Gas Outflows/Inflows at 0.4 < z < 1.4}",
      journal = {\apj},
         year = 2012,
        month = dec,
       volume = {760},
       number = {2},
          eid = {127},
        pages = {127},
          doi = {10.1088/0004-637X/760/2/127},
archivePrefix = {arXiv},
       eprint = {1206.5552},
 primaryClass = {astro-ph.CO},
       adsurl = {https://ui.adsabs.harvard.edu/abs/2012ApJ...760..127M}
}

@ARTICLE{2012ApJ...760L...7K,
       author = {{Kacprzak}, Glenn G. and {Churchill}, Christopher W. and {Nielsen}, Nikole M.},
        title = "{Tracing Outflows and Accretion: A Bimodal Azimuthal Dependence of Mg II Absorption}",
      journal = {\apjl},
         year = 2012,
        month = nov,
       volume = {760},
       number = {1},
          eid = {L7},
        pages = {L7},
          doi = {10.1088/2041-8205/760/1/L7},
archivePrefix = {arXiv},
       eprint = {1205.0245},
 primaryClass = {astro-ph.CO},
       adsurl = {https://ui.adsabs.harvard.edu/abs/2012ApJ...760L...7K}
}

@ARTICLE{2012MNRAS.426..801B,
       author = {{Bouch{\'e}}, N. and {Hohensee}, W. and {Vargas}, R. and {Kacprzak}, G.~G. and {Martin}, C.~L. and {Cooke}, J. and {Churchill}, C.~W.},
        title = "{Physical properties of galactic winds using background quasars}",
      journal = {\mnras},
         year = 2012,
        month = oct,
       volume = {426},
       number = {2},
        pages = {801-815},
          doi = {10.1111/j.1365-2966.2012.21114.x},
archivePrefix = {arXiv},
       eprint = {1110.5877},
 primaryClass = {astro-ph.CO},
       adsurl = {https://ui.adsabs.harvard.edu/abs/2012MNRAS.426..801B}
}

@ARTICLE{2011ApJ...743...10B,
       author = {{Bordoloi}, R. and {Lilly}, S.~J. and {Knobel}, C. and {Bolzonella}, M. and {Kampczyk}, P. and {Carollo}, C.~M. and {Iovino}, A. and {Zucca}, E. and {Contini}, T. and {Kneib}, J. -P. and {Le Fevre}, O. and {Mainieri}, V. and {Renzini}, A. and {Scodeggio}, M. and {Zamorani}, G. and {Balestra}, I. and {Bardelli}, S. and {Bongiorno}, A. and {Caputi}, K. and {Cucciati}, O. and {de la Torre}, S. and {de Ravel}, L. and {Garilli}, B. and {Kova{\v{c}}}, K. and {Lamareille}, F. and {Le Borgne}, J. -F. and {Le Brun}, V. and {Maier}, C. and {Mignoli}, M. and {Pello}, R. and {Peng}, Y. and {Perez Montero}, E. and {Presotto}, V. and {Scarlata}, C. and {Silverman}, J. and {Tanaka}, M. and {Tasca}, L. and {Tresse}, L. and {Vergani}, D. and {Barnes}, L. and {Cappi}, A. and {Cimatti}, A. and {Coppa}, G. and {Diener}, C. and {Franzetti}, P. and {Koekemoer}, A. and {L{\'o}pez-Sanjuan}, C. and {McCracken}, H.~J. and {Moresco}, M. and {Nair}, P. and {Oesch}, P. and {Pozzetti}, L. and {Welikala}, N.},
        title = "{The Radial and Azimuthal Profiles of Mg II Absorption around 0.5 < z < 0.9 zCOSMOS Galaxies of Different Colors, Masses, and Environments}",
      journal = {\apj},
         year = 2011,
        month = dec,
       volume = {743},
       number = {1},
          eid = {10},
        pages = {10},
          doi = {10.1088/0004-637X/743/1/10},
archivePrefix = {arXiv},
       eprint = {1106.0616},
 primaryClass = {astro-ph.CO},
       adsurl = {https://ui.adsabs.harvard.edu/abs/2011ApJ...743...10B}
}

@ARTICLE{2011ApJ...743...46C,
       author = {{Coil}, Alison L. and {Weiner}, Benjamin J. and {Holz}, Daniel E. and {Cooper}, Michael C. and {Yan}, Renbin and {Aird}, James},
        title = "{Outflowing Galactic Winds in Post-starburst and Active Galactic Nucleus Host Galaxies at 0.2 < z < 0.8}",
      journal = {\apj},
         year = 2011,
        month = dec,
       volume = {743},
       number = {1},
          eid = {46},
        pages = {46},
          doi = {10.1088/0004-637X/743/1/46},
archivePrefix = {arXiv},
       eprint = {1104.0681},
 primaryClass = {astro-ph.CO},
       adsurl = {https://ui.adsabs.harvard.edu/abs/2011ApJ...743...46C}
}

@ARTICLE{2012MNRAS.424.1952G,
       author = {{Gauthier}, Jean-Ren{\'e} and {Chen}, Hsiao-Wen},
        title = "{Empirical constraints of supergalactic winds at z{\ensuremath{\gtrsim}} 0.5}",
      journal = {\mnras},
         year = 2012,
        month = aug,
       volume = {424},
       number = {3},
        pages = {1952-1962},
          doi = {10.1111/j.1365-2966.2012.21327.x},
archivePrefix = {arXiv},
       eprint = {1205.4037},
 primaryClass = {astro-ph.CO},
       adsurl = {https://ui.adsabs.harvard.edu/abs/2012MNRAS.424.1952G}
}

@INPROCEEDINGS{2010SPIE.7735E..08B,
       author = {{Bacon}, R. and {Accardo}, M. and {Adjali}, L. and {Anwand}, H. and {Bauer}, S. and {Biswas}, I. and {Blaizot}, J. and {Boudon}, D. and {Brau-Nogue}, S. and {Brinchmann}, J. and {Caillier}, P. and {Capoani}, L. and {Carollo}, C.~M. and {Contini}, T. and {Couderc}, P. and {Daguis{\'e}}, E. and {Deiries}, S. and {Delabre}, B. and {Dreizler}, S. and {Dubois}, J. and {Dupieux}, M. and {Dupuy}, C. and {Emsellem}, E. and {Fechner}, T. and {Fleischmann}, A. and {Fran{\c{c}}ois}, M. and {Gallou}, G. and {Gharsa}, T. and {Glindemann}, A. and {Gojak}, D. and {Guiderdoni}, B. and {Hansali}, G. and {Hahn}, T. and {Jarno}, A. and {Kelz}, A. and {Koehler}, C. and {Kosmalski}, J. and {Laurent}, F. and {Le Floch}, M. and {Lilly}, S.~J. and {Lizon}, J. -L. and {Loupias}, M. and {Manescau}, A. and {Monstein}, C. and {Nicklas}, H. and {Olaya}, J. -C. and {Pares}, L. and {Pasquini}, L. and {P{\'e}contal-Rousset}, A. },
        title = "{The MUSE second-generation VLT instrument}",
    booktitle = {},
         year = 2010,
       editor = {{McLean}, Ian S. and {Ramsay}, Suzanne K. and {Takami}, Hideki},
       series = {SPIE Conference Series},
       volume = {7735},
        month = jul,
          eid = {773508},
        pages = {773508},
          doi = {10.1117/12.856027},
archivePrefix = {arXiv},
       eprint = {2211.16795},
 primaryClass = {astro-ph.IM},
       adsurl = {https://ui.adsabs.harvard.edu/abs/2010SPIE.7735E..08B}
}

@ARTICLE{1999PASP..111...63F,
       author = {{Fitzpatrick}, Edward L.},
        title = "{Correcting for the Effects of Interstellar Extinction}",
      journal = {\pasp},
         year = 1999,
        month = jan,
       volume = {111},
       number = {755},
        pages = {63-75},
          doi = {10.1086/316293},
archivePrefix = {arXiv},
       eprint = {astro-ph/9809387},
 primaryClass = {astro-ph},
       adsurl = {https://ui.adsabs.harvard.edu/abs/1999PASP..111...63F}
}

@ARTICLE{1998ApJ...500..525S,
       author = {{Schlegel}, David J. and {Finkbeiner}, Douglas P. and {Davis}, Marc},
        title = "{Maps of Dust Infrared Emission for Use in Estimation of Reddening and Cosmic Microwave Background Radiation Foregrounds}",
      journal = {\apj},
         year = 1998,
        month = jun,
       volume = {500},
       number = {2},
        pages = {525-553},
          doi = {10.1086/305772},
archivePrefix = {arXiv},
       eprint = {astro-ph/9710327},
 primaryClass = {astro-ph},
       adsurl = {https://ui.adsabs.harvard.edu/abs/1998ApJ...500..525S}
}

@ARTICLE{2010ApJ...710..903M,
       author = {{Moster}, Benjamin P. and {Somerville}, Rachel S. and {Maulbetsch}, Christian and {van den Bosch}, Frank C. and {Macci{\`o}}, Andrea V. and {Naab}, Thorsten and {Oser}, Ludwig},
        title = "{Constraints on the Relationship between Stellar Mass and Halo Mass at Low and High Redshift}",
      journal = {\apj},
         year = 2010,
        month = feb,
       volume = {710},
       number = {2},
        pages = {903-923},
          doi = {10.1088/0004-637X/710/2/903},
archivePrefix = {arXiv},
       eprint = {0903.4682},
 primaryClass = {astro-ph.CO},
       adsurl = {https://ui.adsabs.harvard.edu/abs/2010ApJ...710..903M}
}

@INPROCEEDINGS{2009AIPC.1201..142W,
       author = {{Weiner}, Benjamin J.},
        title = "{Star Formation Driven Galactic Winds at z\raisebox{-0.5ex}\textasciitilde1.4}",
    booktitle = {The Monster's Fiery Breath: Feedback in Galaxies, Groups, and Clusters},
         year = 2009,
       editor = {{Heinz}, Sebastian and {Wilcots}, Eric},
       series = {American Institute of Physics Conference Series},
       volume = {1201},
        month = dec,
    publisher = {AIP},
        pages = {142-145},
          doi = {10.1063/1.3293019},
       adsurl = {https://ui.adsabs.harvard.edu/abs/2009AIPC.1201..142W}
}

@ARTICLE{2001AJ....122..679C,
       author = {{Churchill}, Christopher W. and {Vogt}, Steven S.},
        title = "{The Kinematics of Intermediate-Redshift Mg II Absorbers}",
      journal = {\aj},
         year = 2001,
        month = aug,
       volume = {122},
       number = {2},
        pages = {679-713},
          doi = {10.1086/321174},
archivePrefix = {arXiv},
       eprint = {astro-ph/0106006},
 primaryClass = {astro-ph},
       adsurl = {https://ui.adsabs.harvard.edu/abs/2001AJ....122..679C}
}

@ARTICLE{2009ApJ...703.1394M,
       author = {{Martin}, Crystal L. and {Bouch{\'e}}, Nicolas},
        title = "{Physical Conditions in the Low-ionization Component of Starburst Outflows: The Shape of Near-Ultraviolet and Optical Absorption-line Troughs in Keck Spectra of ULIRGs}",
      journal = {\apj},
         year = 2009,
        month = oct,
       volume = {703},
       number = {2},
        pages = {1394-1415},
          doi = {10.1088/0004-637X/703/2/1394},
archivePrefix = {arXiv},
       eprint = {0908.4271},
 primaryClass = {astro-ph.CO},
       adsurl = {https://ui.adsabs.harvard.edu/abs/2009ApJ...703.1394M}
}

@ARTICLE{2009ApJ...692..187W,
       author = {{Weiner}, Benjamin J. and {Coil}, Alison L. and {Prochaska}, Jason X. and {Newman}, Jeffrey A. and {Cooper}, Michael C. and {Bundy}, Kevin and {Conselice}, Christopher J. and {Dutton}, Aaron A. and {Faber}, S.~M. and {Koo}, David C. and {Lotz}, Jennifer M. and {Rieke}, G.~H. and {Rubin}, K.~H.~R.},
        title = "{Ubiquitous Outflows in DEEP2 Spectra of Star-Forming Galaxies at z = 1.4}",
      journal = {\apj},
         year = 2009,
        month = feb,
       volume = {692},
       number = {1},
        pages = {187-211},
          doi = {10.1088/0004-637X/692/1/187},
archivePrefix = {arXiv},
       eprint = {0804.4686},
 primaryClass = {astro-ph},
       adsurl = {https://ui.adsabs.harvard.edu/abs/2009ApJ...692..187W}
}

@ARTICLE{2010ApJ...719.1503R,
       author = {{Rubin}, Kate H.~R. and {Weiner}, Benjamin J. and {Koo}, David C. and {Martin}, Crystal L. and {Prochaska}, J. Xavier and {Coil}, Alison L. and {Newman}, Jeffrey A.},
        title = "{The Persistence of Cool Galactic Winds in High Stellar Mass Galaxies between z \raisebox{-0.5ex}\textasciitilde 1.4 and \raisebox{-0.5ex}\textasciitilde1}",
      journal = {\apj},
         year = 2010,
        month = aug,
       volume = {719},
       number = {2},
        pages = {1503-1525},
          doi = {10.1088/0004-637X/719/2/1503},
archivePrefix = {arXiv},
       eprint = {0912.2343},
 primaryClass = {astro-ph.CO},
       adsurl = {https://ui.adsabs.harvard.edu/abs/2010ApJ...719.1503R}
}

@ARTICLE{2007NJPh....9..447J,
       author = {{Jullo}, E. and {Kneib}, J. -P. and {Limousin}, M. and {El{\'\i}asd{\'o}ttir}, {\'A}. and {Marshall}, P.~J. and {Verdugo}, T.},
        title = "{A Bayesian approach to strong lensing modelling of galaxy clusters}",
      journal = {New Journal of Physics},
         year = 2007,
        month = dec,
       volume = {9},
       number = {12},
        pages = {447},
          doi = {10.1088/1367-2630/9/12/447},
archivePrefix = {arXiv},
       eprint = {0706.0048},
 primaryClass = {astro-ph},
       adsurl = {https://ui.adsabs.harvard.edu/abs/2007NJPh....9..447J}
}

@ARTICLE{2007ApJ...663L..77T,
       author = {{Tremonti}, Christy A. and {Moustakas}, John and {Diamond-Stanic}, Aleksandar M.},
        title = "{The Discovery of 1000 km s$^{-1}$ Outflows in Massive Poststarburst Galaxies at z=0.6}",
      journal = {\apjl},
         year = 2007,
        month = jul,
       volume = {663},
       number = {2},
        pages = {L77-L80},
          doi = {10.1086/520083},
archivePrefix = {arXiv},
       eprint = {0706.0527},
 primaryClass = {astro-ph},
       adsurl = {https://ui.adsabs.harvard.edu/abs/2007ApJ...663L..77T}
}

@ARTICLE{2023ARA&A..61..131F,
       author = {{Faucher-Gigu{\`e}re}, Claude-Andr{\'e} and {Oh}, S. Peng},
        title = "{Key Physical Processes in the Circumgalactic Medium}",
      journal = {\araa},
         year = 2023,
        month = aug,
       volume = {61},
        pages = {131-195},
          doi = {10.1146/annurev-astro-052920-125203},
archivePrefix = {arXiv},
       eprint = {2301.10253},
 primaryClass = {astro-ph.GA},
       adsurl = {https://ui.adsabs.harvard.edu/abs/2023ARA&A..61..131F}
}

@ARTICLE{2006ApJ...647..853W,
       author = {{Willmer}, C.~N.~A. and {Faber}, S.~M. and {Koo}, D.~C. and {Weiner}, B.~J. and {Newman}, J.~A. and {Coil}, A.~L. and {Connolly}, A.~J. and {Conroy}, C. and {Cooper}, M.~C. and {Davis}, M. and {Finkbeiner}, D.~P. and {Gerke}, B.~F. and {Guhathakurta}, P. and {Harker}, J. and {Kaiser}, N. and {Kassin}, S. and {Konidaris}, N.~P. and {Lin}, L. and {Luppino}, G. and {Madgwick}, D.~S. and {Noeske}, K.~G. and {Phillips}, A.~C. and {Yan}, R.},
        title = "{The Deep Evolutionary Exploratory Probe 2 Galaxy Redshift Survey: The Galaxy Luminosity Function to z\raisebox{-0.5ex}\textasciitilde1}",
      journal = {\apj},
         year = 2006,
        month = aug,
       volume = {647},
       number = {2},
        pages = {853-873},
          doi = {10.1086/505455},
archivePrefix = {arXiv},
       eprint = {astro-ph/0506041},
 primaryClass = {astro-ph},
       adsurl = {https://ui.adsabs.harvard.edu/abs/2006ApJ...647..853W}
}

@ARTICLE{2007MNRAS.376..479S,
       author = {{Swinbank}, A.~M. and {Bower}, R.~G. and {Smith}, Graham P. and {Wilman}, R.~J. and {Smail}, Ian and {Ellis}, R.~S. and {Morris}, S.~L. and {Kneib}, J. -P.},
        title = "{Resolved spectroscopy of a gravitationally lensed L* Lyman-break galaxy at z \raisebox{-0.5ex}\textasciitilde 5}",
      journal = {\mnras},
         year = 2007,
        month = apr,
       volume = {376},
       number = {2},
        pages = {479-491},
          doi = {10.1111/j.1365-2966.2007.11454.x},
archivePrefix = {arXiv},
       eprint = {astro-ph/0701221},
 primaryClass = {astro-ph},
       adsurl = {https://ui.adsabs.harvard.edu/abs/2007MNRAS.376..479S}
}

@ARTICLE{2005ApJS..160..115R,
       author = {{Rupke}, David S. and {Veilleux}, Sylvain and {Sanders}, D.~B.},
        title = "{Outflows in Infrared-Luminous Starbursts at z < 0.5. II. Analysis and Discussion}",
      journal = {\apjs},
         year = 2005,
        month = sep,
       volume = {160},
       number = {1},
        pages = {115-148},
          doi = {10.1086/432889},
archivePrefix = {arXiv},
       eprint = {astro-ph/0506611},
 primaryClass = {astro-ph},
       adsurl = {https://ui.adsabs.harvard.edu/abs/2005ApJS..160..115R}
}

@ARTICLE{2020A&ARv..28....2V,
       author = {{Veilleux}, Sylvain and {Maiolino}, Roberto and {Bolatto}, Alberto D. and {Aalto}, Susanne},
        title = "{Cool outflows in galaxies and their implications}",
      journal = {\aapr},
         year = 2020,
        month = apr,
       volume = {28},
       number = {1},
          eid = {2},
        pages = {2},
          doi = {10.1007/s00159-019-0121-9},
archivePrefix = {arXiv},
       eprint = {2002.07765},
 primaryClass = {astro-ph.GA},
       adsurl = {https://ui.adsabs.harvard.edu/abs/2020A&ARv..28....2V}
}

@ARTICLE{2002AJ....123....1G,
       author = {{Gladders}, Michael D. and {Yee}, H.~K.~C. and {Ellingson}, E.},
        title = "{Discovery of a z =0.77 Galaxy Cluster with Multiple, Bright, Strong-Lensing Arcs}",
      journal = {\aj},
         year = 2002,
        month = jan,
       volume = {123},
       number = {1},
        pages = {1-9},
          doi = {10.1086/324637},
       adsurl = {https://ui.adsabs.harvard.edu/abs/2002AJ....123....1G}
}

@ARTICLE{2000ApJ...539..718C,
       author = {{Charlot}, St{\'e}phane and {Fall}, S. Michael},
        title = "{A Simple Model for the Absorption of Starlight by Dust in Galaxies}",
      journal = {\apj},
         year = 2000,
        month = aug,
       volume = {539},
       number = {2},
        pages = {718-731},
          doi = {10.1086/309250},
archivePrefix = {arXiv},
       eprint = {astro-ph/0003128},
 primaryClass = {astro-ph},
       adsurl = {https://ui.adsabs.harvard.edu/abs/2000ApJ...539..718C}
}

@ARTICLE{1998ARA&A..36..189K,
       author = {{Kennicutt}, Robert C., Jr.},
        title = "{Star Formation in Galaxies Along the Hubble Sequence}",
      journal = {\araa},
         year = 1998,
        month = jan,
       volume = {36},
        pages = {189-232},
          doi = {10.1146/annurev.astro.36.1.189},
archivePrefix = {arXiv},
       eprint = {astro-ph/9807187},
 primaryClass = {astro-ph},
       adsurl = {https://ui.adsabs.harvard.edu/abs/1998ARA&A..36..189K}
}

@ARTICLE{1996A&AS..117..393B,
       author = {{Bertin}, E. and {Arnouts}, S.},
        title = "{SExtractor: Software for source extraction.}",
      journal = {\aaps},
         year = 1996,
        month = jun,
       volume = {117},
        pages = {393-404},
          doi = {10.1051/aas:1996164},
       adsurl = {https://ui.adsabs.harvard.edu/abs/1996A&AS..117..393B}
}

@inproceedings{Weilbacher_2012,
	Adsurl = {http://adsabs.harvard.edu/abs/2012SPIE.8451E..0BW},
	Author = {{Weilbacher}, P.~M. and {Streicher}, O. and {Urrutia}, T. and {Jarno}, A. and {P{\'e}contal-Rousset}, A. and {Bacon}, R. and {B{\"o}hm}, P.},
	Booktitle = {Software and Cyberinfrastructure for Astronomy II},
	Doi = {10.1117/12.925114},
	Eid = {84510B},
	Month = sep,
	Pages = {84510B},
	Series = {\procspie},
	Title = {{Design and capabilities of the MUSE data reduction software and pipeline}},
	Volume = 8451,
	Year = 2012
}

@misc{ESOREX,
	Adsurl = {http://adsabs.harvard.edu/abs/2015ascl.soft04003E},
	Archiveprefix = {ascl},
	Author = {{ESO CPL Development Team}},
	Eprint = {1504.003},
	Howpublished = {Astrophysics Source Code Library},
	Month = apr,
	Title = {{EsoRex: ESO Recipe Execution Tool}},
	Year = 2015}

@article{Soto_2016,
	Adsurl = {https://ui.adsabs.harvard.edu/abs/2016MNRAS.458.3210S},
	Archiveprefix = {arXiv},
	Author = {{Soto}, Kurt T. and {Lilly}, Simon J. and {Bacon}, Roland and {Richard}, Johan and {Conseil}, Simon},
	Doi = {10.1093/mnras/stw474},
	Eprint = {1602.08037},
	Journal = {\mnras},
	Month = {May},
	Number = {3},
	Pages = {3210-3220},
	Primaryclass = {astro-ph.IM},
	Title = {{ZAP - enhanced PCA sky subtraction for integral field spectroscopy}},
	Volume = {458},
	Year = {2016}}

\appendix

\section{Strong lens modelling}
\label{sec:appendix-lens-modeling}
We used the publicly available software \texttt{LENSTOOL} \citep{2007NJPh....9..447J} to perform parametric modelling of the cluster mass distribution. The model includes a large-scale smooth component representing the cluster's dark matter halo and individual mass components associated with cluster galaxies. The mass distributions of both cluster-scale and galaxy-scale halos were modelled as truncated pseudo-isothermal elliptical mass distributions (PIEMDs). For the cluster-scale halo, we adopted a PIEMD potential centred around the brightest cluster galaxy (BCG) at $z_\text{cluster}=0.773$. All model parameters—position, velocity dispersion, ellipticity, and position angle—were allowed to vary with broad priors, except for the cut radius, which was fixed to $1000$~kpc, well beyond the strong-lensing regime. For the galaxy-scale halos, we included sixteen spectroscopically confirmed cluster members detected in the MUSE data. These galaxies were selected to be within $\pm 3000$~\kms{}, of the cluster redshift, and within 30\arcsec\, of the BCG, where strong-lensing constraints are available.

Photometric and morphological properties of the cluster galaxies were measured using \texttt{SExtractor} \citep{1996A&AS..117..393B} in the \textit{HST} imaging. We fixed their centroids, position angles, and ellipticities according to their morphology measured by \texttt{SExtractor}. We scaled their velocity dispersions ($\sigma_{\rm LENSTOOL}$), core ($r_\text{core}$), and cut radius ($r_\text{cut}$) adopting the scaling relations detailed in \cite{2007NJPh....9..447J}. A reference luminosity $L_\ast$ was chosen for a galaxy with apparent magnitude $m_{\mathrm{F160W}}=18$. The reference core radius $r_{\text{core}}^\ast$ was fixed to a negligible but non-zero value. 

The best-fit parameters for the cluster-scale halo are a velocity dispersion $\sigma_{\rm LENSTOOL}=632\pm45$~\kms{}, an ellipticity of $0.65\pm0.1$ with a position angle of $45\pm1.5^\circ$, and a core radius of $r_\text{core}=6.8\pm10$~kpc. For a cluster galaxy with reference magnitude $m_{\mathrm{F160W}}=18$, we obtained a best-fit velocity dispersion $\sigma_{\rm LENSTOOL}=359\pm68$~\kms{} and cut radius $r_\text{cut}=46.2\pm64$~kpc. This model achieves an image-plane RMS of $0.79$\arcsec, with a $\chi^2=13.7$ with $11$ free parameters. 

\section{SED modelling and halo properties}
\label{sec:appendix-sed-halo-properties}
\begin{figure*}[h!]
    \centering
    \includegraphics[width=\linewidth]{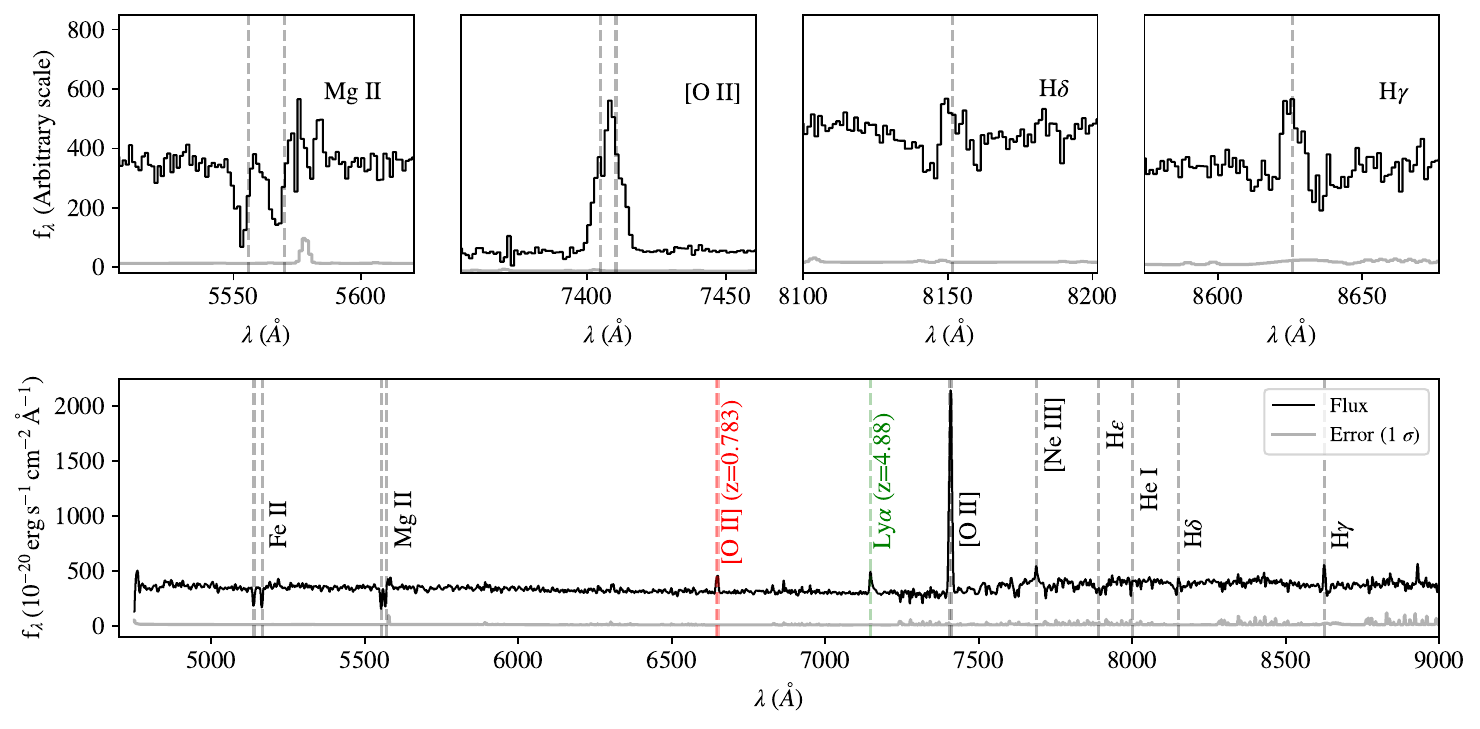}
    \caption{{\it Bottom}: Integrated MUSE spectrum of G1, corrected by Galactic extinction and lensing magnification. Vertical grey dashed lines mark the positions of strong absorption and emission features at $z_{\rm G1}$. Several narrow emission lines are present, including \oii{}, \neiii{}, \hei{}, and Balmer lines. \mgii{} is blue-shifted with respect to the emission lines. \lya{} emission is observed at $z\sim4.88$, corresponding to a counter-image of the arc A, and diffuse \oii{} emission from the galaxy cluster at $\sim\text{\zcluster}$ are also present in the spectrum. {\it Top}: From left to right, zoom-in to \mgii{}, \oii{}, H$\delta$, and H$\gamma$ lines. }
    \label{fig:spec_G1}
\end{figure*}

We first modelled the spectral energy distribution (SED) of G1 using \texttt{BAGPIPES} (\citealt{2018MNRAS.480.4379C},\citealt{2019MNRAS.490..417C}), jointly fitting its MUSE and \textit{HST} photometry. The MUSE spectrum of G1 was extracted on a 1.5\arcsec aperture on the MUSE data, corrected for both lensing magnification and Galactic extinction (using the same procedure as in Sect.~\ref{sec:G1_photometry}), which is shown in Fig.~\ref{fig:spec_G1}. 

For the SED model, we assumed a double power law for the star formation history (SFH), and the \citet{2000ApJ...539..718C} extinction law (CF00), with uniform priors of $0<A_V<2$ and $-4<n<4$ for the dust attenuation. A nebular component is included with ionization parameter $-3 \leq \log U \leq -2$- We adopt uniform priors for the stellar velocity dispersion ($50\text{~\kms{}} < \sigma < 300\text{~\kms{}}$), and metallicity ($0.2<\left[Z/H\right]<4$). The resulting stellar mass and SFR were inferred from the posterior distributions and are listed in Table~\ref{tab:G1_properties}.

Next, we estimate the halo mass $M_h$ using the stellar-mass-halo-mass (SMHM) relation from \citet{2010ApJ...710..903M}, assuming $M_h = M_{200}$. From this, we derived the virial radius via $R_{\rm vir} = \left[3 M_h / (4 \pi 200 \rho_c(z))\right]^{1/3}$ and the maximum circular velocity as $V_\mathrm{max} = \left[10 M_h G H(z)\right]^{1/3}$, where $G$ is the gravitational constant, and $\rho_c(z)$ and $H(z)$ are the critical density and Hubble parameter at redshift $z$. The halo velocity dispersion was estimated using the scaling relation from \citet{2018MNRAS.477..616E}: $\sigma_\mathrm{DM} \simeq 430 \left(M_h / 10^{14} M_\odot\right)^{1/3}$~\kms{}. 

We estimated the escape velocity $V_\text{esc}(r)=\sqrt{2|\phi(r)|}$ and circular velocity $V_\text{circ}(r)=\sqrt{r|d\phi(r)/dr|}$ for G1, assuming a Navarro-Frenk-White \citep{1996ApJ...462..563N} gravitational potential $\phi(r)$ for the dark-matter (DM) halo, with concentration parameter of 5 \citep[for galaxies with similar $M_{200}$ from N-body simulations; ][]{2018ApJ...859...55C}. Maximum circular velocity, and escape velocities at $r=10$~kpc and $r=50$~kpc are indicated in Table~\ref{tab:G1_properties}.

\section{Estimation of N$_\text{\hi{}}$}

\label{sec:column-density}
A column density $N$ of an absorbing species, in the optically thin regime, can be estimated as

\begin{equation}
    \label{eq:logN}
    N > 1.130\times10^{12} \text{cm}^{-1} \frac{W_r}{f \lambda^2}
\end{equation}

\noindent for a given transition, where $f$, $\lambda$ and  $W_r$, are its oscillator strength, rest-frame wavelength, and measured rest equivalent width, respectively \citep{2011piim.book.....D}. Although the \mgii{} absorption is likely saturated and unresolved, we adopt the optically thin limit to estimate a conservative lower bound on the column density. Any degree of saturation or unresolved component structure increases the true $N$ above the linear-regime value, making the optically thin approximation a strict lower limit. For $\text{\mgiiew{}} \sim 1 \text{--} 1.5$~\r{A} we estimate $\log (N_\text{\mgii{}}/\text{cm}^{-2})>13.4\text{--}13.6$. 

Following \citet{2014ApJ...794..156R}, the corresponding $N_\text{\mgii{}}$ can be used to estimate $N_\text{\hi{}}$ as:

\begin{equation}
    \label{eq:logNHI}
    N_{\text{\hi{}}} > \frac{10^{19.3}~\text{cm}^{-2}}{\chi(\text{\mgii{}})} \frac{N_\text{\mgii{}}}{10^{14.4}~\text{cm}^{-2}} \frac{10^{-4.42}}{10^{\log{\text{Mg}/\text{H}}}} \frac{10^{-0.5}}{10^{d\text{(Mg)}}}
\end{equation}

Assuming a solar abundance ratio $\log{\text{Mg}/\text{H}}=-4.42$ \citep{1996ARA&A..34..279S}, unity ionization fraction $\chi(\text{\mgii{}})=1$, and dust depletion factor $d\text{(Mg)}=-0.5$ \citep[Galactic ISM values range between $-0.3$ and $-1.5$][]{2009ApJ...700.1299J}, we get lower limits of  $\log(N_\text{\hi{}}/\text{cm}^{-2})>18.3\text{--}18.5$. 

Self-shielding is required to retain significant amounts of \mgii{} in dense clouds. Photoionization models show that \mgii{} with $\log (N_\text{\mgii{}}/\text{cm}^{-2})\geq 13.4$ can only survive in gas with $\log (N_\text{\hi{}}/\text{cm}^{-2})\gtrsim 18$ \citep{1986A&A...169....1B}. Moreover, empirical \mgiiblue{}-N$_\text{\hi{}}$ relations at $0.4\lesssim z \lesssim 2.5$  based on quasar sightlines \citep{2009MNRAS.393..808M,2017ApJ...850..156L} indicate that $\log (N_\text{\hi{}}/\text{cm}^{-2})$ correlates with \mgiiblue{}, albeit with large scatter, however absorbers with \mgiiblue{} between $\sim1\text{--}1.5$~\r{A} span a around three orders of magnitude in $\log (N_\text{\hi{}}/\text{cm}^{-2})$, between 18.3 and 21.6. The inferred values are consistent with the spread in $N_{\text{\hi{}}}$ from background quasars, and the minimum $N_\text{\hi{}}$ required for self-shielded clouds. We remark that, given our assumptions for saturation, abundance, ionization fraction, and dust depletion, our $N_\text{\hi{}}$ lower limit estimated  Eq.~\ref{eq:logNHI} is extremely conservative, and a direct measurement would almost certainly yield substantially higher column densities.

\onecolumn
\section{G1 spectrum}
\newpage

\section{Spatially resolved absorption-line measurements}
\begin{table*}[ht!]
    \caption{\label{table:ew_mg} Properties of the absorptions measured on the binned MUSE spaxels. }
    \centering
    \begin{tabular}{ccrrrrrcccc}
                \hline
        x & y & $\rho$ & v$_{\rm los}$ & $\sigma_{\rm los}$ & $\phi$ & W$_{2586}$ & W$_{2600}$ & W$_{2796}$ & W$_{2803}$ & \ratio \\
        (pix) & (pix) & (kpc) & ($\text{km}\,\text{s}^{-1}$) & ($\text{km}\,\text{s}^{-1}$) & (deg) & ($\AA$) & ($\AA$) & ($\AA$) & ($\AA$) & dimensionless \\
        (0) & (1) & (2) & (3) & (4) & (5) & (6) & (7) & (8) & (9) & (10) \\
                        \hline
                      \\  
			$46$ & $31$ & $13.3$ & \nodata & \nodata & $175$ & $<0.88$ & $<0.88$ & $<1.05$ & $<1.36$ & \nodata \\
			$46$ & $32$ & $10.1$ & $-239\pm27$ & $92\pm26$ & $169$ & $1.21\pm0.38$ & $1.21\pm0.35$ & $0.78\pm0.34$ & $<1.11$ & $1.55\pm0.81$ \\
			$46$ & $33$ & $9.0$ & $-171\pm18$ & $108\pm17$ & $141$ & $1.68\pm0.37$ & $1.77\pm0.29$ & $2.13\pm0.39$ & $1.62\pm0.55$ & $0.83\pm0.20$ \\
			$47$ & $31$ & $9.0$ & $-62\pm47$ & $179\pm42$ & $-173$ & $1.63\pm0.55$ & $1.89\pm0.42$ & $1.23\pm0.51$ & $<1.23$ & $1.54\pm0.72$ \\
			$47$ & $32$ & $6.0$ & $-164\pm10$ & $64\pm10$ & $166$ & $1.20\pm0.24$ & $1.43\pm0.70$ & $1.62\pm0.24$ & $<1.12$ & $0.88\pm0.45$ \\
			$47$ & $33$ & $4.1$ & $-170\pm12$ & $109\pm11$ & $118$ & $1.89\pm0.27$ & $2.71\pm1.38$ & $1.75\pm0.28$ & $<0.95$ & $1.55\pm0.82$ \\
			$47$ & $34$ & $6.7$ & \nodata & \nodata & $73$ & $<0.73$ & $<0.91$ & $<1.39$ & $<1.68$ & \nodata \\
			$48$ & $30$ & $10.2$ & $-159\pm16$ & $0\pm63$ & $-141$ & $<0.93$ & $1.11\pm0.36$ & $1.03\pm0.36$ & $0.87\pm0.37$ & $1.08\pm0.51$ \\
			$48$ & $31$ & $5.3$ & $-145\pm12$ & $82\pm11$ & $-139$ & $1.50\pm0.30$ & $1.93\pm0.95$ & $1.88\pm0.28$ & $1.31\pm0.38$ & $1.03\pm0.53$ \\
			$48$ & $32$ & $1.0$ & $-103\pm10$ & $98\pm9$ & $-105$ & $1.64\pm0.23$ & $1.97\pm0.19$ & $2.14\pm0.24$ & $1.81\pm0.42$ & $0.92\pm0.14$ \\
			$48$ & $33$ & $3.7$ & $-113\pm14$ & $133\pm13$ & $21$ & $1.73\pm0.26$ & $<0.69$ & $1.96\pm0.27$ & $1.84\pm0.70$ & $<0.18$ \\
			$48$ & $34$ & $8.1$ & $-126\pm27$ & $121\pm26$ & $28$ & $1.49\pm0.44$ & $1.55\pm0.66$ & $1.51\pm0.42$ & $<1.14$ & $1.03\pm0.53$ \\
			$49$ & $31$ & $5.5$ & $-122\pm22$ & $68\pm21$ & $-79$ & $<0.66$ & $1.43\pm0.45$ & $1.18\pm0.38$ & $<1.28$ & $1.20\pm0.55$ \\
			$49$ & $32$ & $5.8$ & $-94\pm12$ & $53\pm11$ & $-30$ & $1.46\pm0.30$ & $1.51\pm0.41$ & $1.49\pm0.30$ & $1.18\pm0.31$ & $1.01\pm0.34$ \\
			$49$ & $33$ & $8.8$ & $-63\pm17$ & $67\pm16$ & $-5$ & $1.06\pm0.30$ & $1.35\pm0.27$ & $1.05\pm0.30$ & $1.04\pm0.44$ & $1.29\pm0.45$ \\
			$49$ & $34$ & $12.4$ & \nodata & \nodata & $6$ & $<0.95$ & $<0.77$ & $<0.80$ & $<1.08$ & \nodata \\
			$\mathbf{44}$ & $\mathbf{39}$ & $\mathbf{32.0}$ & \nodata & \nodata & $\mathbf{82}$ & $\mathbf{<0.80}$ & $\mathbf{<0.89}$ & $\mathbf{<0.81}$ & $\mathbf{<1.46}$ & \nodata \\
			$\mathbf{44}$ & $\mathbf{40}$ & $\mathbf{33.4}$ & \nodata & \nodata & $\mathbf{80}$ & $\mathbf{<1.05}$ & $\mathbf{<0.88}$ & $\mathbf{<1.05}$ & $\mathbf{<1.09}$ & \nodata \\
			$\mathbf{45}$ & $\mathbf{34}$ & $\mathbf{14.6}$ & \nodata & \nodata & $\mathbf{124}$ & $\mathbf{<0.89}$ & $\mathbf{<0.99}$ & $\mathbf{<0.88}$ & $\mathbf{<1.56}$ & \nodata \\
			$\mathbf{45}$ & $\mathbf{35}$ & $\mathbf{16.2}$ & $\mathbf{-177\pm19}$ & $\mathbf{78\pm18}$ & $\mathbf{107}$ & $\mathbf{0.90\pm0.40}$ & $\mathbf{1.70\pm0.70}$ & $\mathbf{1.57\pm0.40}$ & $\mathbf{<1.46}$ & $\mathbf{1.08\pm0.52}$ \\
			$\mathbf{45}$ & $\mathbf{36}$ & $\mathbf{18.9}$ & $\mathbf{-177\pm35}$ & $\mathbf{117\pm34}$ & $\mathbf{93}$ & $\mathbf{1.23\pm0.54}$ & $\mathbf{<0.76}$ & $\mathbf{1.50\pm0.50}$ & $\mathbf{<2.25}$ & $\mathbf{<0.25}$ \\
			$\mathbf{45}$ & $\mathbf{37}$ & $\mathbf{22.6}$ & $\mathbf{-182\pm21}$ & $\mathbf{55\pm20}$ & $\mathbf{84}$ & $\mathbf{<0.91}$ & $\mathbf{1.00\pm0.38}$ & $\mathbf{1.51\pm0.41}$ & $\mathbf{0.97\pm0.40}$ & $\mathbf{0.67\pm0.31}$ \\
			$\mathbf{45}$ & $\mathbf{38}$ & $\mathbf{26.9}$ & \nodata & \nodata & $\mathbf{78}$ & $\mathbf{<0.83}$ & $\mathbf{<0.96}$ & $\mathbf{<0.78}$ & $\mathbf{<0.94}$ & \nodata \\
			$\mathbf{45}$ & $\mathbf{39}$ & $\mathbf{31.5}$ & $\mathbf{-197\pm21}$ & $\mathbf{48\pm20}$ & $\mathbf{75}$ & $\mathbf{<0.73}$ & $\mathbf{<0.81}$ & $\mathbf{1.02\pm0.30}$ & $\mathbf{0.97\pm0.26}$ & $\mathbf{<0.39}$ \\
			$\mathbf{45}$ & $\mathbf{40}$ & $\mathbf{30.7}$ & \nodata & \nodata & $\mathbf{74}$ & $\mathbf{<1.45}$ & $\mathbf{<1.20}$ & $\mathbf{<0.86}$ & $\mathbf{<0.93}$ & \nodata \\
			$\mathbf{45}$ & $\mathbf{43}$ & $\mathbf{45.0}$ & $\mathbf{-134\pm27}$ & $\mathbf{53\pm27}$ & $\mathbf{70}$ & $\mathbf{<0.79}$ & $\mathbf{0.74\pm0.37}$ & $\mathbf{0.94\pm0.38}$ & $\mathbf{<1.04}$ & $\mathbf{0.79\pm0.51}$ \\
			$\mathbf{46}$ & $\mathbf{34}$ & $\mathbf{9.6}$ & $\mathbf{-149\pm15}$ & $\mathbf{98\pm14}$ & $\mathbf{109}$ & $\mathbf{2.08\pm0.39}$ & $\mathbf{<0.77}$ & $\mathbf{1.40\pm0.35}$ & $\mathbf{1.39\pm0.26}$ & $\mathbf{<0.27}$ \\
			$\mathbf{46}$ & $\mathbf{35}$ & $\mathbf{12.4}$ & $\mathbf{-161\pm22}$ & $\mathbf{58\pm21}$ & $\mathbf{87}$ & $\mathbf{<0.78}$ & $\mathbf{1.19\pm0.44}$ & $\mathbf{1.39\pm0.43}$ & $\mathbf{<1.58}$ & $\mathbf{0.86\pm0.41}$ \\
			$\mathbf{46}$ & $\mathbf{40}$ & $\mathbf{30.3}$ & \nodata & \nodata & $\mathbf{61}$ & $\mathbf{<1.10}$ & $\mathbf{<1.03}$ & $\mathbf{<0.76}$ & $\mathbf{<0.82}$ & \nodata  \\
    \hline
    \end{tabular}
    \tablefoot{\\
    \tablefoottext{0}{x coordinate of the binned spaxel.}\\
    \tablefoottext{1}{y coordinate of the binned spaxel.}\\
    \tablefoottext{2}{Impact parameter; projected physical separation between the centre of the spaxel and the centre of G1, in the absorber plane.}\\
    \tablefoottext{3}{Line-of-sight velocity of the absorption centroid with respect to $z_\text{G1}$.}\\
    \tablefoottext{4}{Line-of-sight velocity dispersion after subtracting the instrumental line spread function.$^{\rm a}$}\\
    \tablefoottext{5}{Azimuthal angle of the centre of the spaxel with respect to G1 receding major axis.}\\
    \tablefoottext{6}{\feii{} $2586\r{A}$ rest-frame equivalent width.$^{\rm b}$}\\
    \tablefoottext{7}{\feii{} $2600\r{A}$ rest-frame equivalent width.$^{\rm b}$}\\
    \tablefoottext{8}{\mgii{} $2796\r{A}$ rest-frame equivalent width.$^{\rm b}$}\\
    \tablefoottext{9}{\mgii{} $2803\r{A}$ rest-frame equivalent width.$^{\rm b}$}\\
    \tablefoottext{10}{Ratio between \feiired{} and \mgiiblue{}.}\\
    \tablefoottext{a}{The value is set to 0 when the observed FWHM is larger than the instrumental FWHM.}\\
    \tablefoottext{b}{Non-detections are reported as $2\sigma$ upper limits.}}
\end{table*}
\newpage

\section{Spatial distribution of \mgii{} spectra}
\begin{figure*}[h!]
    \centering
    \includegraphics[width=0.7\linewidth]{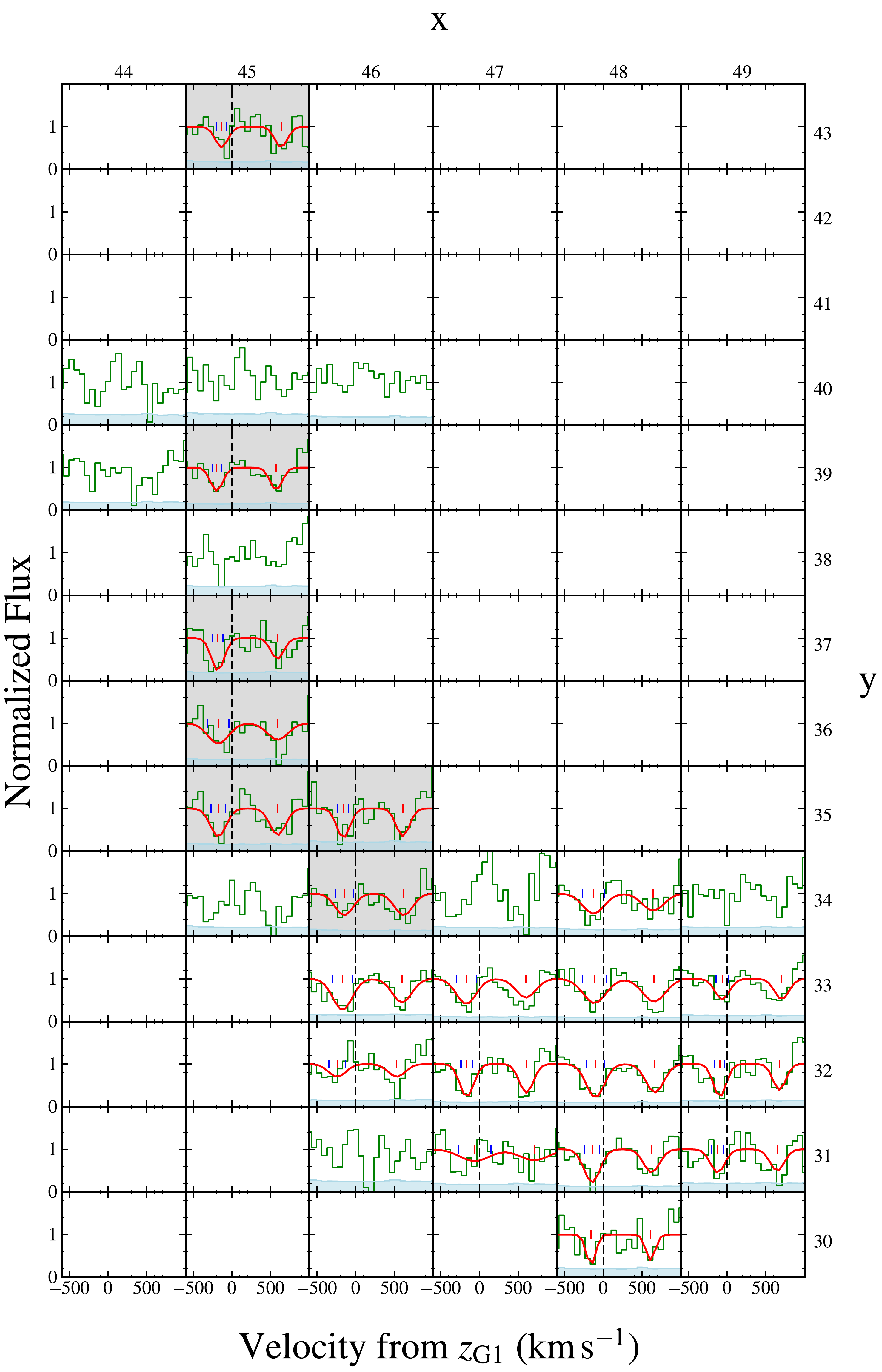}
    \caption{Spatial distribution of the \mgii{} spectra extracted over the arc images B1-B2 meeting the continuum S/N criteria in the image plane. Each panel shows the spectrum in a $4\times4$ ($0.8"\times0.8"$) binned spaxel. Green lines show the observed normalized flux, and the light blue line shows the corresponding $1\sigma$ error. Red curves show the fitted \mgii{} doublet when detected, and the black dashed vertical lines indicate the zero velocity of the \mgii{}~$\lambda 2796\text{\r{A}}$ transition at $z_{\rm G1}$. Spaxels with detected absorption are the same as in Fig.~\ref{fig:ew-map}. Grey-coloured panels indicate transverse sightlines with detected \mgii{} absorption.  } 
    \label{fig:spatial_distrib_fits}
\end{figure*}

\section{Spatial distribution of \feii{} spectra}
\begin{figure*}[h!]
    \centering
    \includegraphics[width=0.7\linewidth]{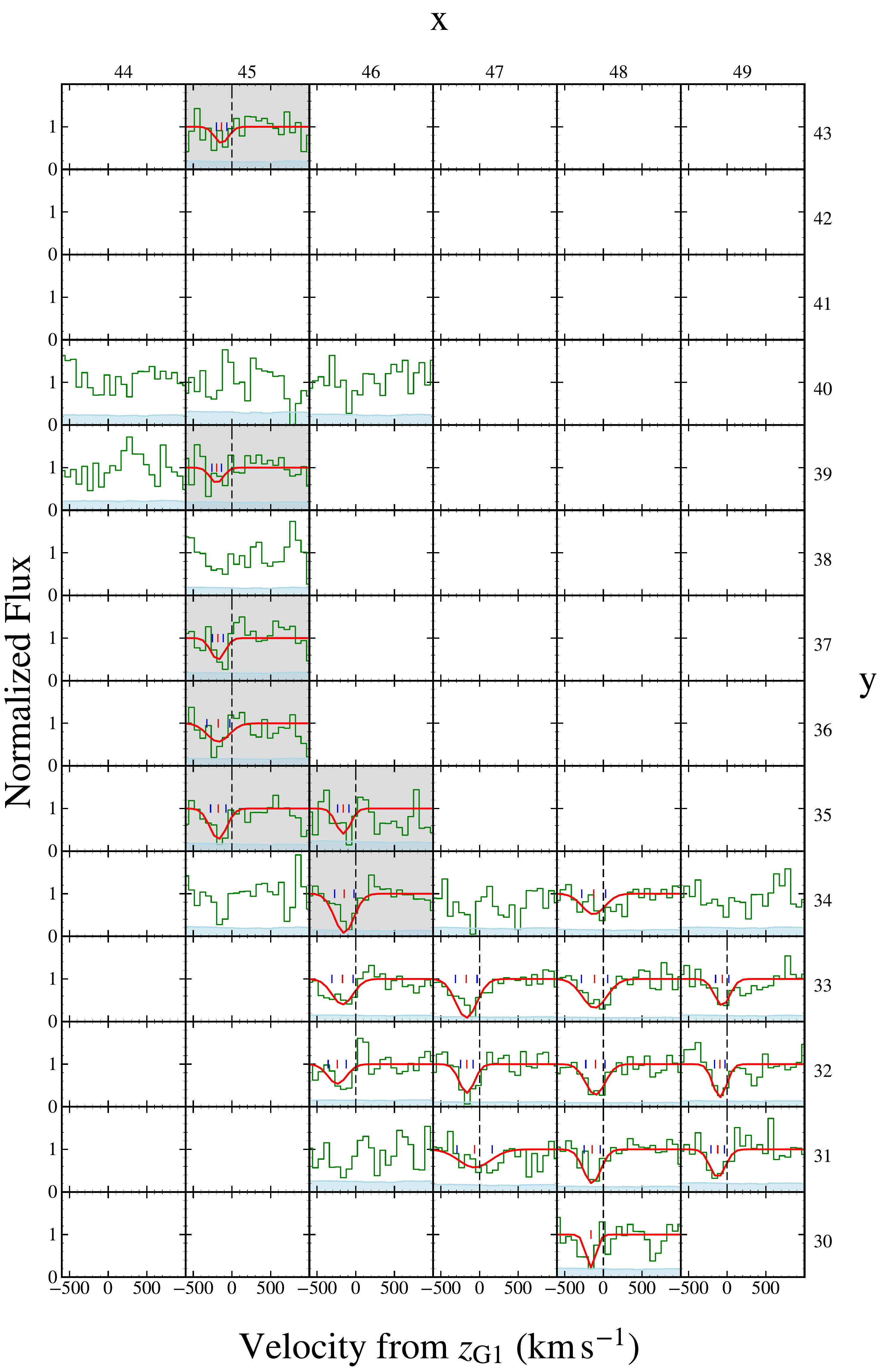}
    \caption{Same as Fig.~\ref{fig:spatial_distrib_fits}, but for the \feii{} spectra. Black dashed vertical lines indicate the zero velocity of the \feii{}~$\lambda 2600\text{\r{A}}$ transition at $z_{\rm G1}$. Grey-coloured panels indicate transverse sightlines with detected \mgii{} absorption. } 
    \label{fig:spatial_distrib_fits_feii}
\end{figure*}

\end{document}